\documentclass[10pt]{smfbook}
\usepackage{color}
\usepackage{amscd}
\usepackage[psamsfonts]{amssymb}
\usepackage{transfig,amssymb,amsmath}
\input{psfig}
\newtheorem{thm}{Theorem}[chapter]
\newtheorem{propos}{Proposition}[chapter]
\newtheorem{lem}{Lemma}[chapter]

\newtheorem{defin}{Definition}[chapter]
\newtheorem{proposdefin}{Definition-Proposition}
\newcounter{remarques}[chapter]
\newtheorem{conjecture}{Conjecture}[chapter]
\renewcommand{\theremarques}{\arabic{chapter}.\arabic{remarques}}
\def\Rm{\refstepcounter{remarques}{\bf Remark \theremarques} }
\def\ali{\hfill\break}
\def\w{\omega}
\def\W{\Omega}
\def\unN{\{1\ldots ,N\}}
\def\rrr{{\mathcal R}}
\def\demi{{1\over 2}}
\def\ddd{{\mathcal D}}
\def\ppp{{\mathcal P}}
\def\hhh{{\mathcal H}}
\def\lll{{\BL^G}}
\def\iii{{\mathcal I}}
\def\jjj{{\mathcal J}}
\def\aaa{{\mathcal A}}
\def\bbb{{\mathcal B}}
\def\eee{{{\mathcal E}}}
\def\nn{{<n>}}
\def\ND{{ND}}
\def\infi{{<\infty>}}
\def\intX{{\stackrel{\circ}{X}}}
\def\intF{{\stackrel{\circ}{F}}}

\def\unN{{\{1,\ldots ,N\}}}

\def\j1p0{{j_1^0,\ldots ,j_p^0}}

\def\jp1{{j_p,\ldots ,j_1}}
\def\j1p{{j_1,\ldots ,j_p}}

\def\BR{{\Bbb R}}
\def\supp{{\hbox{supp}}}
\def\BC{{\Bbb C}}
\def\BN{{\Bbb N}}
\def\BZ{{\Bbb Z}}

\def\BL{{\Bbb L}}

\def\BP{{\Bbb P}}

\def\nonu{\nonumber}
\def\ihat{{\hat\imath }}
\def\jhat{{\hat\jmath}}
\def\oeta{{\overline \eta}}

\def\ord{{\hbox{ord}}}
\def\beq{\begin{eqnarray}}
\def\eeq{\end{eqnarray}}
\def\beqn{\begin{eqnarray*}}
\def\eeqn{\end{eqnarray*}}
\def\bpr{\begin{propos}}
\def\epr{\end{propos}}

\def\loc{{\hbox{loc}}}

\def\cF{{\vert F\vert}}
\def\symF{{\hbox{Sym}_F}}
\def\symFR{{\hbox{Sym}_{F,{{\Bbb R}}}}}

\def\symG{{\hbox{Sym}^G}}
\def\Sym{{\hbox{Sym}}}
\def\sym{{\hbox{Sym}}}
\def\LG{{\BL^G}}
\def\11{{(1,1)}}

\def\Hunl{{H^{1,1} (\lll)}}
\def\Hun {{H^{1,1}}}

\def\Id{{\hbox{Id}}}

\def\ozeta{{\overline{\zeta}}}
\def\im{\hbox{Im}}
\def\re{\hbox{Re}}
\def\ksi{{\xi}}

\def\vect{{\hbox{vect}}}
\def\degree{{\hbox{degree}}}
\def\Qu01{{Q^{u_0,u_1}}}
\def\opsi{{\overline \psi}}
\def\pq{{(p,q)}}

\begin{document}
\centerline{\Large \bf Spectral properties of self-similar}
\centerline{\Large\bf lattices and iteration of rational maps}
\ali\ali
 \centerline{\Large\bf Propri\'et\'es spectrales des r\'eseaux
 auto-similaires}
 \centerline{\Large\bf et it\'eration d'applications rationelles}
\ali \ali \centerline{\Large\bf Christophe Sabot} \ali
\begin{center}
Laboratoire de Probabilit\'es et mod\`eles al\'eatoires,
Universit\'e Paris 6,
\\
4, Place Jussieu, 75252 Paris cedex 5, \\
and
\\
Ecole Normale Sup\'erieure,
\\
DMA, 45, rue d'Ulm, 75005 Paris, \footnote{E-mail address:
sabot@ccr.jussieu.fr}
\end{center}
\ali \ali \ali {\bf Abstract:} In this text we consider discrete
Laplace operators defined on lattices based on finitely-ramified
self-similar sets, and their continuous analogous defined on the
self-similar sets themselves. We are interested in the spectral
properties of these operators. The basic example is the lattice
based on the Sierpinski gasket. We introduce a new renormalization
map which appears to be a rational map defined on a smooth
projective variety (more precisely, this variety is isomorphic to
a product of three types of Grassmannians: complex Grassmannians,
Lagrangian Grassmannians, orthogonal Grassmannians). We relate some
characteristics of the dynamics of its iterates with some
characteristics of the spectrum of our operator. More
specifically, we give an explicit formula for the density of
states in terms of the Green current of the map, and we relate the
indeterminacy points of the map with the so-called
Neumann-Dirichlet eigenvalues which lead to eigenfunctions with
compact support on the unbounded lattice. Depending on the
asymptotic degree of the map we can prove drastically different
spectral properties of the operators. Our formalism is valid for
the general class of finitely ramified self-similar sets (i.e. for
the class of pcf self-similar sets of Kigami, cf \cite{Kigami1}).
Hence, this work aims at a generalization and a better
understanding of the initial work of the physicists Rammal and
Toulouse on the Sierpinski gasket (cf \cite{RammalT},
\cite{Rammal}).
 \ali\ali\ali
{\bf R\'esum\'e:} Dans ce texte, nous consid\'erons le Laplacien
discret, d\'efini sur un r\'eseau construit \`a partir d'un
ensemble auto-similaire finiment ramifi\'e, et son analogue
continu d\'efini sur l'ensemble auto-similaire lui-m\^eme. Nous
nous int\'eressons aux propri\'et\'es spectrales de ces
op\'erateurs. L'exemple le plus classique est celui du triangle de
Sierpinski (Sierpinski gasket) et du r\'eseau discret associ\'e.
Nous introduisons une nouvelle application de renormalisation qui
se trouve \^etre une application rationnelle d\'efinie sur une
vari\'et\'e projective lisse (plus pr\'ecis\'ement, cette
vari\'et\'e est un produit de Grassmaniennes de trois types:
Grassmaniennes classiques, Grassmaniennes Lagrangiennes, Grassmaniennes
orthogonales). Nous
relions certaines propri\'et\'es spectrales de ces op\'erateurs
avec la dynamique des it\'er\'es de cette application. En
particulier, nous donnons une formule explicite de la densit\'e
d'\'etats en termes du courant de Green de l'application, et nous
caract\'erisons le spectre de Neumann-Dirichlet (qui correspond aux
fonctions propres \`a support compact sur l'ensemble infini) \`a
l'aide des points d'ind\'etermination de l'application. Suivant le
degr\'e asymptotique de l'application nous pouvons prouver que les
propri\'et\'es spectrales de l'op\'erateur sont tr\`es
diff\'erentes. Notre formalisme s'applique \`a la classe des
ensembles auto-similaires finiment ramifi\'es (ou autrement dit
\`a la classe des "pcf self-similar sets" de Kigami, cf
\cite{Kigami1}). Ainsi, ce travail g\'en\'eralise et donne une
compr\'ehension plus profonde des r\'esulats obtenus initialement
par Rammal et Toulouse dans le cas du triangle de Sierpinski (cf
\cite{RammalT}, \cite{Rammal}).
 \ali \ali \ali \ali {\bf AMS classification:}
82B44(32H50,28A80)
 \ali\ali\ali {\bf Key words:} Spectral theory
of Schr\"odinger operators, pluricomplex analysis, dynamics in
several complex variables, electrical networks, analysis on
self-similar sets, fractal graphs.

 \vfill\break {\footnotesize
\tableofcontents} \ali \ali

\chapter*{Introduction}
\setcounter{section}{0}

In this text we investigate the spectral properties of Laplace
operators defined on hierarchical lattices based on finitely
ramified self-similar sets, and their continuous analogs.  The
basic example is the lattice based on the Sierpinski gasket. These
operators have much to do with the operators considered in the
context of Schr\"odinger operators with random or quasi-periodic
potential. Here, the disorder is not in the potential but in the
lattice itself. It is well-known that in the context of
Schr\"odinger operators on the line the spectral properties are
intimately related to the dynamics of the propagator of the
underlying differential equation (cf, for example,
\cite{CarmonaL}, \cite{PasturF}). In comparison, in our models we
will show that the characteristics of the spectrum of our operator
are related to the dynamics of the iterates of a certain
renormalization map that we explicitly define and that appears to
be a rational self-map of a compact complex manifold.

The interest in such lattices and in their spectral properties
comes from physicists (cf \cite{RammalT}, \cite{Rammal},
\cite{Alexander} and \cite{Bellissard}) because they present
interesting computable models, with peculiar properties. In
\cite{RammalT}, \cite{Rammal}, on the particular lattice based on
the Sierpinski gasket, Rammal and Toulouse discovered interesting
relations between the spectrum of the discrete Laplace operator
and the dynamics of the iteration of some rational map on $\BC$.
More precisely, they exhibited a polynomial map on $\BC$ that
relates the spectrum of the operator on successive scales: they
remarked that if $\lambda$ is an eigenvalue at level $n+1$ then
$\lambda(5-\lambda)$ is an eigenvalue at level $n$. Traditionally,
this law was called the spectral decimation of the Sierpinski
gasket, i.e. this terminology reflects the existence of a
1-dimensional map that relates the spectrum of the operator on
successive scales. Starting from this, Rammal (\cite{Rammal}) gave
a fairly complete description of the spectrum of the discrete
operator on this lattice. In particular, he computed explicitly
the eigenvalues and showed the existence of the so-called
molecular states (that we call Neumann-Dirichlet eigenfunctions in
this text) which are eigenfunctions with compact support. This was
made rigorous and generalized to the continuous operator defined
on the Sierpinski gasket itself by Fukushima and Shima (cf
\cite{FukuShima1}). The spectral type of the operator on the
Sierpinski lattice, has been analyzed by Teplyaev, cf
\cite{Teplyaev}.

In general, the spectral decimation that works for the Sierpinski
gasket is not valid, and the question of generalizing the initial
work of Rammal remained unsolved. In \cite{FukuShima2} a class of
lattices for which the spectral decimation works is exhibited. In
\cite{Sabot3}, for the particular example of a Sturm-Liouville
operator defined on $\BR$, the author made explicit some relations
between the spectral properties of the operator and the properties
of the dynamics of the iterates of a rational map; this map is no
longer 1-dimensional but is defined on the 2-dimensional
projective space.

This text aims at a generalization of these previous works.
Besides the interest of the generalization, this brings new
understanding of the models. In particular, the renormalization
map involved is now multidimensional and  certain notions which
are specific to the dynamics in higher dimension and which were
hidden in the case of the Sierpinski gasket (where the
renormalization map involved was 1-dimensional), such as the
notion of indeterminacy points (which corresponds to the
singularities of the map), the degree of the iterates, enter the
discussion and play an important role. In comparison with our
previous work, \cite{Sabot3}, the main progress that allows us to
handle the general case is the construction of a new
renormalization map. This renormalization map is a rational map
defined on some compact K\"ahler manifold. It is of the type of
the maps considered in \cite{Favre1}, \cite{DFavre}, and our
techniques rely heavily on recent works of Fornaess Sibony, Diller
Favre, Guedj (cf \cite{Sibony1}, \cite{FSibony2}, \cite{Favre1},
\cite{DFavre}, \cite{FavreG}) on the dynamics of rational maps in
higher dimensions. It is interesting to note that many of the key
notions in this field (such as the degrees of the iterates, the
indeterminacy points, the Green current)  find a significance
related to the spectral properties of our operators. In
particular, we are able to give an explicit expression for the
density of states in terms of the Green current of the map and we
prove that the molecular states of Rammal (called Neumann-Dirichlet
eigenvalues in the text) correspond exactly to the indeterminacy
points of the map.
 \ali

Since the text is long, we first describe the model and our
results on the particular example of the lattice associated with
the Sierpinski gasket. Let $F\subset \BC$, $F=\{0,1,\demi+i{\sqrt
3\over 2}\}$, be the vertices of a unit triangle, and
$\Psi_1,\Psi_2,\Psi_3$ be the three homotheties with ratio $\demi$
and centers the points $0,1,\demi+i{\sqrt 3\over 2}$, repectively.
It is well-known that there exists a unique proper subset $X$ of
$\BC$ self-similar with respect to $\Psi_1,\Psi_2,\Psi_3$, i.e.
such that $X=\cup_{i=1}^3 \Psi_i(X)$, and that it is the
celebrated Sierpinski gasket, represented on the following
picture.
 \ali 
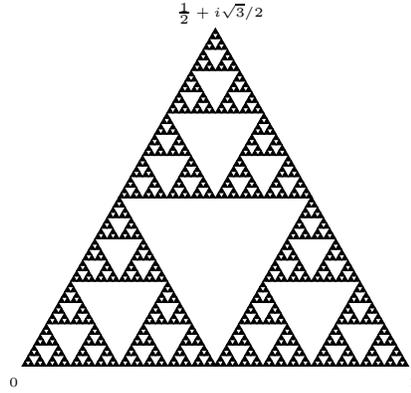
\centerline{\begin{picture}(0,0)%
\includegraphics{gasket.pstex}%
\end{picture}%
\setlength{\unitlength}{1973sp}%
\begingroup\makeatletter\ifx\SetFigFont\undefined%
\gdef\SetFigFont#1#2#3#4#5{%
  \reset@font\fontsize{#1}{#2pt}%
  \fontfamily{#3}\fontseries{#4}\fontshape{#5}%
  \selectfont}%
\fi\endgroup%
\begin{picture}(5025,4770)(3676,-5386)
\put(3676,-5386){\makebox(0,0)[lb]{\smash{\SetFigFont{5}{6.0}{\familydefault}{\color[rgb]{0,0,0}0}%
}}}
\put(8701,-5386){\makebox(0,0)[lb]{\smash{\SetFigFont{5}{6.0}{\familydefault}{\color[rgb]{0,0,0}1}%
}}}
\put(5776,-736){\makebox(0,0)[lb]{\smash{\SetFigFont{5}{6.0}{\familydefault}{\color[rgb]{0,0,0}$\demi+i\sqrt{3}/2$}%
}}}
\end{picture}
}
  \ali
Fix now a sequence $\w\in \{1,2,3\}^\BN$, called the blow-up, and
define $X_{<0>}=X$ and
$$X_\nn =\Psi_{w_1}^{-1}\circ\cdots \circ \Psi_{w_n}^{-1}(X).$$
It is clear that $X_\nn$ is an increasing sequence of sets and
that $X_{<n+1>}$ is a scaled copy of $X$ that contains $X_{\nn}$
as one of the three subcells; more precisely, we have
$X_\nn=\Psi_{\w_1}^{-1}\circ \cdots
\Psi_{\w_{n+1}}^{-1}(\Psi_{\w_{n+1}}(X))$, which is clearly a
subset of $X_{<n+1>}$. Remark that the position of the cell
$X_{<p>}$ in $X_\nn$ for $n>p$ depends on the blow-up $\w$. We
then set
$$X_\infi=\cup_{n=0}^\infty X_\nn.$$
We define the boundary of $X_\nn$ by $\partial X_{<0>}=F$ and
$$\partial X_\nn=\Psi_{w_1}^{-1}\circ\cdots \circ \Psi_{w_n}^{-1}(F).$$

There is a natural discrete sequence of lattices associated with
this structure. The lattice at level 0 is $F_{<0>}=F$, the
vertices of the unit triangle in $X_{<0>}$. The lattice at level
$n$, is the set of vertices of the unit triangles in $X_\nn$. More
precisely,
$$F_\nn=\Psi^{-1}_{\w_1}\circ \cdots \circ
\Psi_{\w_n}^{-1}(\cup_{j_1,\ldots ,j_n} \Psi_{j_1}\circ\cdots\circ\Psi_{j_n}(F)).
$$
The position of $F_{<0>}$ in the lattice at level $n$
depends on $\w$, and
we represent on the following picture
the lattice at level 4, $F_{4}$.
The bolded small triangle  is the set $F_{<0>}$ for the blow-up
starting from $(\w_1,\ldots ,\w_4)=(1,1,1,1)$ on the left and
$(1,3,1,2)$ on the right.
The sequence $F_\nn$ is increasing and we set
$$F_\infi=\cup_{n=0}^\infty F_\nn,$$
and $\partial F_\nn=\partial X_\nn$.
 \ali \ali
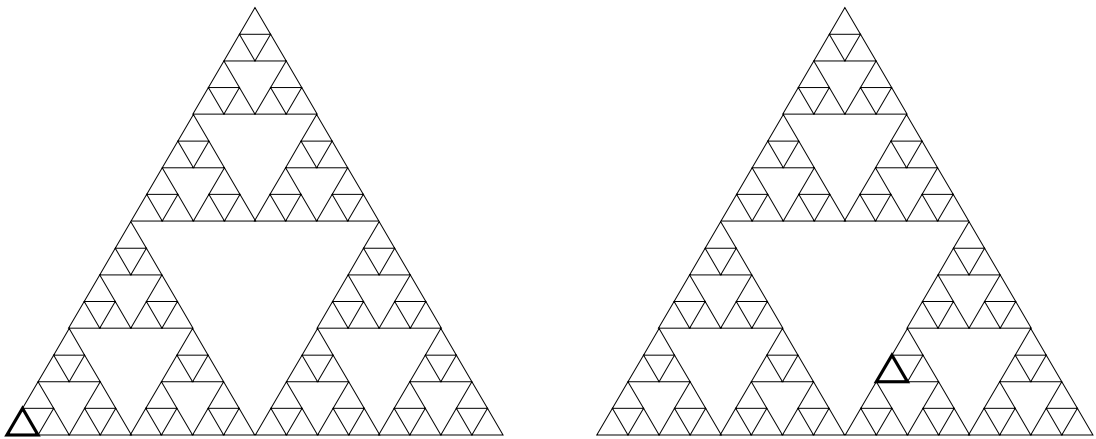
\centerline{\begin{picture}(0,0)%
\includegraphics{figure2.pstex}%
\end{picture}%
\setlength{\unitlength}{2565sp}%
\begingroup\makeatletter\ifx\SetFigFont\undefined%
\gdef\SetFigFont#1#2#3#4#5{%
  \reset@font\fontsize{#1}{#2pt}%
  \fontfamily{#3}\fontseries{#4}\fontshape{#5}%
  \selectfont}%
\fi\endgroup%
\begin{picture}(10539,4178)(1774,-4983)
\end{picture}
}
 \ali
It is important to realize that the infinite lattices
$F_{<\infty>}$ obtained from different blow-ups $\w$ and $\w'$ are
a priori not isomorphic (except when $\w$ and $\w'$ are equal
after a certain level). To understand this, one can compare the
constant blow-up $(1,\ldots ,1,\ldots)$ with a non-stationary
blow-up: the first one contains a point with only 2 neighbors
(which is the point 0, center of the homothetie $\Psi_1$), on the
second one  all points have 4 neighbors (indeed, the boundary
points $\partial F_\nn$ are sent to infinity when $n$ goes to
infinity).

The aim of this text is to investigate the spectral properties of
some natural Laplace operator defined either on the infinite
lattice $F_\infi$ or on the unbounded  set $X_\infi$. The class of
lattices or self-similar sets we consider is issued from the class
of finitely-ramified self-similar sets  (also called p.c.f.
self-similar sets in \cite{Kigami1}) described in section 1.1, and
is much larger than the Sierpinski gasket. Although the classical
examples have a natural geometrical embedding, these sets are
defined abstractly from a very simple finite structure: one starts
from a finite set $F$ and one constructs $F_{<1>}$ as the union of
$N$ copies of $F$, glued together according to a prescribed rule
(represented by an equivalence relation $\rrr$ on $\unN\times F$),
then $F_{<2>}$ is defined as the union of $N$ copies of $F_{<1>}$
glued together according to the same rule, and so on. From this
discrete structure, one can construct an increasing sequence of
sets $F_\nn$, and also a self-similar set $X$ (cf section 1.2 for
precise definitions).

To take into account the eventual symmetries of the picture, we
fix a group of symmetries acting on each $F_\nn$ (but in general
not on $F_\infi$). For the Sierpinski gasket we can see that the
group $G\sim S_3$ ($S_3$ denotes the group of permutation of $F$)
of isometries of the regular triangle $\partial F_\nn$ leaves
globally invariant the lattice $F_{<n>}$. We fix this group $G$ as
the group of symmetries of  the structure (i.e. this means that we
will only consider $G$-invariant objects).
 \ali
Note that for consistency with the notations of the main text, we
denote by $N$ the number of subcells of $F_{<1>}$. Here, we have
$N=3$. \ali

We now define the type of operators we will consider in this text.
We restrict to the discrete setting in this introduction and we
present the definitions only in the case of the Sierpinski gasket.
On $F_\nn$ we define the difference operator $A_\nn$ as the
operator on $\BR^{F_\nn}$ defined by \beq \label{f.0.0.1} A_\nn
f(x)=-\sum_{y\sim x} (f(y)-f(x)),\;\;\;\forall f\in \BR^{F_\nn},
\eeq where $y\sim x$ means that $y$ is in the same triangle of
unit size as $x$. The measure $b_\nn$ is defined as the measure
which gives mass 1 to the points of $\partial F_\nn$ and mass 2 to
the points in $F_\nn\setminus \partial F_\nn$. The choice of this
particular operator and of this measure comes from the fact that
they are self-similar in the sense that $A_\nn$ (resp. $b_\nn$) is
the sum of $N^n$ copies of the operator $A_{<0>}$ (resp. the
measure $b_{<0>}$) on each of the small triangles of $F_\nn$).
\ali The operators we will be interested in are the operators
$H_\nn^+$ defined on $\BR^{F_\nn}$ by \beq \label{f.0.0.2} H_\nn^+
f(x) =-{1\over b_\nn(\{x\})} A_\nn f(x),\;\;\;\forall f\in
\BR^{F_\nn},\;\;\forall x\in F_\nn. \eeq It is clear that
$H_\nn^+$ is self-adjoint on $L^2(F_\nn,b_\nn)$ and semi-negative.
The operator with Dirichlet boundary condition on $\partial
F_\nn$, $H_\nn^-$, is defined as the restriction to $\{f\in
\BR^{F_\nn}, \; f_{|\partial F_\nn}=0\}$ of $H^+_\nn$, and is
self-adjoint on $L^2(F_\nn\setminus \partial F_\nn,b_\nn)$. The
measure $b_\nn$ and the operators $H_\nn^\pm$ naturally extend to
a measure $b_\infi$ on $F_\infi$ and to semi-negative self-adjoint
linear operators $H_\infi^\pm$ on $L^2(F_\infi,b_\infi)$.

There are two measures which play a crucial role in this text. The
first one is the classical density of states: for each $n$, denote
by $\nu_\nn^\pm$ the counting measures of the eigenvalues of
$H_\nn^\pm$. The density of states is defined as the limit
$$\mu=\lim_{n\to\infty} {1\over N^n} \nu_\nn^\pm.$$
(In general, this measure exists and does not depend on the
boundary condition.) The second measure is the measure that counts
the asymptotic number of the so-called Neumann-Dirichlet
eigenvalues: we say that a function is a Neumann-Dirichlet
eigenfunction (or N-D eigenfunction for short) of $H_\nn$ with
eigenvalue $\lambda$ if it is both an eigenfunction of $H_\nn^+$
and $H^-_\nn$ (with eigenvalue $\lambda$), i.e. if it is an
eigenfunction of $H_\nn^+$ null on $\partial F_\nn$. These
particular eigenfunctions play an important role since, when
extended to $F_\infi$ by 0, they are eigenfunctions with compact
support of the operators $H_\infi^\pm$ on the infinite lattice.
One can define the counting measure of N-D eigenvalues
$\nu_\nn^\ND$ and show that $\nu_{<n+1>}^\ND\ge N \nu^\ND_\nn$.
This implies that the limit
$$\mu^\ND=\lim_{n\to\infty} {1\over N^n} \nu_\nn^\ND
$$
exists and is pure point.
We call it the density of N-D eigenvalues.

As we said before, the lattice $F_\infi$ depends on the particular
choice of the blow-up $\w$, hence the topological spectrum
$\Sigma^\pm$ of $H_\infi^\pm$ and the Lebesgue decomposition,
$\Sigma_{ac}^\pm$, $\Sigma_{sc}^\pm$, $\Sigma_{pp}^\pm$,  of the
spectrum of the operator on the infinite lattice $H_\infi^\pm$ a
priori depend on $\w$. By contrast, the measures $\mu$ and
$\mu^\ND$ do not depend on $\w$ (since the measures $\nu^\pm_\nn$
and $\nu^\ND_\nn$ obviously do not depend on $\w$). In
\cite{Sabot6}, we proved basic results regarding the spectral
properties of the operators $H^\pm_\infi$ and, in particular, on
their relations with the measures $\mu$ and $\mu^\ND$. First we
showed that the topological spectrum $\Sigma^\pm(\w)$ and the
Lebesgue decomposition $\Sigma_{ac}^\pm(\w)$,
$\Sigma_{sc}^\pm(\w)$, $\Sigma_{pp}^\pm(\w)$ of the operators
$H_\infi^\pm(\w)$ (we write $\Sigma(\w),\ldots$ to show the a
priori dependence in the blow-up $\w$) are constant almost surely
in the blow-up $\w$, for the measure on $\{1,\ldots ,N\}^\BN$
equal to the product of the uniform measure on $\{1,\ldots, N\}$.
Hence, we can talk about the almost sure spectrum and the almost
sure Lebesgue decomposition of the spectrum of the operator
$H^\pm_\infi$. We also proved that the almost sure spectrum is
equal to the support of the measure $\mu$, i.e. that we have
$\Sigma^\pm (\w)=\supp (\mu)$ for almost all blow-up $\w$
(actually, our result is more precise than this). Finally, we
showed that when the density of states is completely created by
the N-D eigenvalues, i.e. when $\mu^\ND=\mu$, then the spectrum of
$H_\infi^\pm$ is pure point with compactly supported
eigenfunctions, almost surely in $\w$.

So, the measures
$\mu$ and $\mu^\ND$ give important information on the
spectral properties of the operators $H_\infi^\pm$.
Hence, these two measures deserve to be understood and our aim in this
text is to describe the relations that exist between these measures
and the dynamics of a certain renormalization map that we construct.

There are two renormalization maps, closely related, which play an
important role. We do not define precisely these maps in this
introduction, we just describe some of their properties. The first
one, denoted by $T$, is defined on the space $\symG$, the space of
$G$-invariant symmetric operators on $\BC^F$, and is a rational
map (i.e. $TQ$, for $Q$ in $\symG$, is rational in the
coefficients of $Q$). For example, in the case of the Sierpinski
gasket $\symG\sim \BC^2$: indeed, if $W_0$ is the space of
constant functions on $F$ and $W_1$ its orthogonal complement (for
the natural scalar product on $\BC^F$), then any element of
$\symG$ can be written thanks to 2 coordinates $(u_0,u_1)\in
\BC^2$ under the form $u_0p_{|W_0}+u_1p_{|W_1}$ where $p_{|W_0}$
and $p_{|W_1}$ are the orthogonal projections respectively on
$W_0$ and $W_1$. We do not define explicitly the map $T$ in this
introduction (cf section 2 and 3.1), but this map is easy to
compute; for example, in the case of the Sierpinski gasket, in
coordinates $(u_0,u_1)$ we have $T(u_0,u_1)=3({u_0u_1\over
2u_0+u_1},{u_1(u_0+u_1)\over 5u_1+u_0})$. This map is the one that
was considered in earlier work of the author, \cite{Sabot3}; it is
also very closely related to the renormalization map that was
introduced initially by Rammal and Toulouse in the case of the
Sierpinski gasket. The iterates of this map contain some
information on the spectrum of the operators on the n-th level
lattice $H_\nn^\pm$. This explains why it was useful in the
understanding of the spectral properties of these operators.
 \ali
Nevertheless, this map  fails to give enough information. The main
progress in this text is the construction of a new renormalization
map defined on a bigger space: more precisely, it is defined on a
projective space that contains $\symG$ as a smooth subvariety and
it coincides with $T$ on $\symG$. We consider two sets of
variables $(\oeta_x)_{x\in F}$ and $(\eta_x)_{x\in F}$, and the
Grassman algebra generated by these variables (i.e. the algebra
generated by $(\oeta_x)_{x\in F}$ and $(\eta_x)_{x\in F}$, with
the relation of anticommutation between all these variables). We
denote by $\aaa$ the subalgebra generated by the monomials
containing the same number of variables $\oeta$ and $\eta$. We
also denote by $\ppp(\aaa)$ the projective space associated with
$\aaa$ and by $\pi:\aaa\rightarrow \ppp(\aaa)$ the canonical
projection. We set
$$\oeta Q\eta =\sum_{x,y\in F^2} Q_{x,y} \oeta_x \eta_y$$
 for
any
$F\times F$ matrix $Q$.
Then there is a natural embedding of $\symG$ into $\ppp(\aaa)$ given by
$$Q\rightarrow \pi(\exp \oeta Q\eta),$$
where $\exp$ denotes the exponential of the Grassman algebra. We
denote by $\BL^G$ the closure in $\ppp(\aaa)$ of $\symG$ (i.e. of
the points of the type $\pi(\exp \oeta Q\eta)$). We can show (cf
section 2.2) that $\BL^G$ is a smooth analytic subvariety of
$\ppp(\aaa)$ and it defines a compactification of $\symG$. For
example, in the case of the Sierpinski gasket $\BL^G$ is equal to
$\BP^1\times \BP^1$ (and indeed, $\BP^1\times \BP^1$ is a
compactification of $\BC^2$). The key point in our work is the
following: we construct a homogeneous polynomial map
$R:\aaa\rightarrow \aaa$ of degree $N$. This map naturally induces
a rational map on the projective space $\ppp(\aaa)$: a fundamental
property of this map is that it leaves invariant the subvariety
$\BL^G\subset \ppp(\aaa)$, and thus induces by restriction a map
$g:\BL^G\rightarrow \BL^G$. Actually, the map $g$ coincides with
$T$ on the subset $\symG$ of $\BL^G$. More precisely, we have the
following formula
$$g(\pi(\exp(\oeta Q\eta)))=\pi(R(\exp(\oeta Q\eta))),$$
when both expressions are well-defined. Hence, the restriction of
the map $g$ to $\BL^G$ extends $T$ to the compactification $\BL^G$
of $\symG$. The measures $\mu$ and $\mu^\ND$ are related to the
properties of the map $R$ and $g$, and in particular, to the
dynamics of the restriction of $g$ to $\BL^G$.

Let us now state our main results.
We define the Green function of $R$ (introduced by Fornaess and Sibony,
cf \cite{Sibony1} or appendix B) as the function
$G:\aaa\rightarrow \BR\cup\{-\infty\}$ given by
$$G(x)=\lim_{n\to\infty} {1\over N^n}\ln \| R^n(x)\|,
\;\;\;\forall x\in \aaa.$$
The limit exists for all $x$ in $\aaa$ and is a plurisubharmonic
function (this essentially means that
$G$ is subharmonic when restricted to a complex line).
This function $G$
contains important information on the dynamics
of the map induced by $R$ on $\ppp(\aaa)$.
 We denote by $\rho_n(x)$, for $x\in\lll$, the order of vanishing of the
restriction of the function $R^n$ to $\lll$ (cf section 3.2).
Since $R$ is homogeneous of degree $N$, we have $\rho_{n+1}(x)\ge N\rho_n(x)$
and we set
$$\rho_\infty(x)=\lim_{n\to\infty}{1\over N^n} \rho_n(x).
$$
We define $\phi:\BC\rightarrow \aaa$ by $\phi(\lambda)=\exp \oeta
(A_{<0>}-\lambda \Id )\eta$. Theorem (\ref{t.3.1}) gives the
following explicit expressions for $\mu$ and $\mu^\ND$: \beq
\label{f.0.1} \mu={1\over 2\pi} \Delta G\circ \phi,
\\
\label{f.0.2}
\mu^\ND=\sum_\lambda \rho_\infty (\phi(\lambda)) \delta_\lambda,
\eeq
where $\Delta$ in (\ref{f.0.1}) is the distributional Laplacian,
and $\delta_\lambda$ in (\ref{f.0.2}) is the Dirac mass at $\lambda$
($\rho_\infty (\phi(\lambda))$ is null except on a countable
set of $\lambda$'s, so that the sum (\ref{f.0.2}) is well-defined).

In section 4, we investigate the structure of the Green function
on $\pi^{-1}(\lll)$. This is important since the Green function
$G_{|\pi^{-1}(\BL^G)}$ is related to the dynamics of the map $g$.
On the other hand, we see from (\ref{f.0.1}) and (\ref{f.0.2})
that it is also related to the measures $\mu$ and $\mu^\ND$. The
function $G_{|\pi^{-1}(\lll)}$ is the potential of a unique
closed, positive, $(1,1)$-current on $\lll$: precisely, if $s$ is
a local section of the projection $\pi$ on an open subset
$U\subset \lll$ the current $dd^c G\circ s$ does not depend on $s$
and defines a positive closed current on all $\lll$ (cf appendix,
we recall that
$$dd^c G\circ s= {i\over \pi} \partial \overline \partial (G\circ s)=
{i\over \pi} \sum_{i,j} {\partial ^2\over \partial z_i\partial \overline z_j}
(G\circ s)dz_i\wedge d\overline z_j,
$$
where the derivatives are taken in the sense of distributions).
The current $S$ is intimately related to the dynamics of the
map $g$ and to the structure of the measures $\mu$ and $\mu^\ND$.
With the iterates of $g$ we can associate an asymptotic degree $d_\infty$
(called the dynamical degree, cf section 4.3)
and we show in theorem (\ref{t.4.1}) that the
following dichotomy holds:
\begin{itemize}
\item
If $d_\infty<N$, then $S$ is a countable sum of currents of
integration on hypersurfaces of $\lll$. In this case
$\mu^\ND=\mu$, thus the spectrum is pure point with compactly
supported eigenfunctions, almost surely in $\w$. (We also show
that the number of Neumann only eigenvalues, i.e. $\vert
\nu_\nn^+-\nu_\nn^\ND\vert$, grows like $n^{d_\infty}$).
\item
If $d_\infty=N$, then the current $S$ does not charge
hypersurfaces; it is the Green current of the map $g$. In
particular, it is null on the Fatou set of $g$. Moreover,
generically (in a sense made precise in theorem (\ref{t.4.1})), we
have $\mu^\ND=0$, i.e. there does not exist N-D eigenfunctions.
\end{itemize}
Note that a similar dichotomy theorem was shown
in \cite{FukuShima2} for the particular class of decimable fractals,
for which there exists a 1-dimensional renormalization map
that relate the spectrum on different scales (but the relation with
the N-D spectrum was not made).
\ali

Let us now make some remarks. The renormalization map we consider
here is not the same as the one considered in our previous work
\cite{Sabot3}. The introduction of this new map is the key point
that allows us to handle the case of lattices based on general
finitely ramified self-similar sets. In particular, this map is
not defined on a projective space. This induces several
difficulties, for example, the notion of degree is more
complicated and is related to the action of the map on some
cohomology groups. But there are several facts that seem to
indicate that the map we consider is the good renormalization map:
the first one is that the indeterminacy points of $g$, which have
a crucial influence on the dynamical properties of $g$, have a
clear meaning in terms of the Neumann-Dirichlet eigenvalues of our
operator. The second is that the map $g$ behaves well in the
non-degenerate case, i.e. when $d_\infty=N$: to be more precise,
the map $g$ is algebraically stable (cf definition in appendix),
and this allows us to define the Green current; this is not the
case for other maps we could consider (cf section 4.5 where we
compare different renormalization maps) and which are birationally
equivalent to $g$. Finally, let us mention that considering
expressions like $\exp(\oeta Q\eta)$ in the Grassman algebra is
very natural in the context of supersymmetry, and that the theory
of supersymmetry appeared to be very useful in the context of
random Schr\"odinger operators (cf for example \cite{Klein},
\cite{Wang}, \cite{Wang2}).

The important question of determining the type of spectrum of the
operator on the infinite lattice remains largely open: with our
techniques we are only able to characterize a small part of the
spectrum, namely the part of the spectrum which corresponds to
compactly supported eigenfunctions (i.e. N-D eigenfunctions). It
would be very interesting to determine the almost sure type of the
spectrum of the operator (for example, to determine whether it is
continuous or purely punctual) in terms of characteristic of the
dynamics of the map (for example, in the spirit of Kotani's
theorem where the Lyapounov exponent can characterize the type of
the spectrum). There are very few examples where results in this
direction are known: in the case of  the Sierpinski gasket
Teplyaev, cf \cite{Teplyaev}, gave fairly complete results
(depending on the blow-up); for a self-similar Sturm-Liouville
operator on $\BR$ we investigate the type of the spectrum for
different blow-ups, cf \cite{Sabot7}. Another interesting question
would be to consider random potential on $F_{\infi}$ (or random
fractal lattices as in \cite{Hambly}) and to determine whether
Anderson localization occurs as for 1-dimensional Schr\"odinger
operators. \ali

Let us now describe the organization of the paper. In the first
part, we introduce the models and recall three elementary results
obtained in \cite{Sabot6}, concerning the spectrum and the density
of states. In the second part, we give some preliminary results,
which are crucial in the rest of the text. We introduce the
Grassmann algebra and the Lagrangian Grassmanian. The third part
is devoted to the proof of the main formulas (\ref{f.0.1}) and
(\ref{f.0.2}). In the fourth part, we analyze the structure of the
current $S$ on $\BL^G$. Finally, in part 5 we illustrate our
results by several examples. In appendix A, B, C, we recall some of
the results from pluricomplexe analysis and rational dynamics that
we need in the text. In appendix E, we describe the topological structure
of $G$-Lagrangian Grassmanians.

We treat both the lattice case and the case of operators defined
on continuous self-similar sets. For a reader not familiar with
the subject it is better, upon a first reading, to skip the
discussion of the continuous case which is of a more technical
nature.
 \ali

I am very grateful to Nessim Sibony and Charles Favre for many
very helpful discussions. I also thank P. Bougerol for a useful
reference and the referee for many valuable advices.

\chapter{Definitions and basic results}
\setcounter{section}{0}
\section{Finitely ramified self-similar sets and associated lattices}
We introduce here an abstract definition of the notion of finitely
ramified self-similar sets and of the associated lattices.
Although all classical examples have a natural geometrical
representation we choose to present the abstract setting since the
procedure of construction is simple and natural.  Let us describe
it briefly: first choose two integers $N$ and $N_0$, $1< N_0\le
N$. The basic cell is $F_{<0>}=F=\{1,\ldots ,N_0\}$. The set as
level 1 is defined as the union of $N$ copies of $F$, glued
together according to a prescribed law $\rrr$ (formally defined as
an equivalent relation on $\unN\times F$). In $F_{<1>}$ we define
the boundary set $\partial F_{<1>}$ as the set of points of the
type $(x,x)$ for $x\in F$ which can be identified with $F$ (if
$\rrr$ satisfies some minor properties). Then we define the set at
level 2, $F_{<2>}$, as the union of $N$ copies of $F_{<1>}$ glued
together by their boundary points, $\partial F_{<1>}$, according
to the law $\rrr$, and so on. To define the infinite lattice we
blow-up the structure, according to a sequence $\w$ in $\unN^\BN$,
i.e. at each level, $F_{<n>}$ is the sublattice $\w_{n+1}$ of
$F_{<n+1>}$. To construct the self-similar set we refine the
structure instead of blowing it up. Let us now present precise
definitions.

\subsection{The lattice case}
Let $N$ and $N_0$ be two  positive integers such that $1<N_0\le
N$. We set $F=\{1,\ldots ,N_0\}$. The set $F$ will represent the
basic cell and $N$ the number of cells at level 1. We suppose
given an equivalence relation $\rrr$ on $\unN\times F$ (this
equivalence relation will describe the connections in the set at
level 1).  For some reasons that will appear clearly later we
assume that $\rrr$ satisfies
\begin{itemize}
\item
$(i,x)\rrr (i,y)$ implies $x=y$.
\item
The class of $(i,i)$ for $i$ in $F$ is a singleton.
\item
For any $(i,i')$ in $\unN$ there exists a sequence $i_1=i, \ldots i_p=i'$ of $\unN$ such that
for all $k\le p-1$ there exists $j$ and $j'$ such that  $(i_k,j)\rrr (i_{k+1},j')$.
\end{itemize}
We first give the formal definition of the infinite set $F_\infi$
and of its subsets $F_\nn$. The lattice structure on these sets
will be apparent only in section 1.2 and will be inherited from
the discrete Laplace operator we construct on these sets. Let us
fix an element $\w=(\w_1,\ldots ,\w_k,\ldots)$ in $\{1,\ldots
,N\}^\BN$ called the blow-up. As explained in the introduction,
the blow-up $\w$ describes how the sets $F_\nn$ fit into one each
other. Denote by $\tilde F_\infi$ the set of backward sequences
$(\ldots ,j_{-k},\ldots ,j_{-1}, x)$ in $\{1,\ldots ,N\}^\BN\times
F$ such that $j_{-k}=\w_k$ after a certain level. On $\tilde
F_\infi$ we define the relation $\rrr_\infi$ given by $(\ldots
,j_{-k},\ldots ,j_{-1},x)\rrr_\infi (\ldots ,j'_{-k},\ldots
,j'_{-1},x')$ if and only if there exists $k_0$ such that \beq
\label{f.1.0.0.1} j_{-k}=j'_{-k},\;\;\hbox{ for $k\ge k_0+1$},
\\
\label{f.1.0.0.2}
j_{-k}=x,\; j'_{-k}=x',\;\;\hbox{ for $k\ge 1$ and $k\le k_0-1$},
\\
\label{f.1.0.0.3}
\hbox{and } \;
(j_{-k_0},x)\rrr (j'_{-k_0},x').
\eeq
Using the second property of $\rrr$, we easily check that
$\rrr_\infi$ is an equivalence relation.
Then we define $F_\infi$ as the set $\tilde F_\infi$ quotiented by
$\rrr_\infi$.
The increasing sequence of
subsets $F_{<0>}\subset \cdots \subset F_\nn\subset \cdots \subset F_\infi$ is
defined by
$$F_\nn=\{(\cdots ,j_{-k},\cdots ,j_{-1},x)\in F_\infi, \hbox{ s.t.
$j_{-k}=\w_k$ for all $k\ge n$}\}.
$$
(N.B.: in the last expression and in the following, we simply write
$(\cdots ,j_{-k},\cdots ,j_{-1},x)$ for the class of $(\cdots ,j_{-k},\cdots ,j_{-1},x)$
in $F_\infi$).
It is clear that
$$F_\infi=\cup_{n=0}^\infty  F_\nn.$$
Denote by $\rrr_\nn$ the equivalence relation on $\unN^n\times F$
exactly as $\rrr$, i.e. by $(j_{-n},\ldots ,j_{-1},x)\rrr_\nn
(j'_{-n},\ldots ,j'_{-1},x')$ iff there exists $k_0\le n$ for
which (\ref{f.1.0.0.1}), (\ref{f.1.0.0.2}), (\ref{f.1.0.0.3}) are
satisfied. In the definition of $F_\nn$ we see that only the terms
$(j_{-n},\ldots ,j_{-1},x)$ count - the others are fixed to $\w_k$
- and it is easy to see that $F_{\nn}$ can be identified with
$\unN^n\times F/\rrr_\nn$. (More precisely, if $\tilde F_\nn$ is
defined as the set of points $(\ldots ,j_{-k},\ldots ,x)$ such
that $j_{-k}=\w_k$ for all $k\ge n$, then it is clear, thanks to
the first property of $\rrr$, that the natural bijection $\Theta:
\tilde F_\nn\rightarrow \unN^n\times F$, commutes with
$\rrr_\infi$ and $\rrr_\nn$, i.e. that $X\rrr_\infi Y$ iff $\Theta
(X)\rrr_\nn \Theta (Y)$. Hence, $F_\nn$ can be identified with
$\unN^n\times F/\rrr_\nn$.) For example, the set $F_{<0>}$ is
equal to $F$ and the set $F_{<1>}$ is equal to $\{1,\ldots
,N\}\times F/\rrr$. The boundary of the set $F_\nn$ is defined as
$$\partial F_\nn=\{(x,\ldots ,x)\in F_\nn
 ,\;\;\hbox{for some $x$ in $F=\{1,\ldots ,N_0\}$}\}.
$$
Thanks to the second property of $\rrr$, we see that
$\partial F_\nn$ can be identified with $F$
(i.e. the map $x\in F\rightarrow (x,\ldots ,x)\in \partial F_\nn$
is bijective).
The boundary set $\partial F_\infi$ is defined as
\beqn
\partial F_\infi
&=&\cap_{n=0}^\infty \cup_{m\ge n} \partial F_{<m>}
\\
&=& \{ (\ldots ,j_{-k},\ldots ,j_{-1},x)\in F_\infi \hbox{ s.t.
$j_{-k}=x$ for all $k\ge 0$}\}. \eeqn We set
$\intF_\nn=F_\nn\setminus \partial F_\nn$ and
$\intF_\infi=F_\infi\setminus \partial F_\infi$. \ali\ali
 \Rm: By
definition $\partial F_\infi\ne \emptyset$ if and only if $\w_k$
is stationary to a certain $x$ in $F$: in this case $\partial
F_\infi$ contains the unique point $(\ldots ,x ,\ldots ,x)$. \ali

For all $n$, $p$ and $\{i_1,\ldots ,i_p\}$ in $\{1,\ldots ,N\}^p$
we set
\begin{eqnarray*}
&&F_{<n+p>,i_1,\ldots ,i_p} \\
&=& \{(j_{-(n+p)},\ldots ,j_{-1},x)\in F_{<n+p>}, \;\hbox{ s.t.
$j_{-(n+p)}=i_1,\ldots ,j_{-(n+1)}=i_p$}\}.
\end{eqnarray*}
It is clear that
$$F_\nn=F_{<n+p>,\w_{n+p},\ldots ,\w_{n+1}}$$
and that
$$F_{<n+p>}=\cup_{i_1,\ldots ,i_p=1}^N F_{<n+p>,i_1,\ldots ,i_p}.$$
In this sense, $F_{<n+p>}$ is the (non-disjoint) union of $N^p$
copies of $F_{<n>}$. The subsets $F_{<n+p>,i_1,\ldots ,i_p}$ are
called the $<n>$-cells of $F_{<n+p>}$. It is clear that the
$<n>$-cells can only intersect by their boundary sets $\partial
F_{<n+p>,i_1,\ldots ,i_p}$ (with the obvious definition $\partial
F_{<n+p>,i_1,\ldots ,i_p}= \{(i_1,\ldots ,i_p,x,\ldots ,x)\in
F_{<n+p>},\hbox{ for $x\in F$}\}$).
 \ali\ali
 \Rm: At this point we did
not construct any lattice structure: the lattice structure will be
induced by the discrete Laplace operator we shall construct on
$F_\nn$ and $F_\infi$ (in fact, as we shall see, two points will
be neighbors for this operator if they belong to the same
$<0>$-cell). \ali

To take into account the eventual symmetries of the structure we suppose given
a finite group $G$ (eventually trivial) acting on $\{1,\ldots ,N\}$ and leaving
invariant the subset $F=\{1,\ldots ,N_0\}$. We suppose that the relation
$\rrr$ is $G$-invariant, for the action of $G$ on the product $\{1,\ldots ,N\}\times F$.
The relation $\rrr_\nn$ is then clearly $G$-invariant for the action of
$G$ on $\{1,\ldots ,N\}^n\times F$. Thus, the group $G$ acts  on the
quotient $F_\nn$, leaving globally invariant its boundary set $\partial F_\nn$
(remark that if we consider the action of $G$ on $F_{<n+1>}$ then it does not
leave the subset $F_\nn$ invariant in general. For this reason there is no
natural action of $G$ on the lattice $F_\infi$).
This symmetry group will play the following role: all the objects we will
consider will be $G$-invariant, in particular the discrete Laplace operator
we will construct on $F_\nn$.

\subsection{The continuous (or fractal) case.}
It is easy to construct a self-similar set
from the previous discrete structure.
The definition we introduce here is a bit less general
than the classical definition of p.c.f. self-similar sets
introduced by Kigami (cf \cite{Kigami1}), but a bit more
constructive.
Formally, we define $X$ as the set $\{1,\ldots ,N\}^\BN$ quotiented
by the equivalence relation
$\sim$ given by:
$(j_1,\ldots ,j_k,\ldots )\sim (j'_1,\ldots ,j'_k,\ldots )$
if and only if there exists an integer $k_0$ and
two elements $x$ and $x'$ in $F=\{1,\ldots ,N_0\}$ such that
\beqn
(j_1,\ldots ,j_{k_0-1})=(j'_1,\ldots ,j'_{k_0-1}),
\\
\hbox{$j_k=x$ and $j'_k=x'$
for all $k\ge k_0+1$,}
\\
\hbox{ and $(j_{k_0},x)\rrr (j'_{k_0},x')$.}
\eeqn
If we equip $\unN^\BN$ with the usual metric $d$ given by
$d((j_k),(j'_k))={1\over 2^{\inf\{k, j_k\ne j'_k\}}}$,
then
$X$ is compact for the quotient topology (and metrizable).
It is also easy to check that the third property of
$\rrr$ implies that $X$ is connected.
The boundary set of $X$ is define as the set
$$\partial X=\{(x,\ldots ,x,\ldots),\;\hbox{ for a $x$ in $F$}\}.$$
It is clear that $\partial X$ can be identified with $F$ thanks to the
second hypothesis on $\rrr$ (so we usually write $\partial X=F$).
We denote by
$\Psi_i\;:\; X\rightarrow X$ the application
$$\Psi_i((i_1,\ldots ,i_n,\ldots))= (i,i_1,\ldots ,i_n,\ldots).$$
Thanks to the first hypothesis on $\rrr$,
each $\Psi_i$ is injective.
We see that
$$X=\cup_{i=1}^N \Psi_i(X),$$
$$\Psi_i(X)\cap \Psi_j(X)=\Psi_i(F)\cap \Psi_i(F),\;\;\;\forall i\ne j.$$
Hence the set $X$ is self-similar with respect to the injections
$\Psi_i$. From the second relation we see that the connections
between the subsets $\Psi_i(X)$ of $X$  are contained in the image
of $F$ by the application $\Psi_i$: this justifies that we
consider $F$ as the boundary of the set $X$. We now construct an
infinite sequence of sets $X_{<0>}\subset  \cdots \subset
X_{<n>}\subset \cdots\subset X_\infi $ as in the discrete case.
Remind that we fixed $\w$ in $\unN^\BN$ called the blow-up. We
set:
$$\tilde X_\infi=\{ (j_k)\in \unN^\BZ,\hbox{ s.t. there exists $k_0\ge 0$
s.t. $j_{-k}=\w_k$ for all $k\ge k_0$}\}.
$$
Then we define $X_\infi$ as the quotient of $\tilde X_\infi$ by
the equivalence relation $\sim_\infi$ defined exactly as $\sim$, i.e.
by: $(j_k)_{k\in \BZ}\sim_\infi (j'_k)_{k\in \BZ}$ if and only if
there exists $k_0\in\BZ$ and two elements $x$ and $x'$ in $F$
\beqn
j_k=j_{k}' \hbox{ for all $k\le k_0-1$},
\\
\hbox{$j_k=x$ and $j'_k=x'$
for all $k\ge k_0+1$,}
\\
\hbox{ and $(j_{k_0},x)\rrr (j'_{k_0},x')$.}
\eeqn
Then we set for all $n\ge 0$
$$X_\nn=\{(j_k)\in X_\infi,\;\hbox{ s.t. $j_{-k}=\w_k$ for all $k\ge n$}\}.
$$
It is clear that for $X_\nn$ only the terms $(j_{-n},\ldots )$ counts, thus
the set $X_\nn$ can be considered as the set $\unN^{[-n,\infty)}$
quotiented by the equivalent relation induced by $\sim_\infi$
(and in the following we will represent the points in $X_\nn$
only by the sequence $(j_{-n},\ldots)$).
It is then clear that $X_{<0>}=X$
(actually all $X_\nn$ can be identified with $X$, just by shifting the
indices), and that
$$X_\infi=\cup_{n=0}^\infty X_\nn.$$
As in the discrete case we set
 \begin{eqnarray*}
 &&X_{<n+p>,i_1,\ldots ,i_p}
 \\
 &=&
\{(j_k)_{k=-(n+p)}^\infty \in X_{<n+p>},\;\hbox{ s.t.
$(j_{-(n+p)},\ldots ,j_{-(n+1)})= (i_1,\ldots ,i_p)$ }\}.
\end{eqnarray*}
It is clear that
$$X_\nn=X_{<n+p>,\w_{n+p},\ldots ,\w_{n+1}}$$
and that the set $X_{<n+p>}$ is the non-disjoint union
$$X_{<n+p>}=\cup_{i_1,\ldots ,i_p} X_{<n+p>,i_1,\ldots ,i_p}.$$

The boundary of $X_\nn$, is defined by
$$\partial X_\nn=\{(x,\ldots ,x,\ldots), \hbox{ for a $x$ in $F$}\}.$$
We set $\intX_\nn=X_\nn\setminus \partial X_\nn$.
Finally we set
\beqn
\partial X_\infi
&=&
\cup_{n=0}^\infty \cap_{m\ge n} \partial X_{<m>}
\\
&=& \{(j_k) \in X_\infi,\;\hbox{ s.t. $j_k=x$ for some $x$ in
$F$}\}, \eeqn and $\intX_\infi=X_\infi\setminus \partial X_\infi$.
 \ali\ali
 \Rm: The set $X_\nn$ (resp. $X_\infi$) contains the discrete
set $F_\nn$ (resp. $F_\infi$) as the set of sequences $(j_k)$ such
that $j_k=x$ for a $x$ in $F$ and all $k\ge 0$. With this
identification we clearly have $\partial X_\nn =\partial F_\nn$
and $\partial X_\infi=\partial F_\infi$.

\subsection{Geometric embedding.}
All classical examples come from self-similar sets which have a
natural embedding in $\BR^d$. We now describe how such a structure
appears in geometrical examples of self-similar sets. It is
essentially related to the property of finite ramification.
Suppose given $\Psi_1,\ldots ,\Psi_N$, $N$ strictly contractive
similitude of $\BR^d$ with different fixed points $x_1,\ldots
,x_N$. It is well-known that there exists a unique proper subset
$X$ of $\BR^d$ such that \beq \label{f.1.1} X=\cup_{i=1}^N
\Psi_i(X). \eeq The set $X$ is compact and actually equal to the
set of limits $\lim_{n\to\infty} \Psi_{i_1}\circ \cdots \circ
\Psi_{i_n} (y)$ for $(i_k)\in \unN^\BN$. It is easy to see that
the former limit does not depend on $y$, so that it defines a
mapping from $\unN^\BN$ onto $X$. Hence, $X$ can be written
$\unN^\BN/\sim$ for a certain equivalence relation $\sim$. We can
easily check
 that $\sim$ can be constructed
as previously if the set $X$ is connected and if there exists a
subset $F$ of the set of fixed points of $\Psi_1,\ldots ,\Psi_N$
such that \beq \label{f.1.2} \Psi_i(X)\cap \Psi_j
(X)=\Psi_i(F)\cap \Psi_j(F),\;\;\;\forall i\ne j. \eeq (This last
condition is usually called the condition of finite ramification.)
Indeed we can as well suppose that $F=\{x_1,\ldots ,x_{N_0}\}$ for
$N_0=\vert F\vert \le N$ and identify $F$ with $\{1,\ldots
,N_0\}$. We define the relation $\rrr$ on $\unN\times F$ by
$(i,j)\rrr (i',j')$ if and only if
$\Psi_{i}(x_{j})=\Psi_{i'}(x_{j'})$. It is then easy to see that
the relation $\sim $ we just defined is also the relation obtained
as in section 1.1.2 from the relation $\rrr$. \ali In this case
the sequences $X_\nn$ and  $F_\nn$  are naturally embedded in
$\BR^d$ as
$$X_\nn =\Psi_{\w_1}^{-1}\circ\cdots \circ\Psi_{\w_n}^{-1} (X),$$
$$F_\nn =
\Psi^{-1}_{\w_1}\circ\cdots \circ \Psi_{\w_n}^{-1}
(\cup_{j_1,\ldots ,j_n} \Psi_{j_1}\circ \cdots
\circ \Psi_{j_n} (F)).$$
The sets $\partial F_\nn$ and $\partial X_\nn$ are just scaled copies of
$F$, given by
$$\partial F_\nn=\partial X_\nn=\Psi_{\w_1}^{-1}\circ \cdots \circ \Psi_{\w_n}^{-1}(F).
$$

\subsection{Examples}
{\it The Sierpinski gasket} \ali In this case $N_0=N=3$,
$F=\{1,2,3\}$ and the relation $\rrr$ is given on the following
picture:
 \ali \ali
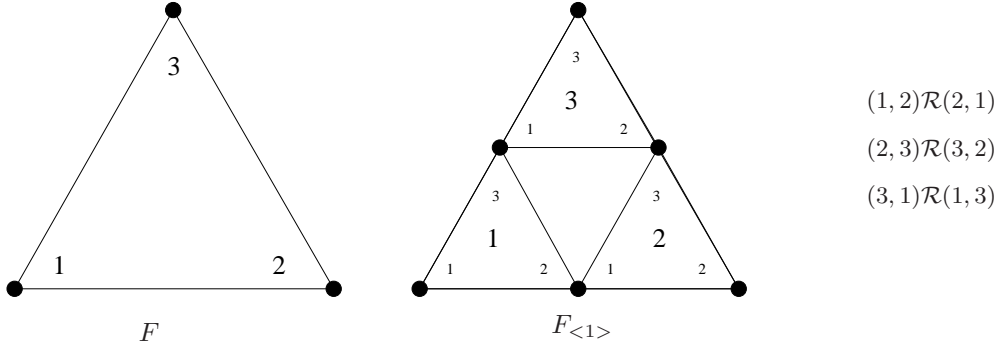
{\begin{picture}(0,0)%
\includegraphics{fig3-1f.pstex}%
\end{picture}%
\setlength{\unitlength}{3158sp}%
\begingroup\makeatletter\ifx\SetFigFont\undefined%
\gdef\SetFigFont#1#2#3#4#5{%
  \reset@font\fontsize{#1}{#2pt}%
  \fontfamily{#3}\fontseries{#4}\fontshape{#5}%
  \selectfont}%
\fi\endgroup%
\begin{picture}(6758,2662)(293,-2116)
\put(1336,-2116){\makebox(0,0)[lb]{\smash{\SetFigFont{10}{12.0}{\familydefault}$F$}}}
\put(7051,-661){\makebox(0,0)[lb]{\smash{\SetFigFont{9}{10.8}{\familydefault}$(2,3)\rrr (3,2)$}}}
\put(7051,-1036){\makebox(0,0)[lb]{\smash{\SetFigFont{9}{10.8}{\familydefault}$(3,1)\rrr (1,3)$}}}
\put(7051,-286){\makebox(0,0)[lb]{\smash{\SetFigFont{9}{10.8}{\familydefault}$(1,2)\rrr (2,1)$}}}
\put(4576,-2056){\makebox(0,0)[lb]{\smash{\SetFigFont{10}{12.0}{\familydefault}$F_{<1>}$}}}
\end{picture}
}
 \ali The
set at level 4 is represented on figure 2 in the introduction: the
initial cell $F_{<0>}$ is represented by the bolded triangle for
the blow-up starting from $(1,1,1,1)$ on the left and for the
blow-up starting from $(1,3,1,2)$ on the right. Usually we take
for $G$ the group of permutations of $F=\{1,2,3\}$ (i.e.
geometrically, $G$ acts on $F_\nn$ as the group of isometries of
the boundary triangle $\partial F_\nn$). But we could also
consider the trivial group as the group of symmetries (this is
considered in section 6). \ali As it is well-known, the Sierpinski
gasket is traditionally considered as a self-similar subset of
$\BC$. Let us now describe this and the relations with the
discrete structure we just introduced. Consider the 3 homotheties
\beqn \Psi_1(x)={x\over 2} ;\;\;\; \Psi_2(x)=\demi(x-1)+1;
\\
\Psi_3(x)=\demi(x-(\demi+i{\sqrt{3}\over 2}))+\demi+i{\sqrt{3}\over 2}.
\eeqn
There exists a unique proper subset of $\BC$ that satisfies
equation (\ref{f.1.1}) and it is the celebrated Sierpinski gasket
represented on picture 1 of the introduction.
Remark that the set of fixed points $F=\{0,1,\demi+i{\sqrt{3}\over 2}\}$
of $\Psi_1,\Psi_2,\Psi_3$ satisfies (\ref{f.1.2}) and defines the relation
$\rrr$ as explained in section 1.1.3 by $(i,j)\rrr(i',j')$
if and only if $\Psi_i(x_j)=\Psi_{i'}(x_{j'})$
(if we denote by $x_1,x_2,x_3$ the fixed points of
$\Psi_1,\Psi_2,\Psi_3$).
The natural geometric representation of $X_\nn$ is given by
the sequence of preimages
$$X_\nn =\Psi_{\w_1}^{-1}\circ\cdots \circ\Psi_{\w_n}^{-1} (X).$$
The set $F_\nn$ is the set of vertices of the triangles of size 1
in $X_\nn$. \ali {\it The unit interval} \ali Let us first
describe the continuous model, by its geometric representation.
Consider $X=[0,1]$ and a real $0<\alpha <1$. It is clear that $X$
is self-similar with respect to the $N=2$ homotheties
$\Psi_1(x)=\alpha x$ and $\Psi_2(x)=1+(x-1)(1-\alpha)$. Remark
that $F=\{0,1\}$ satisfies equation (\ref{f.1.2}): the abstract
self-similar set would be constructed from the equivalence
relation $\rrr$ such that $(1,2)\rrr(2,1)$. (Remark that when
$\alpha=\demi$ the abstract definition of the self-similar set
$[0,1]$ as a quotient of $\{1,2\}^\BN$ corresponds exactly to the
expression of a point in $[0,1]$ in base 2). We take for $G$ the
trivial group $G=\{\Id\}$. We remark that for different values of
$\alpha$ we just have different geometric representations of the
same self-similar structure as defined in 1.1.2. We will consider
on these sets different operators depending on $\alpha$, which
have a natural expression in these geometric representations.
Concerning the blow-up we remark that $\partial X_\infi $ is non
empty if and only if $\w$ is stationary. More precisely, if $\w$
is stationary to 1 (resp. to 2) then $X_\infi$ is a half-line
bounded from the left (resp. from the right) (for example, if
$\w_n=1$ then $X_\nn=[0,\alpha^{-n}]$). If $\w$ is not stationary
then $X_\infi =\BR$. \ali {\it The nested fractals} \ali The
nested fractals define a class of finitely ramified self-similar
sets, introduced by Lindstr\"om (cf \cite{Lindstrom}),
 embedded in $\BR^d$, which are invariant
by a large group of symmetries.
We refer to \cite{Lindstrom} for the definitions.
Note that the Sierpinski gasket is the basic example
of nested fractals.

\section{Construction of a self-similar Laplacian.}
We fix for the rest of the text two $N$-tuples
$(\alpha_1,\ldots ,\alpha_N)$
and $(\beta_1,\ldots ,\beta_N)$ of positive real numbers.
The $N$-tuple $(\alpha_1,\ldots ,\alpha_N)$, resp. $(\beta_1,\ldots ,\beta_N)$
will represent the scaling in energy, resp. in measure in our structure.
We suppose moreover that $(\alpha_1,\ldots ,\alpha_N)$, $(\beta_1,\ldots ,\beta_N)$
are $G$-invariant, i.e. that $(\alpha_{g\cdot 1},\cdots ,\alpha_{g\cdot N})=(\alpha_1,\ldots ,\alpha_N)$,
$(\beta_{g\cdot 1},\cdots ,\beta_{g\cdot N})=(\beta_1,\ldots ,\beta_N)$.
We set $\gamma_i=(\alpha_i \beta_i)^{-1}$ and
we make the following assumption
\ali

(H) We suppose that $(\beta_1,\ldots ,\beta_N)$ is
proportional to $(\alpha_1^{-1},\ldots ,\alpha_N^{-1})$
so that $\gamma_i$ does not depend on $i$. We denote by
$\gamma$ the common value of the $\gamma_i$.
\subsection{Discrete difference operators}
We suppose given $A$,  a semi-positive symmetric endomorphism  of
$\BR^F$ of the form \beq \label{f.1.5} Af(x)=-\sum_{y\in F, y\ne x
} a_{x,y} (f(y)-f(x)), \;\;\;\forall f\in \BR^F, \forall x\in F,
\eeq where $a_{x,y}$, $x\ne y$, are non negative reals such that
$a_{x,y}=a_{y,x}$. We suppose moreover that $A$ is irreducible,
i.e. that the graph on $F$ defined by strictly positive $a_{x,y}$
is connected and that $A$ is $G$-invariant, i.e. that $a_{g\cdot
x, g\cdot y}=a_{x,y}$ for all $g$ in $G$. We suppose also given a
$G$-invariant positive measure $b$ on $F$.
 \ali\ali
 \Rm: A typical
example is the discrete Laplace operator $Af (x)=-\sum_{y\ne x}
(f(y)-f(x))$ and $b$ the uniform measure on $F$. \ali \ali We
denote by $A_{<n>,i_1,\ldots ,i_n}$ the symmetric operator on
$\BR^{F_\nn}$ defined as the copy of $A$ on the cell $F_{\nn,
i_1,\ldots ,i_n}$, i.e. the operator defined for $f$ in
$\BR^{F_\nn}$ by
$$\left\{
\begin{array}{l}
(A_{<n>,i_1,\ldots ,i_n}f)_{|F_{\nn, i_1,\ldots ,i_n}}
=A(f_{|F_{\nn, i_1,\ldots ,i_n}}),
\\
A_{<n>,i_1,\ldots ,i_n}f(x)=0 \;\hbox{ if $x \not\in F_{\nn, i_1,\ldots ,i_n}$}.
\end{array}
\right.
$$
N.B.: In the first line we considered $f_{|F_{\nn, i_1,\ldots ,i_n}}$
as a function on $F$ since $F_{\nn, i_1,\ldots ,i_n}$ can be identified
with $F$.
\ali
We denote by $b_{\nn,i_1,\ldots ,i_n}$ the measure on $F_\nn$ defined as
the copy of $b$ on $F_{\nn, i_1,\ldots ,i_n}$, i.e. given by
$$\int_{F_\nn} f db_{\nn,i_1,\ldots ,i_n}=
\int f_{|F_{\nn, i_1,\ldots ,i_n}} db, \;\;\;\forall f\in
\BR^{F_\nn}.$$ Then we set
 \beq \label{f.1.6}
A_\nn =\sum_{i_1,\ldots ,i_n=1}^N \alpha_{\w_n}\cdots
\alpha_{\w_1} \alpha^{-1}_{i_1}\cdots \alpha_{i_n}^{-1}
A_{\nn,i_1,\ldots ,i_n},
\\
\label{f.1.7} b_\nn =\sum_{i_1,\ldots ,i_n=1}^N
\beta_{\w_n}^{-1}\cdots \beta_{\w_1}^{-1} \beta_{i_1}\cdots
\beta_{i_n} b_{\nn,i_1,\ldots ,i_n}. \eeq
 \Rm: \label{r.1.0} We
see from the definition that the value of $A_\nn$ and $b_\nn$
depend on the $N$-tuples $(\alpha_1,\ldots ,\alpha_N)$ and
$(\beta_1,\ldots ,\beta_N)$ only up to a constant. \ali \ali
Remark that $A_\nn$ and $b_\nn$ form an inductive sequence since,
for $n\ge p$, if $\supp (f)\subset \intF_{<p>}\cup(\partial
F_{<p>}\cap \partial F_\nn)$ then
$$A_\nn f = A_{<p>} f \;\;\;\hbox{and}\;\;\; \int f db_\nn =\int f db_{<p>}.$$
(Indeed, this comes from the fact that $F_{<p>}=F_{\nn,
\w_n,\ldots ,\w_{p+1}}$). Therefore $b_\nn$ can be extended to a
measure $b_\infi$ on $F_\infi$. Similarly, the linear operators
$A_\nn$ can be extended to a linear operator $A_\infi$ on
$\BR^{F_\infi}$ (a priori $A_\infi$ is defined on compactly
supported functions of $F_\infi$, but since there is only "local
interactions", $A_\infi$ can be extended to a linear operator on
$\BR^{F_\infi}$ itself). Remark that thanks to the third property
of $\rrr$ the operator $A_\nn$ is conservative, i.e. $A_\nn f=0$
is equivalent to $f$ constant. \ali

Denote by $<\cdot ,\cdot >$ the usual scalar product on
$\BR^{F_\nn}$.
Let $H_\nn^+$ be the operator on $L^2(F_\nn, b_\nn)$ defined by:
\beq
\label{f.1.8}
<A_\nn f,g>=- \int H_\nn^+ f g db_\nn\;\;\; \forall f,g\in \BR^{F_\nn}.
\eeq
The operator $H_\nn^+$ is semi-negative,
self-adjoint and must be viewed as a discrete difference operator
with Neumann boundary condition on $\partial F_\nn$ (since no condition is
imposed on the value of the functions on the boundary points).
The operator with Dirichlet boundary condition, denoted $H_\nn^-$, is the
self-adjoint operator on $\BR^{\intF_\nn}$ defined as the restriction
of $H_\nn^+$ to
$\BR^{\intF_\nn}\simeq \{f \in \BR^{F_\nn},\;\; f_{|\partial F_\nn}=0\}$.
We sometimes write $\ddd_\nn^+=\BR^{F_\nn}$ and $\ddd_\nn^-=\BR^{\intF_\nn}$
for the domain of $H_\nn^\pm$.

If $K>0$ is such that $<Af,f>\le K \int f^2 db$ for all $f$ in
$\BR^F$ then it is easy to see from (\ref{f.1.6}) and (\ref{f.1.7}) and
assumption (H)
that the same inequality is true for
$A_\nn$ and $b_\nn$. Thus the sequence $H_\nn^\pm$ is uniformly bounded
for the operator norm on  $L^2(b_\nn)$ and can be extended into a
semi-negative, self-adjoint operator $H_\infi^+$ on $\ddd_\infi^+=L^2(b_\infi)$.
We define $H^-_\infi$ as the restriction of $H_\infi^+$ to
$\ddd_\infi^-=\{f\in \ddd_\infi^+,\; f_{|\partial F_\infi}=0\}$.
Clearly, we have
$$<A_\infi f,g>=-\int H_\infi^\pm f g db_\infi,\;\;\; \forall f,g\in \ddd_\infi^\pm.
$$
Finally note that if $\partial F_\infi =\emptyset $ then the operators $H_\infi^+$
and $H_\infi^-$ are equal and in this case we simply write $H_\infi$ for
$H_\infi^+=H_\infi^-$.
\ali

Let us now explain the consequences of condition (H). Let $f$ be a
function with support contained in $\intF_\nn$. Denote by $\tilde
f$ the function on $F_\infi$ with support in $F_{<n+p>,i_1,\ldots
,i_p}$ and which is a copy of $f$ on $F_{<n+p>,i_1,\ldots ,i_p}$.
Then from formula (\ref{f.1.6}), (\ref{f.1.7}) and (\ref{f.1.8})
we see that \beq \label{f.1.9} (H_\infi^\pm \tilde
f)_{|F_{<n+p>,i_1,\ldots ,i_p}}= \gamma_{\w_{n+1}}\cdots
\gamma_{\w_{n+p}}(\gamma_{i_1}\cdots \gamma_{i_p})^{-1}
(H^\pm_\infi f)_{|F_\nn} \eeq and that $H_\infi^\pm \tilde f $ is
null on the complement of $F_{<n+p>,i_1,\ldots ,i_p}$. By (H) the
coefficients
 $\gamma_{\w_1}\cdots \gamma_{\w_n}(\gamma_{i_1}\cdots \gamma_{i_p})^{-1}$
are equal to 1, which means that $H^\pm_\infi$ is locally invariant by translation.
This property is the counterpart of the property of  statistical translation invariance
traditionally assumed in the case
of Schr\"odinger operator with random potential.
\subsection{The continuous situation}
In this section we define a "natural" Laplace operator on the
continuous sets $X_\nn$. The problem of the construction of such
an operator is not easy (cf for example, \cite{Lindstrom},
\cite{Kigami1}, \cite{Kusuoka}, \cite{Sabot1}) and it is now clear
that the best framework to use is the framework of Dirichlet
spaces. We essentially follow the definitions of \cite{Sabot1}. We
suppose here that $\sum_{i=1}^N \beta_i=1$ and that $\alpha_i<1$
for all $i$. We know that there exists a unique probability
measure $m$ on $X$ such that \beq \label{f.1.10.1} \int_X fdm
=\sum_{i=1}^N \beta_i \int_X f\circ \Psi_i dm. \eeq We suppose
given a $G$-invariant self-similar Dirichlet  form $(a,\ddd)$ in
the sense of \cite{Sabot1}, i.e. an irreducible, local,
conservative, regular  Dirichlet form on $(X,m)$ satisfying the
conditions of theorem 2.6. of \cite{Sabot1}. In particular,
$(a,\ddd)$ is self-similar with respect to the weights
$(\alpha_1,\ldots ,\alpha_N)$, i.e. for all $f\in\ddd$, $f\circ
\Psi_i$ is in $\ddd$ and \beq \label{f.1.10} a(f,f)=\sum_{i=1}^N
(\alpha_i)^{-1} a(f\circ \Psi_i,f\circ \Psi_i ), \eeq and is
$G$-invariant, i.e. $a(g\cdot f,g\cdot f)=a(f,f)$. In
\cite{Sabot1}, a criterion is given for the existence and
uniqueness of such a Dirichlet form. Remark that the weights
$(\alpha_1,\ldots ,\alpha_N)$ can be chosen only up to a constant,
since the scaling factor is determined by equation (\ref{f.1.10}).
The problem of the existence and uniqueness of such a Dirichlet
form is not trivial and has been investigated in \cite{Lindstrom},
\cite{Sabot1} (and references therein). This is related to the
existence of a fixed point for a certain renormalization map (that
corresponds to the restriction of the renormalization map $T$ that
we introduce in section 3 to a certain subset, cf remark
\ref{r.3.c.1}).

Remind that $X_\nn$ is isomorphic to $X$ so that we can
define the Dirichlet form $(a_\nn,\ddd_\nn)$ by
$\ddd_\nn=\ddd$ and $a_\nn=\alpha_{\w_1}\cdots \alpha_{\w_n} a$
and the measure
$b_\nn$ on $X_\nn$ by
$b_\nn=\beta_{\w_1}^{-1}\cdots \beta_{\w_n}^{-1} b$.
If $f$ in $\ddd_{<n+p>}$ is such that
$\supp (f)\subset \intX_\nn$ then using (\ref{f.1.10.1}) and (\ref{f.1.10}) we see that
$a_{<n+p>} (f,f)= a_\nn(f,f)$ and $\int f dm_\nn =\int fdm_{<n+p>}$.
Hence, we see that $m_{\nn}$ can be extended to a measure $m_{<\infty>}$ on
$X_\infi$, and we set
$$\ddd_\infi =\{f\in L^2(X_\infi, m_\infi),\;\; \sup_n a_\nn (f_{|X_\nn}, f_{|X_\nn})<\infty\}.
$$
On $\ddd_\infi$ we define $a_\infi$ by
$a_\infi (f,f)=\lim_{n\to\infty} a_\nn (f_{|X_\nn}, f_{|X_\nn})$.
One can check that $a_\infi$ is a local, regular, conservative and irreducible
Dirichlet form (cf \cite{Fukushima2}).
We set $\ddd_\nn^-=\{f\in \ddd_\nn,\;\; f_{|\partial X_\nn} =0\}$
and $\ddd^+_\nn =\ddd_\nn$ (and idem for $\ddd^\pm_\infi$).
We define $H_\nn^\pm$ and $H^\pm_\infi$ as the infinitesimal generators of
$(a_\nn,\ddd_\nn^\pm)$ and $(a_\infi,\ddd_\infi^\pm)$.
Note that they satisfy the same property of local invariance by translation
as in the discrete case, i.e. formula (\ref{f.1.9}) is still valid.
\subsection{Examples}
{\it The Sierpinski gasket} \ali If we take $G\sim S_3$ the group
of isometries of the unit triangle $F$, then the values of
$(\alpha_i)$ and $(\beta_i)$ are determined up to a constant, so
we can as well take in the lattice case $\alpha_i=1$, $\beta_i=1$.
There is only one possible choice for $A$ and $b$, up to a
constant: we take for $A$ the canonical discrete Laplace operator
on $F$ given by formula (\ref{f.1.5}) with $a_{x,y}=1$ if $x\ne
y$, and for $b$ the measure that gives a mass 1 to the points of
$F$. The operator $A_\nn$ is  obviously given by formula
(\ref{f.0.0.1}) of the introduction and $b_\nn$ is the measure
that gives a mass 1 to the points of $\partial F_\nn$ and 2 to the
points of $\intF_\nn$. Hence, the operator $H_\nn^\pm$ is the
operator defined by (\ref{f.0.0.2}) in the introduction. \ali In
the continuous case the construction of the Laplace operator was
initially done in \cite{BarlowP} by probabilistic means. There is
uniqueness of such an operator. The value of $(\alpha_i)$ and
$(\beta_i)$ is determined by equation (\ref{f.1.10.1}) and
(\ref{f.1.10}) to $\alpha_i={3\over 5}$, $\beta_i={1\over 3}$ (cf
for example, \cite{Kusuoka}). \ali {\it The unit interval} \ali In
the lattice case we can take any $\alpha_i$ and $\beta_i$ but
since they matter only up to a constant we take $\alpha_1=\alpha$,
$\alpha_2=(1-\alpha)$ and $\beta_1=1-\alpha$, $\beta_2=\alpha$
(hence, assumption (H) is satisfied). In the lattice case the only
possible choice for $A$ is, up to a constant, the discrete Laplace
operator
$$A=\left(\begin{array}{cc} 1&-1\\-1&1\end{array}\right).
$$
The measure $b$ is determined by two positive reals $m_0,m_1$ by
$b=m_0\delta_0+m_1\delta_1$.

In the continuous case we can explicitly construct the
self-similar structure. Let $m$ be the unique self-similar measure
on $X$ satisfying (\ref{f.1.10.1}). Consider on $[0,1]$ the
canonical Dirichlet form $a(f,f)=\int_0^1 (f')^2 dx$ defined on
$\ddd=\{ f\in L^2(m),\; f'\in L^2([0,1],dx)\}$. By a simple change
of variables it is clear that $a$ satisfies (\ref{f.1.10}) with
$\alpha_1=\alpha$ and $\alpha_2=1-\alpha$. Hence we have
explicitly constructed the self-similar Dirichlet space
$(a,\ddd)$, for all possible values of $\alpha$. The operator
$H^+=H^+_{<0>}$ is the operator ${d\over dm}{d\over dx}$ defined
on \beqn \{f\in L^2(X,m),\; \;  \exists g\in L^2(X,m), f(x)=ax+b+
&\int _0^x \int _0^y g(z) dm(z) dy, \nonumber \\
&f'(0)=f'(1)=0\},
\nonumber
\\
\hbox{by\;\;\; $H^+f =g$.} \eeqn Similarly $H^-$ is the operator
${d\over dm}{d\over dx}$ with Dirichlet boundary condition on
$\{0,1\}$. The author considered this case in \cite{Sabot3} and
\cite{Sabot7}.
 \ali
  {\it The nested fractals}
  \ali In general, for
nested fractals we take all the $\alpha_i$ equal, and the
$\beta_i$ equal (in the continuous case the exact value of the
$\alpha_i$ is given by the self-similar structure and the
$\beta_i$ must be equal to ${1\over N}$). There is nothing special
to say about the lattice case. In the continuous case, Lindstr\"om
and Kusuoka constructed the self-similar Dirichlet space (cf
\cite{Lindstrom}, \cite{Kusuoka}) and the author proved the
uniqueness of such a self-similar Dirichlet space (cf
\cite{Sabot1}).

\section{The density of states}
\subsection{Definition}
{\it The lattice case}

Denote by $0=\lambda^+_{<n>,1}>\lambda^+_{<n>,2}\ge \cdots\ge
\lambda_{<n>,\vert F_\nn\vert}^+$
the eigenvalues of
$H_{<n>}^+$.
Denote by
$0>\lambda^-_{<n>,1}\ge \cdots\ge
\lambda_{<n>,\vert \intF_\nn\vert}^-$
the Dirichlet eigenvalues,
i.e. the eigenvalues of
$H_{<n>}^-$.
\ali
Let $\nu_{<n>}^+$ (resp. $\nu_\nn^-$) be the counting
measures of the Neumann (resp. Dirichlet spectrum) defined
by:
\begin{eqnarray}
\label{f.1.15}
\nu_{<n>}^\pm=\sum_k \delta_{\lambda_{<n>,k}^\pm},
\end{eqnarray}
where $\delta_x$ stands for the Dirac mass at $x$.
We write $\nu^\pm_\nn(\lambda) =\int_\lambda^0 \nu_\nn^\pm (d\lambda)$,
$\lambda\le 0$,
for its repartition function.
\ali
It is clear by construction that the counting measures do
not depend on the blow-up $\w$
(since the operators $H_\nn^\pm$ are isomorphic
for different blow-up
$\w$).
\begin{defin}
\label{d.1.1}
If the limit
\begin{eqnarray}
\label{f.1.16}
\lim_{n\to \infty} {1\over N^n}
\nu_{<n>}^\pm
\end{eqnarray}
exists and does not depend on the choice of the boundary condition then it
is called the density of states
and denoted by $\mu$.
\end{defin}
\noindent\Rm The existence of the limit is proved in
\cite{Fukushima2} and in \cite{KigamiL} but will also be a
consequence of theorem (\ref{t.3.1}). \ali \Rm: The reader must be
careful that our terminology is not coherent with the classical
terminology of \cite{CarmonaL}, \cite{PasturF} and with the
terminology of our previous paper \cite{Sabot3} where the measure
$\mu$ is called the integrated density of states. Remark that
despite the terminology, $\mu$ is a measure which may have no
density. \ali \ali {\it The continuous case}

In this case the operator $H_\nn^\pm$ has compact resolvent (cf
for example, \cite{Kusuoka}) and the eigenvalues form a
non-increasing sequence going to $-\infty$. \ali We adopt the same
definition for the counting measures $\nu_{<n>}^\pm$ and for the
density of states as in the lattice case.
\begin{propos}
\label{p.1.1}
If the density of states exists then its
repartition function  $\mu(\lambda) =\int^0_\lambda \mu(\lambda) (d\lambda)$
satisfies:
\begin{eqnarray}
\label{f.1.17}
\mu(\gamma \lambda)=N \mu (\lambda).
\end{eqnarray}
\end{propos}
\noindent
Proof: This is clear from the scaling relations satisfied by $a_\nn$ and $m_\nn$.
$\Box$

\subsection{The density of Neumann-Dirichlet eigenvalues.}
We say that a function f is a Neumann-Dirichlet (N-D for short) eigenfunction
of $H_{<n>}$ with eigenvalue $\lambda$ if it is both a Dirichlet
and a Neumann eigenfunction (therefore we forget the supscript $\pm$
in $H_\nn$ since it is at the same time an eigenfunction of $H_\nn^+$ and $H_\nn^-$),
i.e. in the lattice case this means that
\begin{itemize}
\item
$f$ is in $\ddd_\nn^-$, i.e.
$f\in \BR^{F_{<n>}}$ and $f_{|\partial F_{\nn}}=0$,
\item
$<A_\nn f, g>=-\lambda \int f g db_{<n>},$ for all function $g$
in $\ddd_\nn^+=\BR^{F_{<n>}}$.
\end{itemize}
and in the continuous case that
\begin{itemize}
\item
$f$ is in $\ddd_{<n>}^-$,
\item
$a_\nn (f, g)=-\lambda \int f g dm_{<n>},$ for all function $g$ in $\ddd_{<n>}^+$.
\end{itemize}
We denote by $\nu^\ND_\nn$ the counting measure of the N-D eigenvalues of $H_\nn$
(counted with multiplicity) and by $E_\nn^\ND$ the subspace of $\ddd_\nn^+$
generated by the N-D eigenfunctions.

Remark that any function $f$ of $E_\nn^\ND$, when extended by 0 to
$F_{<n+p>}$ (resp. $X_{<n+p>}$) is a N-D eigenfunction of
$H_{<n+p>}$. When extended by 0 to $F_\infi$ (resp.  $X_\infi$)
it is an eigenfunction of $H_\infi^+$ and $H_\infi^-$, with
compact support. Hence $E^\ND_\nn$ is an increasing sequence of
subspaces of $\ddd_\infi^-$ and we denote by $\hhh_\ND$ the
closure in $\ddd_\infi^-$ of the space $\cup_n E_{<n>}^\ND$.
 \ali\ali
 \Rm: By definition, the restriction of $H_\infi^\pm$ to the
Hilbert subspace $\hhh_\ND$ is purely punctual, more precisely,
the set of Neumann-Dirichlet eigenfunctions form a Hilbert basis of
compactly supported eigenfunctions.
 \ali

It is easy to see that
$$\nu^\ND_{<n+1>}\ge N \nu_\nn^\ND.$$
Indeed, if $f$ is a N-D eigenfunction of $H_\nn$ then we can
construct $N$ copies of $f$ on the $N$ $\nn$-cells of $F_{<n+1>}$.
Precisely, for all $i=1,\ldots ,N$ we consider the function
$f_{i}$ on $\BR^{F_{<n+1>}}$ which is the copy of $f$ on
$F_{<n+1>,i}$ and equal to $0$ on $F_{<n+1>}\setminus
F_{<n+1>,i}$. These functions form an orthogonal family of N-D
eigenfunctions of $H_{<n+1>}$ with same eigenvalues (by the
hypothesis (H) and formula (\ref{f.1.9})).
\begin{defin}
\label{d.1.3}
The limit $${1\over N^n}\nu_\nn^\ND$$ exists and is called the density
of N-D eigenvalues and denoted by $\mu^\ND$.
\end{defin}
\noindent \Rm: Obviously, the measure $\mu^\ND$ is purely
punctual. It is clear that $\supp \mu^\ND$ is the topological
spectrum of the restriction of $H_\infi^\pm$ to the Hilbert
subspace $\hhh_\ND$. \ali \Rm: In the continuous case the
repartition function $\mu^\ND(\lambda)=\int_{\lambda}^0 d\mu^\ND$,
$\lambda\le 0$, satisfies the same scaling relation as
$\mu(\lambda)$: $\mu^\ND(\gamma \lambda)=N\mu^\ND(\lambda)$.

\section{Some basic results}
We recall from \cite{Sabot6} three basic results on the spectrum of the
operators $H_\infi^\pm$ and their relations with the measures
$\mu$ and $\mu^\ND$.
For convenience, we suppose here
the existence of the density of states.
We denote by $\Sigma^\pm$ the spectrum of the operators $H_\infi^\pm$
(and we simply write $\Sigma$ when $\partial F_\infi=\emptyset$).
 We recall that the essential spectrum is
obtained from the spectrum by removing all isolated points corresponding
to eigenvalues with finite multiplicity, we denote it by $\Sigma^\pm_{ess}$.
\begin{propos}
{\it (proposition 1, \cite{Sabot6})}
\ali
For both the lattice and the continuous case we have the following:

i) If the boundary set $\partial X_\infi=\partial F_\infi$ is
empty  then
$\supp \mu = \Sigma=\Sigma_{ess}$.

ii) Otherwise we just have
$\supp \mu =\Sigma^+_{\hbox{ess}}=\Sigma^-_{\hbox{ess}}$.
Moreover, the eigenvalues eventually lying in
$\Sigma^\pm\setminus \supp (\mu)$ have multiplicity 1.
\end{propos}
\noindent \Rm: In \cite{Sabot7}, we show that in the case of the
unit interval blown-up to the half-line $\BR_+$ (by the constant
blow-up $\w_k=1$) the spectrum of the operator can be pure point
with isolated eigenvalues of multiplicity  1 lying in the
complement of $\supp \mu$ and accumulating on $\supp \mu$.
Therefore in this case the equality $\Sigma_{\hbox{ess}}^\pm=\supp
\mu$ is satisfied by not $\Sigma^\pm=\supp \mu$. \ali

We endow $\W=\unN^\BN$ with the product of the uniform measure on
$\unN$. The next two propositions give almost sure results
on the blow-up.
Remind that the lattices $F_\infi$ (and the sets $X_\infi$)
are not isomorphic for different blow-ups.
Hence, to show the dependence of the operator $H_\infi^\pm$ and of the
spectrum $\Sigma^\pm$ on the blow-up we write $H_\infi^\pm(\w)$
and $\Sigma^\pm(\w)$.
We denote by $\Sigma_{ac}^\pm(\w)$, $\Sigma_{sc}^\pm(\w)$,
and $\Sigma_{pp}^\pm(\w)$ resp. the absolutely continuous, singular
continuous and pure point part of the Lebesgue decomposition
of the spectrum of $H_\infi^\pm(\w)$ (cf \cite{CarmonaL}, \cite{PasturF}
 for definition,
or \cite{Sabot6}).
The first result is the analogous of a classical result for
ergodic families of Schr\"odinger operators (cf \cite{CarmonaL} or
\cite{PasturF}).
\begin{propos}
{\it (proposition 2, \cite{Sabot6})}
\label{p.1.2.1}
\ali
There exists deterministic $\Sigma$, $\Sigma_{ac}$, $\Sigma_{sc}$,
and $\Sigma_{pp}$ such that for almost all $\w$ in $\W$
(for the product of the uniform measure on $\unN$)
we have $\Sigma_\cdot^\pm (\w)=\Sigma_\cdot $.
\end{propos}
\begin{propos}
\label{p.1.3}
{\it (proposition 3, \cite{Sabot6})}
\ali
If the density of states is completely created by the N-D eigenvalues, i.e. if
$\mu^\ND =\mu$ then for almost all $\w$ in $\W$
the set of N-D eigenfunctions is
complete i.e. $\hhh_\ND = \ddd_\infi^+(\w)=\ddd_\infi^-(\w)$.
Thus, the spectrum is pure point with compactly supported eigenfunctions.
\end{propos}

\chapter{Preliminaries.}
\setcounter{section}{0}
\section{The notion of trace on a subset}
Let $F$ be a finite set and $F'\subset F$ a subset.
\begin{defin}
Let $Q$ be a complex symmetric $F\times F$ matrix.
We denote by $Q_{|F'}$ the  restriction of $Q$ to $F'$, i.e.
the $F'\times F'$ matrix  defined by
$(Q_{|F'})_{x,y}=Q_{x,y}$ for $x,y$ in $F'$.
We call trace on $F'$ of the matrix $Q$,
the $F'\times F'$  matrix $Q_{F'}$, given,
when the expression is defined, by
$$Q_{F'}=\left( (Q^{-1})_{|F'}\right)^{-1}.$$
\end{defin}
\noindent N.B.: One must be careful that the close notations
$(Q_{|F'})$ and $(Q_{F'})$ represent two different types of
restriction.
 \ali N.B.: These definitions
could of course be given for non symmetric matrices but we will
only be concerned with the symmetric case.
 \ali \Rm: $Q_{F'}$ is
sometimes called the Schur complement and appears in several
circumstances, cf for example, \cite{Metz2}, \cite{Carlson}. In
\cite{Collin1} the properties of this operation are carefully
investigated, this operation is called "la reponse du r\'eseau".
\begin{propos}
\label{p.2.1}
i)
If $Q$ has the following block decomposition on
$F'$ and $F\setminus F'$
\begin{eqnarray*}
Q=
\left(
\begin{array}{cc}
Q_{|F'}&B\\
B^t& Q_{|F\setminus F'}
\end{array}
\right)
\end{eqnarray*}
then
\begin{eqnarray}
\label{f.2.1} Q_{F'}=Q_{|F'}-B(Q_{|F\setminus F'})^{-1}B^t.
\end{eqnarray}
Therefore the map $Q\rightarrow Q_{F'}$ is rational in the
coefficients of $Q$ with poles included in the set $\det
(Q_{|F\setminus F'})=0$.

ii) If $\det(Q_{|F\setminus F'})\neq 0$, then for any function $f$
in $\BC^{F'}$, we denote by $Hf$, the function of $\BC^F$ given by
 \beqn \left\{
\begin{array}{l}
Hf=f \;\;\hbox{on $F'$,}
\\
Hf=-(Q_{|F\setminus F'})^{-1}B^t f\;\;\hbox{on $F\setminus F'$.}
\end{array}
\right.
 \eeqn
We call $Hf$ the harmonic prolongation of $f$ with respect to $Q$
and we have $Q_{F'}(f)=(Q(Hf))_{|F'}$.

iii) If moreover $Q$ is real semi-positive then $Q_{F'}$ is
characterized by
\begin{eqnarray}
\label{f.2.2}
<Q_{F'}f,f>= \inf_{g\in \BR^F, \; g_{|F'}=f} <Qg,g>, \;\;\; \forall f\in \BR^{F'},
\end{eqnarray}
where $<\cdot ,\cdot >$ denotes the usual scalar product resp. on
$\BR^{F'}$ and $\BR^F$. The infimum is reached at the unique point
$Hf$.
\end{propos}
\Rm: The terminology comes from the theory of Dirichlet forms: if
$Q$ is semi-positive and such that $<Q\cdot, \cdot >$ is a
Dirichlet form (i.e. $Q$ is Markovian, cf for example,
\cite{Sabot1}) then $<Q_{F'}\cdot, \cdot>$ is a Dirichlet form
called the trace of $<Q\cdot, \cdot>$ on $F'$ (cf
\cite{Fukushima1}). In particular, we remark that the Markov
property  is preserved by the operation of taking the trace. \ali
Proof: i) Let $f$ be a function in ${\BC}^F$ null on $F\subset
F'$. Set $g=Q^{-1}f$. We easily get
$$g_{|F\setminus F'}=-(Q_{|F\setminus F'})^{-1} B^t g_{|F'},$$
and
 \beqn f_{|F'}&=&Q_{|F'} g_{|F'} +Q_{|F\setminus F'}
g_{|F\setminus F'}
\\&=&
(Q_{|F'}-B(Q_{|F\setminus F'})^{-1} B^t)g_{|F'}.
 \eeqn
By definition $((Q^{-1})_{|F'}f_{|F'})=g_{|F'}$. This implies that
$f_{|F'}=Q_{F'}(g_{|F'})$ and thus formula (\ref{f.2.1}).
 \ali
  ii) It is an immediate consequence of i).
  \ali
  iii) Classically, $g$ realizes the infimum in (\ref{f.2.2}) if
and only if $(Qg)_{|F\setminus F'}=0$, i.e. if $B^t f+
Q_{|F\setminus F'} g_{|F\setminus F'}=0$. If $Q_{|F\setminus F'}$
is invertible then $g$ is unique and given on $F\setminus F'$ by
$g_{|F\setminus F'} =-(Q_{|F\setminus F'})^{-1}B^t f$. This
implies that $g=Hf$ and thus concludes the proof of the
proposition. $\Box$
\section{The Grassmann algebra}
The operation of taking the trace of a symmetric matrix on a subset
is central in our problem. It is
a complicated operation since it is rational.
However we can embed the space of symmetric matrices in a Grassmann algebra
in such a way that this operation becomes linear. This will be crucial
in our work in order to construct a good renormalization
map. This is the key tool we use to generalize some of our previous works
(cf \cite{Sabot3}).
\subsection{Definition}
As in 2.1, $F$ is a finite set and $\vert F\vert $ denotes its
cardinality, most of the time we identify $F$ with $\{1, \ldots,
\vert F\vert \}$. Consider $\overline E$ and $E$ two copies of
$\BC^F$, with canonical basis $(\oeta_x)_{x\in F}$ and
$(\eta_x)_{x\in F}$. We consider the Grassmann algebra $\bigwedge
(\overline E\oplus E)$ defined by
$$\bigwedge (\overline E\oplus E)=\bigoplus_{k=0}^{2\vert F\vert}
(\overline E\oplus E)^{\wedge k},$$ where $\wedge$ denotes the
exterior product. We denote  by $\aaa$ the subalgebra generated by
the monomials containing the same number of variables $\oeta$ and
$\eta$, i.e.
$$
\aaa =\oplus_{k=0}^\cF \overline E^{\wedge k} \wedge E^{\wedge k}.
$$
A canonical basis of $\aaa$ is
$$(1, \oeta_{i_1}\wedge \cdots \wedge \oeta_{i_k}\wedge \eta_{j_1}\wedge \cdots
\wedge \eta_{j_{k}},\; i_1<\cdots <i_k,\; j_1<\cdots <j_{k},\;
1\le k\le \cF ).$$ We endow $\aaa$ with $<\cdot ,\cdot >$ the
scalar product which makes this basis an orthonormal basis. To
simplify notations we will forget the sign $\wedge$ to denote the
exterior product and simply write $\eta_i\eta_j$ for $\eta_i\wedge
\eta_j$. Remark that the elements of $\aaa$ commute since $\aaa$
is generated by the monomials of even degrees.

If $Q$ is a $F\times F$ matrix then we denote $\oeta Q\eta$ the element
of $\aaa$:
$$\oeta Q\eta =\sum_{i,j\in F} Q_{i,j} \oeta_i \eta_j.$$
We will be particularly interested in terms of the type
\begin{eqnarray}
\nonu
\exp (\oeta Q \eta )&=&
\sum_{k=0}^n
{1\over k!} \left(\sum_{i,j} Q_{i,j} \oeta_i \eta_j \right)^k
\\ \label{f.2.2002}
&=&
\sum_{k=0}^n \sum_{{i_1<\cdots <i_k \atop j_1<\cdots <j_k}}
\det \left( (Q)_{{i_1,\ldots ,i_k \atop j_1,\ldots ,j_k}}
\right)
\oeta_{i_1} \eta_{j_1}\cdots \oeta_{i_k} \eta_{j_k},
\end{eqnarray}
where $(Q)_{{i_1,\ldots ,i_k \atop j_1,\ldots ,j_k}}$ is the
$k\times k$ matrix obtained from $Q$ by keeping only the lines
$i_1,\ldots ,i_k$ and the columns $j_1,\ldots ,j_k$.
\begin{lem}
\label{l.2.0}
Let $Q$ be a complex $\vert F\vert \times \vert F\vert$ matrix, then
\begin{eqnarray}
\label{f.2.3} \| \exp \oeta Q\eta \|^2 &=& \det (\Id +QQ^*)
\\
&=&  \Pi_{i=1}^\cF (1+\rho_i^2)
 \eeq
where $\rho_1\le \cdots \le \rho_{\vert F\vert}$ are the
characteristic roots of $Q$, i.e. the eigenvalues of
$\sqrt{Q^*Q}$, and $\|\;\|$ is the norm induced by the canonical
scalar product $<\cdot,\cdot>$ on $\aaa$.
\end{lem}
Proof:
It is well-known that we can find unitary matrices $U$, $W$ such that
$$
Q=W\left(\begin{array}{ccc} \rho_1 &&
\\
&\ddots&
\\
&&\rho_{\vert F\vert }
\end{array}
\right) U .$$ Denote $\ozeta = \oeta W$ (i.e. $\ozeta_k=\sum_{i}
W_{i,k} \oeta_i$ for all $k$) and $\psi=U\eta$ (i.e.
$\psi_k=\sum_j U_{k,j} \psi_j$), we have \beqn \exp (\oeta Q \eta)
&= &\exp \ozeta \left(\begin{array}{ccc} \rho_1 &&
\\
&\ddots&
\\
&&\rho_{\vert F\vert }
\end{array}
\right) \psi
\\
&=& 1 +\sum_{i=1}^{\vert F\vert} \rho_i\, \ozeta_i \psi_i+
\sum_{i_1<i_2} \rho_{i_1}\rho_{i_2} \,\ozeta_{i_1} \psi_{i_1}
\ozeta_{i_2} \psi_{i_2} +\cdots
\\
&&+\rho_1\cdots \rho_{\vert F\vert} \,\ozeta_{1} \psi_{1} \cdots
\ozeta_{\vert F\vert }\psi_{\vert F\vert }
 \eeqn
  But the family of
vectors $\left(1, (\ozeta_i\psi_i), (\ozeta_{i_1}\psi_{i_1}
 \ozeta_{i_2}\psi_{i_2}),
\ldots ,\ozeta_{1} \psi_{1}\cdots
\ozeta_{\vert F\vert } \psi_{\vert F\vert }\right)$
is orthonormal, so we proved (\ref{f.2.3}). $\Box$
\ali

If $Y$ is in $\aaa$ we denote by $i_Y$ the interior product
by $Y$, i.e. the linear operator $i_Y\;:\;\aaa\rightarrow \aaa$
defined by
\beq
\label{f.2.4}
<i_Y(X), Z>=<X, Y Z>, \;\;\; \forall X,Z\in \aaa.
\eeq
In particular, remark that
$$i_{\Pi_{x\in F} \oeta_x\eta_x} (\exp \oeta Q\eta ))=\det Q.
$$

Suppose now that $F'$ is a subset of $F$, and denote by $\aaa_{F'}$ the
subalgebra of $\aaa$ generated by the variables $(\oeta_x)_{x\in F'}$,
$(\eta_x)_{x\in F'}$.
We define the linear operator
\begin{eqnarray}
R_{F\rightarrow F'}\;:\; \aaa&\rightarrow & \aaa_{F'}
\\
X&\rightarrow & i_{\Pi_{x\in F\setminus F'} \oeta_x \eta_x} (X).
\end{eqnarray}
\Rm:
The operator $R_{F\rightarrow F'}$ is often presented as an antisymmetric
integral.
More precisely, $R_{F\rightarrow F'}(X)$ coincides with the antisymmetric
integral of $X$ with respect to $\Pi_{x\in F\setminus F'} d\eta_x\oeta_x$,
i.e. $R_{F\rightarrow F'}(X) =\int X\Pi_{x\in F\setminus F'} d\eta_x\oeta_x$,
as defined in \cite{Berezin} (cf also \cite{Wang}).
The antisymmetric integral appears in the context of
supersymmetry, as an antisymmetric counterpart of the Gaussian
integral.
It is interesting to note that supersymmetry has been used in the
context of random Schr\"odinger operators several times
(cf for instance \cite{Klein}, \cite{Wang}, \cite{Wang2}
 and references therein).
\begin{propos}
\label{p.2.2}
Let $Q$ be a complex symmetric $F\times F$ matrix, we have
\beq
\label{f.2.5}
\det(Q)=
<R_{F\rightarrow F'} (\exp \oeta Q\eta ), \Pi_{x\in F'} \oeta_x \eta_x>,
\eeq
\beq
\label{f.2.6}
\det (Q_{|F\setminus F'}) =<R_{F\rightarrow F'} (\exp \oeta Q\eta ), 1>,
\eeq
and
\begin{eqnarray}
\label{f.2.7}
\exp (\oeta Q_{F'} \eta )=
{R_{F\rightarrow F'} (\exp \oeta Q \eta )\over \det (Q_{|F\setminus F'})},
\end{eqnarray}
when $\det ( Q_{|F\setminus F'})\ne 0$.
\end{propos}
Proof: The first two formulas are simple consequences of the
definitions. For a matrix $Q$ we denote by $\det(Q_{i_1,\ldots
,i_k\atop j_1,\ldots ,j_k})$ the determinant of the matrix where
we keep only the lines $i_1,\ldots ,i_k$ and columns $j_1,\ldots
,j_k$, and by $\det(Q^{i_1,\ldots i_k\atop j_1,\ldots ,j_k})$ the
determinant of the matrix where we removed the lines $i_1,\ldots
i_k$ and the columns $j_1,\ldots ,j_k$. It is well-known that
$$\det(Q^{i_1,\ldots ,i_k\atop j_1,\ldots ,j_k})
=\det ( (Q^{-1})_{i_1,\ldots ,i_k\atop
j_1,\ldots ,j_k})\det Q.
$$
Let $i_1<\cdots <i_k$, $\ihat_1<\cdots <\ihat_{\vert F'\vert -k}$
be elements of $F'$ such that
$$\{i_1,\ldots ,i_k,\ihat_1,\ldots ,\ihat_{\vert F'\vert -k}\}=F',$$
and $j_1<\cdots <j_k$, $\jhat_1<\cdots <\jhat_{\vert F'\vert -k}$
be elements of $F'$ such that
$$\{j_1,\ldots ,j_k,\jhat_1,\ldots
,\jhat_{\vert F'\vert -k}\}=F'.
$$
We have
\begin{eqnarray*}
 && <R_{F\rightarrow F'} (\exp \oeta
Q\eta ), \oeta_{i_1}\eta_{j_1} \cdots \oeta_{i_k}\eta_{j_k}>
\\
&=& \det (Q^{\ihat_1,\ldots ,\ihat_{\vert F'\vert -k}\atop
\jhat_1,\ldots ,\jhat_{\vert F'\vert -k}} )
\\
&=& \det Q\det ((Q^{-1})_{\ihat_1,\ldots ,\ihat_{\vert F'\vert -k}
\atop \jhat_1,\ldots ,\jhat_{\vert F'\vert -k}})
\\
&=&
\det((Q_{F'})_{i_1,\ldots ,i_k,\atop j_1,\ldots ,j_k})
{\det Q\over \det (Q_{F'})}
\\
&=& <{\det Q\over \det (Q_{F'})}\exp \oeta Q_{F'}\eta ,\oeta_{i_1}
\eta_{j_1} \cdots \oeta_{i_k}\eta_{j_k}>.
\end{eqnarray*}

Evaluating the equality for $k=0$ we get
$$<R_{F\rightarrow F'} (\exp \oeta Q\eta ),1>=\det(Q_{|F\setminus F'})=
{\det Q\over \det Q_{F'}},
$$
and this is enough to conclude the proof of the proposition. $\Box$
\ali

Let us introduce a notation: if $f$ is a holomorphic function from
a domain $D\subset \BC^n$ to $\BC^m$ then we denote by
$\ord(f,x_0)$ the order of vanishing of $f$ at the point $x^0\in D$,
i.e. the maximal integer $p$ such that one can find an open set $U$ containing $x_0$ and
holomorphic functions $h_{i_1,\ldots ,i_p}$, $1\le i_1\le \ldots \le i_p\le n$
on $U$
such that
$$f=\sum_{i_1\le \cdots \le i_p}
(x_{i_1}-x_{i_1}^0)\cdots (x_{i_p}-x_{i_p}^0) h_{i_1,\ldots ,i_p}(x),\;\;\; \hbox{on $U$
.}
$$
If $Q$ is a $F\times F$ symmetric matrix then we denote by
$\ker^\ND(Q)$ (for the Neumann-Dirichlet kernel) the subspace
$\ker^\ND(Q)=\{f\in \ker(Q),\; f_{|F'}=0\}$.
\begin{propos}
\label{p.2.3}
i) If $Q_0$ is a symmetric $F\times F$ real matrix then:
\begin{eqnarray}
\label{f.2.8}
\dim (\ker^\ND (Q_0))=
\ord \left(Q\rightarrow R_{F\rightarrow F'}(\exp \oeta Q\eta ),Q_0\right),
\end{eqnarray}
where in the right hand side $Q$ is taken from the
set of complex symmetric $F\times F$ matrices.
\ali
ii)
Moreover, if $B$ is a real positive definite symmetric
$F\times F$ matrix then
\begin{eqnarray}
\label{f.2.8.1}
\dim (\ker^\ND (Q_0))=
\ord \left(\lambda \rightarrow R_{F\rightarrow F'}
(\exp \oeta (Q_0-\lambda B)\eta ),0\right).
\end{eqnarray}
\end{propos}
\noindent \Rm: The fact that the matrix $Q_0$ is real is crucial
since this result is essentially related to the fact that this
matrix is diagonalizable.
 \ali
 \Rm: A priori the order of
vanishing of the function
$$\lambda \rightarrow
R_{F\rightarrow F'} (\exp \oeta (Q_0-\lambda B)\eta )
$$
at $\lambda=0$ is greater or equal to the order of vanishing of
$Q\rightarrow R_{F\rightarrow F'}(\exp \oeta Q\eta )$ at $Q_0$.
The point ii) tells us that there is actually equality, i.e. that
the line $\{Q_0-\lambda B, \lambda \in \BC\}$ intersects
transversally at $\lambda =0$ the analytic set $\{Q \hbox{ s.t. }
R_{F\rightarrow F'}(\exp\oeta Q\eta) =0\}$.
 \ali\ali
  Proof: i) The
point i) is a consequence of ii). Indeed, the order of vanishing
of $Q\rightarrow R_{F\rightarrow F'}(\exp \oeta Q\eta )$ is equal
to the order of vanishing in a generic direction. Otherwise
stated, this means that there is a proper analytic subset
$\aaa\subset \symF$ such that for any direction $B$ in
$\symF\setminus \aaa$ the order of vanishing in i) is equal to the
order of vanishing of the function $\lambda\rightarrow
R_{F\rightarrow F'} (\exp \oeta (Q_0-\lambda B)\eta )$ at the
point $\lambda=0$. Denote by $\symFR$ the space of real symmetric
$F\times F$ matrices, regarded as the set of real directions in
the tangent vector space $\symF$. Since $\aaa$ is analytic,
$\aaa\cap \symFR$ is of empty interior. Considering that the
subset of $B$ in $\symFR$ which are positive definite is open in
$\symFR$, we know that this set cannot be contained in $\aaa$,
hence if we assume ii) we know that $\dim\ker(Q_0)$ is the order
of annulation in a generic direction. Hence ii) implies i).
 \ali ii) We first derive
an explicit expression for $T(Q_0-\lambda B)$. Since $Q_0$ is real
and $B$ real positive definite we can diagonalize
$(Q_0)_{|F\setminus F'}$ in an orthonormal basis for
$B_{|F\setminus F'}$, i.e. we can find eigenvalues
$\lambda_1^-,\ldots , \lambda_{\vert F\setminus F'\vert}^-$ and a
family of functions $f_1^-,\ldots ,f_{\vert F\setminus F'\vert}^-$
in $\BR^{F\setminus F'}$ such that
\begin{eqnarray*}
& <f_k^-, B f_{k'}^->_{F\setminus F'}=\delta_{k,k'} &
\\
& (Q_0)_{|F\setminus F'} f_k^-=\lambda_k^- (B_{|F\setminus F'})f_k^-,
&
\end{eqnarray*}
where $<,>_{F\setminus F'}$ is the usual scalar product on
$\BR^{F\setminus F'}$. For a real function $f$ on $F'$ we denote
by $H_\lambda f$ the harmonic prolongation of $f$ with respect to
$Q_0-\lambda B$. The function $H_\lambda f$ can be written
$$
H_\lambda f= f +\sum_{k=1}^{\vert F\setminus F'\vert } c_k f_k^-,
$$
and we easily get
$$c_k= {<f,((Q_0-\lambda B)f_k^-)_{|F'}>_{F'}\over
\lambda-\lambda_k^-}.
$$
 We set $\phi_{k,\lambda}=((Q_0-\lambda
B)f_k^-)_{|F'}$ and we denote by
$p_{k,\lambda}:\BR^{F'}\rightarrow \BR^{F'}$ the projector given
by $p_{k,\lambda}(f)=<f,\phi_{k,\lambda}>_{F'} \phi_{k,\lambda}$.
Thus we have from proposition \ref{p.2.1} ii)
$$
TQ=(Q_0-\lambda B)_{|F'} +\sum_{k=1}^{\vert F\setminus F'\vert }
{p_{k,\lambda}\over \lambda -\lambda_k^-}.
$$
Denote now by $n_0$ the dimension of $\ker^\ND(Q_0)$, and by
$n_0'$ the dimension of $\ker((Q_0)_{|F\setminus F'})$, so that
$n_0'\ge n_0$. This means that $n_0'$ of the eigenvalues
$\lambda_k^-$ are null. We can as well suppose that
$\lambda_1^-=\cdots =\lambda_{n'_0}^-=0$ and that $f_1^-,\ldots
,f_{n_0}^-$ form an orthonormal basis (w.r. to $B$) of $\ker^\ND
(Q_0)$. This implies that for $k\le n_0$, $Q_0 f_k^-=0$ and thus
that $\phi_{k,\lambda} = \lambda (Bf_k^-)_{|F'}$. For $k\le n_0$
we denote by $\tilde p_k:\BR^{F'}\rightarrow \BR^{F'}$ the
projector given by
$$
\tilde p_k(f)= <f,(Bf_k^-)_{|F'}> (Bf_k^-)_{|F'}
$$
and we have $p_{k,\lambda}=\lambda^2 \tilde p_k.$ For $k\ge n_0+1$
we simply denote $\phi_k =\phi_{k,0}$ and $p_k=p_{k,0}$. The
functions $\{\phi_k\}_{k=n_0+1,\ldots ,n'_0}$ are linearly
independent. Indeed, otherwise there would exit a linear
combination of the $f_k^-$, $k=n_0+1,\ldots ,n'_0$, belonging to
$\ker^\ND(Q_0)$. This is not possible since the dimension of
$\ker^\ND(Q_0)$ is $n_0$.

Considering relation (\ref{f.2.7}), we see that for small
$\lambda$'s the function $R_{F\rightarrow F'}(\exp(\oeta
(Q_0-\lambda B)\eta))$ can be written
\begin{eqnarray*}
&&(-\lambda)^{n'_0}\left(\prod_{k=n'_0+1}^{\vert
F\vert}(\lambda_k^- -\lambda)\right) \exp(\oeta (Q_0-\lambda
B)_{|F'}\eta)
\\
&&\exp(\oeta (\sum_{k=n_0+1}^{n'_0}{1\over \lambda}
p_{k,\lambda})\eta) \exp(\oeta (\sum_{k=n'_0+1}^{\vert F\setminus
F'\vert } {1\over\lambda- \lambda_k^-} p_{k,\lambda})\eta)
\exp(\lambda \oeta(\sum_{k=1}^{n_0} \tilde p_k)\eta)
\\
&=&C\lambda^{n_0'}\exp(\oeta Q_0\eta)\exp(\oeta
(\sum_{k=n_0+1}^{n'_0}{1\over \lambda} p_k)\eta) \exp(\oeta
(\sum_{k=n'_0+1}^{\vert F\setminus F'\vert } {-1\over \lambda_k^-}
p_k)\eta)(1+o(\lambda))
\end{eqnarray*}
where $C$ is a non null constant. Considering that the operators
$p_k$ have rank 1, the last expression equals
$$
C \lambda^{n_0} \prod_{k=n_0+1}^{n'_0} (\lambda +\oeta p_k\eta)
\exp(\oeta(\sum_{k=n_0'+1}^{\vert F\setminus F'\vert} {1\over
\lambda_k^-} p_k)\eta)(1+o(\lambda)).
$$
From this we deduce that $\lambda^{n_0}$ can be factorized in the
last expression, hence that the order of vanishing of the function
$\lambda \rightarrow R_{F\rightarrow F'} (\exp \oeta (Q_0-\lambda
B)\eta )$ is at least $n_0$. In the last expression, the term of
degree $n'_0-n_0$ in the variables $\oeta$ and in the variables
$\eta$ is
$$
\lambda^{n_0} \prod_{k=n_0+1}^{n_0'} (\oeta p_k\eta)+(\hbox{terms
of order $\lambda^k$, $k>n_0$}).
$$
Since the $p_k$'s are linearly independent
$\prod_{k=n_0+1}^{n'_0}(\oeta p_k\eta)$ is not null. This proves
that the order of vanishing of $\lambda \rightarrow
R_{F\rightarrow F'} (\exp \oeta (Q_0-\lambda B)\eta )$ is exactly
$n_0$. $\Box$

\subsection{The Lagrangian Grassmannian}
We denote by $\symF$ the space of complex symmetric $\cF \times
\cF$ matrices. We denote by $\ppp(\aaa)$ the projective space
associated with $\aaa$ and by $\pi:\aaa\rightarrow \ppp(\aaa)$ the
canonical projection. It is clear that the  map $Q\rightarrow
\pi(\exp (\oeta Q\eta ))$ is injective and hence defines an
embedding of $\symF$ in $\ppp(\aaa)$. In this section we describe
the subvariety defined as the closure of the set of points of the
type $\pi(\exp (\oeta Q\eta ))$ for $Q$ in $\symF$. This
subvariety defines a compactification of the set $\symF$, and we
will see that it is a Lagrangian Grassmaniann. This type of
compactification already appeared in the context of electrical
network, cf \cite{Collin1}, \cite{Collin2}.

We first recall some classical notions. Let $n$ be an integer
and $(\cdot ,\cdot )$ be the bilinear form on
$\BC^{2 n}$ given by $(X,Y)=\sum X_iY_i$ and $J$ be the $2 n\times 2 n$ matrix given
by:
$$J=\left(
\begin{array}{cc}
0& I_n
\\
-I_n& 0
\end{array}
\right)
$$
where $I_n$ is the $n \times n$ identity matrix.
\ali
Obviously $(\cdot , J\cdot )$ is an antisymmetric non-degenerate bilinear form
(usually called the symplectic form when considered on $\BR^{2 n}$).
\begin{defin}
A linear subspace $L$ of $\BC^{2n}$ is called Lagrangian if for all $x,y$ in $L$,
$(x,Jy)=0$.
\end{defin}
We denote by $\BL^n$ the set of $n$-dimensional  Lagrangian
subspace of $\BC^{2n}$. It is a homogeneous space. It can be
indeed described as the quotient of the complex symplectic group
by the stabilizer of a point. Therefore $\BL^n$  is a
${n(n+1)\over 2}$ compact smooth manifold. To precise the
situation we describe explicitly a local parameterization. At a
point $L$ the set $\BL^n$ can be parameterized explicitly by the
space $\sym_n$ of symmetric $n\times n$ complex matrices. Indeed,
if $(v_1,\ldots ,v_n)$ and $(v_1',\ldots ,v_n')$ are orthonormal
basis of respectively $L$ and $L^\perp$ (for the usual scalar
product on $\BC^{2n}$ and $L^\perp$ the orthogonal subspace of $L$
for this scalar product) then the map
\begin{eqnarray}
\nonu
\sym_n &\rightarrow & \BL^n
\\
\label{f.2.12}
Q&\rightarrow & \hbox{Vect}\{ v_i+\sum_j Q_{i,j} v'_j \}_{i=1\ldots n}
\end{eqnarray}
defines a local set of coordinates. Indeed, it is easy to check
that the subspace $\hbox{Vect}\{ v_i+\sum_j Q_{i,j} v'_j
\}_{i=1\ldots n}$ is Lagrangian and that any Lagrangian subspace
in a neighborhood of $L$ can be represented in such a form. \ali

Let $(e_1,\ldots ,e_n,e'_1,\ldots e'_n)$ denote the canonical
basis of $\BC^{2n}$. We denote by $\ppp (\wedge^n \BC^{2n})$ the
projective space associated with $\wedge^n \BC^{2n}$ and by
$\pi\;:\; \wedge^n \BC^{2n} \rightarrow \ppp(\wedge^n \BC^{2n})$
the canonical surjection. Classically, the manifold $\BL^n$ can be
embedded in the projective space $\ppp (\wedge^n \BC^{2n})$ by the
Pl\"ucker embedding
\begin{eqnarray}
\nonu
\BL^n &\rightarrow & \ppp (\wedge^n \BC^{2n})
\\
\label{f.2.13}
L= \hbox{Vect}\{ x_1,\ldots ,x_n\} &\rightarrow & \pi ( x_1\wedge \cdots \wedge x_n).
\end{eqnarray}

We come back now to the Grassmann algebra introduced in section 2.2.1.
Take $n=\cF$, the subalgebra $\aaa$
  can be easily identified with $\wedge^\cF  \BC^{2\vert F\vert}$
by the isomorphism defined on monomials by: \beq \nonumber
\wedge^\cF \BC^{2\vert F \vert} & \rightarrow &\aaa
\\
\label{f.2.14}
e_{i_1}\wedge \cdots \wedge e_{i_{\vert F\vert-k}}\wedge e'_{j_1}\wedge \cdots \wedge e'_{j_k}
&\rightarrow
&
(-1)^{\sum_{p=1}^{\cF-k} i_p-p}
 \oeta_{\ihat_1}\eta_{j_1} \cdots \oeta_{\ihat_k}\eta_{j_k},
\eeq
where $i_1<\cdots < i_{\cF-k}$, $j_1<\cdots <j_k$ and $\ihat_1, \ldots , \ihat_k$ is defined
by $\ihat_1< \cdots < \ihat_k$ and
$\{ \ihat_1, \ldots , \ihat_k,i_1,\ldots , i_{\cF-k}\}=\{1,\ldots ,\cF\}$.
\ali
Thanks to the embedding (\ref{f.2.13}) and the isomorphism (\ref{f.2.14})
the manifold $\BL^\cF$ can be considered as a smooth subvariety of $\ppp(\aaa)$.
It is easy to see that by the isomorphism (\ref{f.2.14}) the point
$$\wedge_{i=1}^\cF (e_i+\sum Q_{i,j}e_j)$$
is sent to the point $\exp (\oeta Q \eta )$ and thus that
$\pi(\exp(\oeta Q\eta))$ is in $\BL^\cF$. Hence we deduce the
following
\begin{propos}
The application $Q\rightarrow \pi(\exp \oeta Q\eta )$ defines an embedding
of the space $\symF$ of $F\times F$ symmetric matrices into the smooth projective
subvariety $\BL^\cF$.

More precisely, the set $\symF$ is sent onto the subset
$\BL^\cF \setminus\pi\{X\in \aaa,\; <X,1>=0\}$.
Hence, the closure of the set of points of the type
$\pi(\exp\oeta Q\eta)$ is equal to $\BL^\cF$.
\end{propos}
\noindent\Rm: Therefore the set $\BL^\cF$ defines a
compactification of $\symF$. There are many different
compactifications of $\symF$ (for example, in section 4.4 we
consider the compactification by a projective space) but this one
seems to be the best-suited to our problem.
 \ali

We will need some results on the dimension of the cohomology
groups of $\BL^\cF$. We recall from \cite{SankaranV} that the
first and second betti numbers are given by:
 \beq \label{f.2.14.1}
b_1=\dim(H^1(\BL^\cF,\BC))=0,\;\; b_2 =\dim(H^2(\BL^\cF,\BC))=1.
\eeq
 The manifold $\BL^\cF$ is a K\"ahler manifold, as a smooth
projective subvariety. A natural K\"ahler form on $\BL^\cF$ is the
restriction of the Fubini-Study form on $\ppp(\aaa)$ (cf appendix
A.5): we call this form the canonical K\"ahler form on $\BL^\cF$.
By definition, the K\"ahler form is in $\Hun (\BL^\cF)$ the
$(1,1)$ Dolbeault cohomology group of $\BL^\cF$ (which coincides
for K\"ahler manifold with the subspace of $H^2(\BL^\cF,\BC)$
generated by the forms of the type $(1,1)$, cf appendix C). Thus
we have by (\ref{f.2.14.1})  and general results on K\"ahler
manifold that $\dim(H^{2,0}(\BL^\cF))=\dim(H^{0,2}(\BL^\cF))=0$
and $\dim(\Hun (\BL^\cF))=1$.
 \hfill\break \hfill\break
 {\it $G$-invariant Lagrangian Grassmannian}
 \ali
 We suppose now given a finite group
$G$ acting on $F$. 
We denote by ${\Bbb L}^G$ the closure in ${\Bbb L}^\cF$ of the subset 
$\hbox{Sym}^G$, the space of $G$-invariant complex $F\times F$
matrices. As we shall see in appendix E, ${\Bbb L}^G$ is a smooth
projective variety, whose structure can be explicitely described.
It is also clear that for the
isomorphism (\ref{f.2.14}) the submanifold $\lll$ is the closure
in $\ppp(\aaa)$ of the points of the type $\pi(\exp\oeta Q\eta)$
for $Q$ in $\symG$.
 \hfill\break \hfill\break
 \Rm: \label{r.2.100} Assume that $F'$ is
a subset of $F$, invariant by the group $G$. The element
$\pi(R_{F\rightarrow F'} (\exp \oeta Q\eta))$ is then in the
$G$-invariant Lagrangian Grassmannian associated with $F'$.
Moreover, remark that  formula (\ref{f.2.8}) of  proposition
(\ref{p.2.3}) remains valid for $Q_0$ in $\symG$ if we let $Q$ run
in $\symG$ instead of $\symF$. Indeed, using ii) of proposition
\ref{p.2.3}, we know that for $B$ real symmetric positive,
$G$-invariant
\begin{eqnarray*}
\dim\ker^\ND(Q_0)&=& \ord(\lambda\rightarrow R_{F\rightarrow F'}(
\exp\oeta(Q_0-\lambda B)\eta), 0)
\\
&\ge & \ord( Q\rightarrow R_{F\rightarrow F'}(\exp(\oeta Q\eta)),
Q_0)
\end{eqnarray*}
for $Q$ running in $\symG$. This last expression is bounded from
below by the same expression for $Q$ running in $\symF$ instead of
$\symG$, which is equal to $\dim\ker^\ND(Q_0)$ by proposition
\ref{p.2.3}, i).

\section{Trace of a Dirichlet form in the continuous situation}
We recall here some results from \cite{Sabot3} that will be useful
for the continuous case. Let $X$ be a locally compact denumerable
metric space and $m$ a finite positive Radon measure on $X$ such
that $\hbox{supp}(m)=X$. \ali Let $(a,\ddd)$ be a regular
Dirichlet form on $L^2(X,m)$ such that:

(i) $a$ is irreducible (i.e. $a(f)=0$ implies that $f$ is constant).

(ii) $(a,\ddd)$ has a compact resolvent.

(iii) There exists $c>0$ such that $\hbox{cap}_1(\{x\})\ge c$ for
all $x\in X$.
\ali
N.B.: $\hbox{cap}_1(\{x\})$ stands for the 1-capacity of the point
$\{x\}$ (cf \cite{Fukushima1}, section 2).
\ali
\ali
The assumption (iii) implies in particular that the functions of the
domain have a continuous modification, so that the value at one point can
be defined (cf \cite{Fukushima1}, theorem 2.1.3).
A second implication of assumption (iii)
is that the resolvent $R_\lambda$ is
trace-class (cf \cite{Sabot3}).

Let $F$ be a finite subset of $X$.
The regularity of the form and assumption (iii) imply that for any
$f\in {\Bbb R}^F$ there exists $g\in \ddd$ such that $g_{|F}=f$.
\ali
We define the trace of $(a,\ddd)$ on the subset $F$ as the bilinear
form on ${\Bbb R}^F$ defined by:
\begin{eqnarray}
\label{f.2.15}
a_{F}(f,f)=\inf\{a(g,g),g\in\ddd,g_{|F}=f\},\;\;\;\forall f\in{\Bbb R}^F.
\end{eqnarray}
The irreducibility of $(a,\ddd)$ implies that the infimum in
(\ref{f.2.15}) is reached at unique point called the
harmonic continuation of $f$ with respect to $a$.
\ali
If $F$ is endowed with a positive measure $b$ with full
support then $(a_{F},{\Bbb R}^F)$
is a regular, irreducible Dirichlet form on
$L^2(F,b)$
(the process associated with $a_{F}$ and $b$
on states space $F$ can
be represented by a time changed of the initial process
associated with $(a,\ddd)$ on $L^2(X,m)$ (cf \cite{Fukushima1},
theorem 6.2.1).
\ali
For $\lambda\ge 0$, let
$a_{\lambda}(f)=a(f)+\lambda\int_X f^2dm$ for $f\in \ddd$.
The bilinear form
$a_{\lambda}$ is a regular irreducible Dirichlet form satisfying
(i), (ii) and (iii).
We denote by $A_{(\lambda)}$ the $F\times F$ symmetric matrix given by
$<A_{(\lambda)}\cdot ,\cdot>=
({a_{\lambda}})_{F}(\cdot, \cdot)$ where $<\cdot ,\cdot>$ is the
usual scalar product on $\BR^F$, and by
$H_\lambda f$ the harmonic continuation of $f\in {\Bbb R}^F$  with
respect to $a_\lambda$, so that $<A_{(\lambda)}(f), f> =a_\lambda (H_\lambda f, H_\lambda f)$.

Set $\ddd^-=\{f\in \ddd,\;\; f_{|F}=0\}$ (N.B.:
$\ddd^-$ is the
domain with Dirichlet boundary conditions on $F$;
$(a,\ddd^-)$ is a regular Dirichlet form on
$L^2(X\setminus\{F\},m)$).
\ali
We denote by $0>\lambda_1^+\ge\cdots\ge\lambda_k^+\ge\cdots$ the
negative eigenvalues of the infinitesimal generator
associated with $(a,\ddd)$ and by $\sigma_0$ the
multiplicity of the eigenvalue 0 (which can be 0 or 1, indeed
$\sigma_0=1$ if $1\in \ddd$ and 0 otherwise).
\ali
We also denote by $0>\lambda_1^{-}\ge\cdots\lambda_k^-\ge\cdots$ the
eigenvalues of the infinitesimal generator of
$(a,\ddd^-)$ (in this case 0 is not eigenvalue because of
the boundary condition and assumption (i)).
Let $f_1^-,\ldots,f_k^-,\ldots$ be an orthonormal basis of eigenfunctions
associated with the preceding eigenvalues.
\begin{lem}
(\cite{Sabot3}, lemma 2.1.)
\label{l.2.2}
For any $f\in{\Bbb R}^F$, $\lambda\ge 0$:
\begin{eqnarray}
\label{f.2.16}
<A_{(\lambda)}(f),f> =<A_{(0)}(f),f>
+\lambda\int (H_0f)^2dm-\lambda^2
\sum_{k=1}^\infty{(\int H_0 f f_k^-dm)^2\over
\lambda-\lambda_k^-}.
\end{eqnarray}
In particular $A_{(\lambda)}$ is meromorphic on ${\Bbb C}$
with at worst simple poles at the points
\ali
$\{\lambda_1^-,\ldots,\lambda_k^-\ldots\}$.
\end{lem}

We define some infinite dimensional determinants
by the following
formula: for $\lambda\in {\Bbb C}$ we set
\begin{eqnarray}
\label{f.2.17}
d^+(\lambda)=\lambda^{\sigma_0}\Pi_{k=1}^\infty
(1-{\lambda\over\lambda_k^+}),
\end{eqnarray}
\begin{eqnarray}
\label{f.2.18}
d^-(\lambda)=\Pi_{k=1}^\infty
(1-{\lambda\over\lambda_k^-}).
\end{eqnarray}
The existence of these functions comes from the fact that the
resolvent of $(a,\ddd)$ and $(a,\ddd^-)$ are trace class.
\begin{lem}
\label{l.2.3}
There exists a constant $C>0$
such that for
all
\ali
$\lambda\in{\Bbb C}\setminus\{\lambda_1^-,\ldots,\lambda_k^-,\ldots\}$:
\begin{eqnarray}
\label{f.2.19}
\det(A_{(\lambda)})=C {d^+(\lambda)\over d^-(\lambda)}.
\end{eqnarray}
\end{lem}
A proof of this result is given in \cite{Sabot3}. A more general
version of this result, but valid only for differential operators
on $\BR^d$, can be found in \cite{Forman1}.
\chapter{The renormalization map. Expression of the density of states.}
\setcounter{section}{0}
\section{Construction of the renormalization map.}
In section 3 we will only be interested in the density of states.
Therefore, since the counting measures $\nu_\nn^\pm$ do not
depend on the particular blow-up, we suppose to simplify notations
that $\w_k=1$ for all $k$.
\subsection{On the set of symmetric $F\times F$ matrices.}
We come back to the situation described in section 1 and denote by
$\symF$ the space symmetric $F\times F$ matrices.
We denote by $\symG$ the subspaces of $\symF$ of $G$-invariant
matrices,
i.e. of symmetric matrices $Q$ satisfying:
$$g.Q f=Q(g.f),\;\;\;\forall g\in G,\forall f\in \BC^F.
$$
Starting from $Q$ in $\symF$ we can construct a $F_{<1>} \times F_{<1>}$
symmetric matrix $Q_{<1>}$ by:
$$
Q_{<1>}(f)=\sum_i \alpha_1\alpha_i^{-1} Q_{<1>,i},
$$
where $Q_{<1>,i}$ is a copy of $Q$ on the cell $F_{<1>,i}$, i.e. it is
a $F_{<1>}\times F_{<1>}$ symmetric matrix defined by
$$
\left\{
\begin{array}{l}
(Q_{<1>,i} f)_{|F_{<1>,i}}= Q(f_{|F_{<1>,i}}),
\\
Q_{<1>,i}f(x)=0,\;\;\hbox{if $x\not\in F_{<1>,i}$},
\end{array}
\right.
$$
for all $f$ in $\BR^{F_{<1>}}$.
\ali
On the set $\{\det (Q_{<1>})_{|F_{<1>}\setminus \partial F_{<1>}}\ne 0\}$
we consider the trace on $\partial F_{<1>}$ of
$Q_{<1>}$, which is an element of $\symF$.
So we define:
\begin{eqnarray}
\nonu
T\;:\; \symF &\rightarrow &\symF
\\
Q &\rightarrow & (Q_{<1>})_{\partial F_{<1>}}.
\end{eqnarray}
Considering the symmetries of the structure we see that $TQ$ is
$G$-invariant if $Q$ is $G$-invariant, i.e. $T(\symG)\subset
\symG$. In all the following we will rather consider $T$ as a map
on $\symG$ than on $\symF$. Using proposition (\ref{p.2.1})  we
know that the map $T$ is rational with poles included in the set
$\{\det (Q_{<1>})_{|F_{<1>}\setminus \partial F_{<1>}}= 0\}$.

Let $S_+$ denote the set of complex symmetric $F\times F$ matrices
with positive definite imaginary part:
 \beq \label{f.3.0.0}
&S_+=\{ Q \hbox{ complex symmetric $F\times F$ matrix,} ,\;\;\;
\hbox{ $\im (Q)$ is positive definite}\},&
 \eeq
and set $S_+^G=S_+\cap \symG$. Usually, $S_+$ is called the Siegel
upper half-space (cf \cite{Siegel}, \cite{Terras}). In the next
lemma we prove a key property of $T$, which is that $S_+$ is left
invariant by $T$. We also give some estimates, useful in lemma
\ref{l.3.1}, to bound the speed at which the iterates of $T$
approach the boundary of $S_+$. In appendix D, we present a
different approach which avoid all explicit estimates, but only
uses properties of contraction of holomorphic maps on $S_+$.

For a matrix $Q$ we denote by $\underline \rho (Q)$ and $\overline
\rho (Q)$ the minimal and maximal characteristic root of $Q$ (i.e.
the minimal and maximal eigenvalues of $\sqrt{Q^*Q}$).
\begin{lem}
\label{l.3.0.0} The map $T$ is well-defined and holomorphic on
$S_+$ (resp. on $S_+^G$) and $S_+$ (resp $S_+^G$) is
$T$-invariant. Moreover for all $Q$ in $S_+$ we have
 \beq \label{f.3.0.1}
\underline \rho (\im (TQ))\ge \alpha_1 \overline \alpha^{-1}
\underline \rho (\im Q),
\\
\label{f.3.0.2} \underline \rho (\im ((TQ)^{-1})) \ge
\alpha_1^{-1} \underline \alpha \underline \rho (\im (Q^{-1})),
 \eeq
where $\underline \alpha =\inf \{\alpha_i\}$, $\overline
\alpha=\sup \{\alpha_i\}$.
\end{lem}
 \ali
Proof: It is clear by proposition
(\ref{p.2.1})
 that $T$ is well-defined on  $S_+$. We first derive an expression
for $\im (TQ)$. Let $Q=\re Q +i\im Q$ be in $S_+$. Let $f$ be a
real function on $F$ (which can also be considered as a real
function on $\partial F_{<1>}$), and $Hf$ its harmonic
prolongation with respect to $Q_{<1>}$ (which can be defined for
$Q$ in $S_+$ using proposition (\ref{p.2.1})). We have:
 \beqn <TQ f,f>&=& <Q_{<1>}Hf, Hf>
\\
&=& <Q_{<1>} Hf, \re Hf>
\\
&=&<\re Q_{<1>} \re Hf, \re Hf> +i <\re Q_{<1>} \im Hf, \re Hf>
\\
&& + i<\im Q_{<1>} \re Hf, \re Hf> - <\im Q_{<1>} \im Hf, \re Hf>
 \eeqn
Considering that $<Q_{<1>} Hf, \im Hf> = 0$ we deduce the
following identities:
 \beqn
  <\re Q_{<1>} \re Hf, \im Hf>=<\im Q_{<1>} \im Hf, \im Hf >,
\\
<\im  Q_{<1>} \re Hf, \im Hf>=-<\re  Q_{<1>} \im Hf, \im Hf>.
\eeqn
 Replacing in the expression of $<TQ f,f>$ we get
 \beqn <\im TQ f, f>& =& <\im Q_{<1>} \re Hf, \re Hf> + <\im Q_{<1>} \im Hf,
\im Hf>
\\
&\ge &
\sum_{i=1}^N \alpha_1\alpha^{-1}_i <\im Q\re Hf_{|F_{<1>,i}}, \re Hf_{|F_{<1>,i}}>
\end{eqnarray*}
Since $\im(Q)$ is positive definite we deduce from the last
expression that $<\im(TQ)f,f>>0$ since $\re(Hf)$ is not zero, and
thus that $TQ$ is indeed in $S_+$. From the last inequality we can
also deduce
\begin{eqnarray*}
<\im TQ f,f> &\ge& \underline \rho(\im Q) \alpha_1\overline
\alpha^{-1} \sum_{i=1}^N <\re Hf_{|F_{<1>,i}}, \re
Hf_{|F_{<1>,i}}>
\\
&\ge&\underline \rho(\im Q) \alpha_1\overline \alpha^{-1} \| f
\|^2.
\end{eqnarray*}
Thus formula (\ref{f.3.0.1}) is proved.

To prove equation (\ref{f.3.0.2}) we need to express
$Q_{<1>}^{-1}$ in terms of $Q^{-1}$ by a kind of harmonic
prolongation (just like $TQ$ is expressed in terms of $Q$ by a
harmonic prolongation). This is done in \cite{Sabot2} in a
slightly less general context. We state without proof the
following lemma (the proof is  a simple algebraic manipulation and
essentially similar to the proof of proposition 1.1 of
\cite{Sabot2}. Note that this technical lemma can be avoided, cf
appendix D.)
\begin{lem}
\label{l.3.0} Let $\nu$ be in $\BC^{F_{<1>}}$ and $\ddd_\nu$ be
the set
$$\ddd_\nu=\{(\nu_1,\ldots ,\nu_N)\in \BC^{F_{<1>,1}}\times \cdots \times
\BC^{F_{<1>,N}}, \;\;\; \sum_{i=1}^N \nu_i =\nu\}.
$$
If $Q$ is invertible then there exists a unique $(\nu_1,\ldots
,\nu_N)$ in $\ddd_\nu$ such that
$$\sum_{i=1}^N \alpha_1^{-1} \alpha_i <Q^{-1} \nu_i,\tilde \nu_i> =0
$$
for all $(\tilde \nu_1,\ldots ,\tilde \nu_N)$ in $\ddd_0$, and we
have:
$$
((Q_{<1>})^{-1} \nu )_{|F_{<1>,i}} = \alpha_1^{-1} \alpha_i Q^{-1}
\nu_i.
$$
\end{lem}
Let $Q$ be in $S_+$. We have $\underline \rho( \im ((TQ)^{-1}))\ge
\underline \rho (\im  ((Q_{<1>})^{-1}))$. Let $\nu$ be  in
$\BR^{F_{<1>}}$. Proceeding just as previously we can prove: \beqn
<\im((Q_{<1>})^{-1}) \nu,\nu>&= & \sum_{i=1}^N \alpha^{-1}_1
\alpha_{i} (<\im (Q^{-1}) \re \nu_i,\re \nu_i> + <\im (Q^{-1}) \im
\nu_i,\im \nu_i>),
\\
&\ge& \underline \rho (\im (Q^{-1})) \sum_{i=1}^N \alpha^{-1}_1
\alpha_{i} (<\re \nu_i,\re \nu_i> + <\im \nu_i,\im \nu_i>)
\\&\ge &
\underline \rho (\im (Q^{-1})) \alpha_1^{-1} \underline \alpha \|
\nu\|^2, \eeqn where $(\nu_1,\ldots ,\nu_N)$ is the element of
$\ddd_\nu$ obtained from lemma (\ref{l.3.0}). Thus we have proved
formula (\ref{f.3.0.2}). $\Box$

\subsection{The map $R$ defined on the
Grassmann algebra.}
Consider the Grassmann algebra with generators $\{\oeta_x,\eta_x\}_{x\in F}$
 and $\aaa$ its subspace generated by the monomials
containing the same number of variables $\eta$ and $\oeta$ (cf
section 2.2). We denote by $\aaa_{<1>}$ the counterpart for the
set $F_{<1>}$. Remind that the elements of $\aaa$ and $\aaa_{<1>}$
commute since they contain only monomials of even degrees.
 \ali The canonical injections $s_i\;:\; F\rightarrow F_{<1>}$
given by $s_i(x)=(i,x)$ naturally induce the morphism $s_i\;:\;
\aaa\rightarrow \aaa_{<1>}$ defined on the generators by:
$(\oeta_x,\eta_x)\rightarrow (\oeta_{s_i(x)},\eta_{s_i(x)})$.

For a real $\beta>0$ we denote by $\tau_\beta$ the linear map
defined on monomials by: \beq \nonu
\begin{array}{cccc}
\tau_\beta \;:\; &\aaa&\rightarrow &\aaa
\\
&\oeta_{i_1}\cdots \oeta_{i_k} \eta_{j_1}\cdots \eta_{j_k}
&\rightarrow &\beta^k \oeta_{i_1}\cdots \oeta_{i_k} \eta_{j_1}\cdots \eta_{j_k}.
\end{array}
\eeq
Then we set
\begin{eqnarray}
\nonu
\tau_{\beta_1,\ldots ,\beta_N} \;:\; \aaa&\rightarrow & \aaa_{<1>}
\\
\nonu
X
&\rightarrow & s_1(\tau_{\beta_1}(X)) \cdots s_N(\tau_{\beta_N} (X)) .
\end{eqnarray}
Using the commutativity of the subalgebra $\aaa_{<1>}$, with these
definitions, we get:
$$
\exp \oeta Q_{<1>} \eta =
\Pi_{i=1}^N \exp(\alpha_1\alpha_i^{-1} \oeta Q_{<1>,i} \eta )=
\tau_{\alpha_1\alpha_1^{-1},\ldots ,\alpha_1\alpha_N^{-1}}
(\exp \oeta Q \eta )
$$
Finally, using the construction of section 2,
 we define the map $R:\aaa\rightarrow \aaa$ by
$$ R=R_{F_{<1>}\rightarrow \partial F_{<1>}}
\circ \tau_{\alpha_1\alpha_1^{-1},\ldots ,\alpha_1\alpha_N^{-1}}.
$$
For $Q$ in $\symF$, we denote by $Q_\nn$, the $F_\nn\times F_\nn$,
$G$-invariant, symmetric matrix defined, as in formula
(\ref{f.1.6}), by
$$
Q_\nn=\sum_{i_1,\ldots ,i_n=1}^N \alpha_{1}^n
\alpha^{-1}_{i_1}\cdots \alpha_{i_n}^{-1} Q_{\nn,i_1,\ldots ,i_n},
$$
where $Q_{\nn, i_1, \ldots ,i_n}$ is the copy of $Q$ on $F_{\nn,
i_1, \ldots ,i_n}$, as in section 1.2.1 (N.B.: remind that here we
suppose $\w_i=1$ for all $i$). Similarly $Q_{\nn,k}$ denotes the
$F_{\nn,k}\times F_{\nn, k}$ matrix defined by
$$
Q_{\nn,k}=\sum_{i_2,\ldots ,i_n=1}^N \alpha_{1}^{n-1}
\alpha^{-1}_{i_2}\cdots \alpha_{i_n}^{-1} Q_{\nn,k,i_1,\ldots
,i_n}.
$$
\begin{propos}
\label{p.3.2}
i) The map $R$ is polynomial homogeneous of degree $N$.

ii) We have the following relation:
 \beq \label{f.3.1}
R^n(\exp \oeta Q\eta)=C_\nn  \det\left(
(Q_{<n>})_{|\intF_\nn}\right)\exp(\oeta T^n Q \eta ),
 \eeq
where $C_\nn$ is a constant depending only on the $\alpha_i$'s
$$
C_\nn=(\prod_{k=1}^N \alpha_1\alpha_k^{-1})^{\sum_{i=0}^{n-1}
\vert\intF_{<i>}\vert N^{n-1-j}}.
$$
  \epr

\noindent N.B.: For any matrix $Q$ in $\symG$, $Q_\nn$ is the
$G$-invariant, symmetric $F_\nn\times F_\nn$ matrix with complex
coefficients defined by formula (\ref{f.1.6}), where $A$ is
replaced by $Q$.
 \ali
Proof: i) The map $\tau_{\beta_1,\ldots ,\beta_N}(X)$ is obviously
homogeneous polynomial of degree $N$ in the coefficients of $X$.
The map $R_{F_{<1>}\rightarrow \partial F_{<1>}}$ is linear.

ii) We clearly have, using, for example, the variational formula
of proposition \ref{p.2.1} that $T^nQ$ is the trace on $\partial
F_\nn$ of $Q_\nn$, i.e.
\begin{eqnarray}\label{f.Z.3.1}
T^nQ= (Q_\nn)_{\partial F_\nn}
\end{eqnarray}
(where as usual we identify $\partial F_\nn$ and $F$).
 Iterating the map $R$ we have
 \beq \label{f.3.1.1}
R^n (\exp (\oeta Q\eta ))= C_\nn i_{\Pi_{x\in \intF_{<n>}} \oeta_x
\eta_x} \left( \exp (\oeta Q_\nn \eta )\right),
 \eeq
 where as usual we identify the points of $\partial F_\nn$ with
 the points of $F$.
Let us prove the last formula by recurrence. Suppose that for
$n>0$
$$
R^{n-1} (\exp (\oeta Q\eta ))= C_{<n-1>} i_{\Pi_{x\in
\intF_{<n-1>}} \oeta_x \eta_x} \left( \exp (\oeta Q_{<n-1>} \eta
)\right).
$$
By definition, for all $k$ in $\unN$, if we identify $F_{<1>,k}$
with $\partial F_{\nn,k}$ we get
\begin{eqnarray*}
&&(\alpha_1\alpha_k^{-1})^{\vert \intF_{<n-1>}\vert}
s_k(\tau_{\alpha_1\alpha_k^{-1}}(R^{n-1}(\exp\oeta Q\eta)))
\\
&=& C_{<n-1>} i_{\Pi_{x\in \intF_{<n>,k}} \oeta_x \eta_x} \left(
\exp (\alpha_1\alpha_k^{-1} \oeta Q_{<n>,k} \eta )\right).
\end{eqnarray*}
 Considering that the term $\exp\oeta Q_{\nn,k}\eta$ does not
contain any term $\oeta_x,\eta_x$ for $x$ in $\intF_{\nn,k'}$ and
$k'\neq k$, we see that
$$
\prod_{k=1}^N \left( i_{\prod_{x\in
\intF_{\nn,k}}\oeta_x\eta_x}\left( \exp \alpha_1\alpha_k^{-1}\oeta
Q_{\nn,k}\eta\right)\right) = i_{\prod_{k=1}^N\prod_{x\in
\intF_{\nn,k}}\oeta_x\eta_x}(\exp\oeta Q_\nn \eta)
$$
since $Q_\nn=\sum_{k=1}^N \alpha_1\alpha_k^{-1} Q_{\nn,k}$.
Formula (\ref{f.3.1.1}) follows directly this last equality since
$C_\nn=C_{<n-1>}^N(\prod_k\alpha_1^{-1}\alpha_k)^{\vert\intF_{<n-1>}\vert}$.
We see then that formula (\ref{f.3.1}) is a direct consequence of
formula (\ref{f.3.1.1}), (\ref{f.Z.3.1})  and proposition
(\ref{p.2.2}). $\Box$
 \ali

We denote by $\ppp(\aaa)$ the projective space associated with
$\aaa$ and by $\pi:\aaa\rightarrow \ppp(\aaa)$ the canonical
projection.
We denote by $\lll$ the closure in $\ppp(\aaa)$ of the elements of the
type $\pi(\exp(\oeta Q\eta))$ for $Q\in \symG$.
We know from section 2.2.2 that $\lll$ is a smooth subvariety of the
Lagrangian Grassmannian $\BL^\cF$, with the same dimension as $\symG$.
We remark from (\ref{f.3.1}) that $\pi^{-1}(\lll)\cup\{0\}$ is invariant by
$R$.

\section{The main theorem in the lattice case}
We define the Green function (cf appendix B) associated with $R$ as the
map $G\;:\; \aaa \rightarrow \BR\cup \{-\infty \}$ given by
$$ G(x)=\lim_{n\to\infty}
{1\over N^n} \ln \| R^n (x) \|,\;\;\; x\in \aaa.
$$
This limit always exists and is a plurisubharmonic function
(cf Appendix A \& B).
This function is related to the dynamics of the map on $\ppp(\aaa)$
induced by $R$ (cf appendix B).

For $x\in \LG$ we denote by $\rho_n(x)$ the order of vanishing at
$x$ of the restriction of the map $R^n$ to the submanifold $\LG$:
precisely if  $s:U\rightarrow \aaa$ is a local holomorphic section
of the projection $\pi$ on an open subset $U\subset \LG$
containing $x$, then we denote by $\rho_n(x)=\ord(R^n\circ s,x)$
the order of vanishing  of the function $R^n\circ s$ at the point
$x$ (cf section 2.2). The functions $\rho_n$ satisfy:
$$\rho_{n+1} (x)\ge N \rho_n (x).
$$
since $R$ is homogeneous of degree $N$.
Thus the limit
$$
\rho_\infty (x)=\lim_{n\to\infty}  {1\over N^n} \rho_n (x),
$$
exists since $\rho_n$ is bounded by $N^n$, the
degree of $R^n$.

Remind that in section 1.2.1  we fixed a difference operator $A$
with real coefficients and a positive measure $b$ on $F$. We
denote by $I_{b_\nn}$ (and simply $I_b=I_{b_{<0>}}$ for $n=0$) the
diagonal $F_{\nn}\times F_\nn$ matrix with diagonal terms
$(I_{b_\nn} )_{x,x}= b_\nn(x)$. We denote by $\phi\;:\;
\BC\rightarrow \aaa$ the map
 \beq \label{f.3.2.0}
  \phi(\lambda)=
\exp \oeta (A-\lambda I_b )\eta, \;\;\; \lambda \in \BC.
 \eeq
 \noindent\Rm:
Remark that the map $\phi$ is polynomial (cf formula
(\ref{f.2.2002})).
\begin{propos}
\label{p.3.3}
i) The Neumann and Dirichlet spectrum are related to the map $R$ by
the following formulas:
\beq
\label{f.3.2.1}
\nu_\nn^- &=&{1\over 2\pi}
\Delta \ln \vert <R^n\circ \phi(\lambda ),1> \vert
\\
\label{f.3.3.1} \nu_\nn^+ &=& {1\over 2\pi} \Delta \ln \vert
<R^n\circ \phi(\lambda ),\Pi_{x\in F} \oeta_x\eta_x> \vert \eeq
where $\Delta$ denotes the distributional Laplacian.

ii) The Neumann-Dirichlet spectrum is related to the zeroes of
$R^n$ by the following formula:
\begin{eqnarray}
\label{f.3.4} \nu^\ND_\nn =\sum_\lambda  \rho_n (\pi(\phi(\lambda
))) \delta_\lambda.
 \eeq \epr
 \noindent N.B.: $\delta_\lambda$ is the Dirac
mass at $\lambda$. The terms in the sum (\ref{f.3.4})  are non
null only for a finite set of points $\lambda$.
 \ali N.B.: we
recall that $<\cdot ,\cdot >$ appearing in formulas
(\ref{f.3.2.1}), (\ref{f.3.3.1}) is the scalar product on $\aaa$
defined in section 2.2. \ali Proof: Using (\ref{f.3.1.1}) and
proposition (\ref{p.2.2}) we get
 \beqn
C_\nn \det ((A_\nn -\lambda I_{b_{<n>}})_{|\intF_\nn}) =<R^n\circ
\phi(\lambda ),1>,
\\
C_\nn \det (A_\nn -\lambda I_{b_{<n>}}) =<R^n\circ \phi(\lambda
),\Pi_{x\in F} \oeta_x\eta_x>. \eeqn Hence, formulas
(\ref{f.3.2.1}) and (\ref{f.3.3.1}) come from the classical
${1\over 2 \pi} \Delta \ln \vert \lambda \vert =\delta_0$.
 \ali
ii) The map $Q\rightarrow \pi(\exp(\oeta Q\eta))$ is locally
invertible from $\symG$ to $\BL^G$. Hence, for any $Q_0$ in
$\symG$, we have
$$\rho_n(\pi(\exp\oeta Q_0\eta))=\ord(Q\rightarrow R^n(\exp(\oeta
Q\eta)), Q_0),
$$
 for $Q$ running in $\symG$. By formula
(\ref{f.3.1.1}) we have
$$
R^n(\exp\oeta Q\eta)=C_\nn R_{F_\nn\rightarrow \partial
F_\nn}(\exp\oeta Q_\nn \eta),
$$
for any $Q$ in $\symG$. This implies that
\begin{eqnarray*}
\rho_n(\pi(\exp\oeta Q_0\eta))\ge \ord(Q\rightarrow
R_{F_\nn\rightarrow
\partial F_\nn}(\exp\oeta Q \eta), (Q_0)_\nn),
\end{eqnarray*}
where in the right-hand side $Q$ runs in the set of $F_\nn\times
F_\nn$, complex symmetric matrices. For $Q_0$ real, the right-hand
side equals $\dim\ker^\ND ((Q_0)_\nn)$ by proposition \ref{p.2.3}.
On the other hand, we have
\begin{eqnarray*}
\rho_n(\pi(\exp\oeta Q_0\eta))&\le& \ord(\lambda \rightarrow
R^n(\exp\oeta(Q_0-\lambda I_b)\eta), 0)
\\
&=& \ord(\lambda\rightarrow R_{F_\nn\rightarrow \partial F_\nn}(
\exp\oeta((Q_0)_\nn-\lambda I_{b_\nn})), 0)
\end{eqnarray*}
and this last expression equals $\dim\ker^\ND((Q_0)_\nn)$ when
$Q_0$ is real, by proposition \ref{p.2.3}, ii). Thus we proved
that for all $Q_0$ real we have
\begin{eqnarray} \label{f.3.2002}
\rho_n(\pi(\exp\oeta Q_0\eta))=\dim\ker^\ND((Q_0)_\nn).
\end{eqnarray}
Formula (\ref{f.3.4}) is a direct consequence of this last formula
since for any $\lambda_0$ in $\BR$ we have
$$
\nu_\nn^\ND(\{\lambda_0\}) =\dim\ker^\ND((A-\lambda_0I_b)_\nn)=
\rho_n(\pi(\phi(\lambda_0))).
 \Box
 $$
\begin{thm}
\label{t.3.1}
i)
The density of states is given by the following formula
\beq
\label{f.3.6}
\mu={1\over 2\pi} \Delta (G\circ \phi ).
\eeq

ii) The density of Neumann-Dirichlet eigenvalues is given by \beq
\label{f.3.7} \mu^\ND =\sum_\lambda \rho_\infty(\pi(\phi (\lambda
)))\delta_\lambda. \eeq
\end{thm}
\noindent \Rm: Remark from formula (\ref{f.3.2.0}) that $\phi$ is
holomorphic,  thus $G\circ \phi$ is subharmonic and $\Delta
(G\circ \phi)$ defines a positive measure.
 \ali \Rm: By
construction, we have $\supp \mu\subset \BR$. This implies that
$G\circ \phi$ is harmonic on $\BC\setminus \BR$. This property can
be seen directly from the dynamics of $g$. We know from lemma
\ref{l.3.1} that $g$ is holomorphic on the Siegel upper-half space
$S^+$, and that $S^+$ is left invariant by $g$. On the other hand,
$S^+$ is hyperbolic in the sense of Kobayashi, cf for example,
\cite{Sibony1}, definition 2.1. This implies that $S_+^G+$ is in
the Fatou set of $g$, that $G$ is pluriharmonic on $S_+^G$, and
thus that $G\circ \phi$ is harmonic on $\{\lambda, \;
\im(\lambda)>0\}$, since $\phi(\lambda)\in S_+^G$ for $\im
(\lambda)<0$. \ali \ali
 Proof: The proof of ii) is a direct
application of proposition (\ref{p.3.3}), ii).
 \ali
  i) By general
results on subharmonic functions (cf for example,
\cite{Hormander}) w.e know from formulas (\ref{f.3.2.1}) and
(\ref{f.3.3.1}) that the weak convergence ${1\over N^n}
\nu_\nn^\pm \rightarrow {1\over 2\pi} \Delta (G\circ \phi)$ would
be implied by the following convergence in $L^1_\loc(\BC)$:
 \beq \label{f.3.8.0}
\lim_{n\to\infty}
 {1\over N^n} \ln \vert <R^n\circ \phi,1>\vert &=& G\circ \phi,
\\
\label{f.3.9.0}
\lim_{n\to\infty}
 {1\over N^n} \ln \vert <R^n\circ \phi,\Pi_{x\in F}\oeta_x\eta_x>\vert
&=& G\circ \phi.
 \eeq
This will follow from
\begin{lem}
\label{l.3.1}
For all $Q$ in $S_+$
\beq
\label{f.3.8}
\lim_{n\to\infty}
 {1\over N^n} \ln \vert <{R^n(\exp \oeta Q\eta )\over
\|R^n(\exp \oeta Q\eta )\|} ,1>\vert &=&0,
\\
\label{f.3.9}
\lim_{n\to\infty}
 {1\over N^n} \ln \vert <{R^n(\exp \oeta Q\eta )\over
\|R^n(\exp \oeta Q\eta )\|},\Pi_{x\in F}\oeta_x\eta_x>\vert
&=&
0.
\eeq
\end{lem}
Let us finish the proof of the theorem before starting the proof
of lemma (\ref{l.3.1}). We only prove (\ref{f.3.8.0}), the proof
of (\ref{f.3.9.0}) is strictly identical. We write: \beq \nonu
{1\over N^n} \ln \vert <R^n\circ \phi(\lambda),1>\vert = {1\over
N^n} \ln \vert <{R^n\circ \phi(\lambda)\over \|R^n\circ
\phi(\lambda)\|} ,1>\vert + {1\over N^n} \ln \|R^n\circ
\phi(\lambda) \| . \eeq The first term of the right-hand side is
negative since $\hbox{$\vert <{R^n\circ \phi(\lambda)\over
\|R^n\circ \phi(\lambda)\|} ,1>\vert$} \le 1$, and converges to 0
for $\lambda \in \BC\setminus \BR$ since $A-\lambda I_\w$ is in
$S_+$ or in $-S_+$ for $\lambda $ in $\BC\setminus \BR$, using
lemma (\ref{l.3.1}). Therefore, we know that the sequence of psh
functions ${1\over N^n} \log\vert <R^n\circ \phi ,1>\vert$ is
uniformly bounded from above and converges pointwise to $G\circ
\phi$ for $\lambda $ in $\BC\setminus \BR$. Using proposition
(\ref{ap.1}) of appendix (or proposition 3.2.12 of
\cite{Hormander}) we know that (\ref{f.3.8.0}) is true for the
convergence in $L^1_\loc(\BC)$. $\Box$
 \ali
  Proof of lemma (\ref{l.3.1}).
We first remark that the terms of the sequences in formulas
(\ref{f.3.8}) and (\ref{f.3.9}) are non-positive. By proposition
(\ref{p.3.2}) and lemma (\ref{l.2.0}) we have:
 \beqn \vert
<{R^n(\exp \oeta Q\eta )\over \|R^n(\exp \oeta Q\eta )\|},1>\vert
&=& \vert <{\exp \oeta T^n Q\eta \over \|\exp \oeta T^n
Q\eta \|} ,1>\vert
\\
&=& {1\over  \|\exp \oeta T^n Q\eta \|}\\
&\ge&{1\over (1+\left(\overline \rho (T^n Q)\right)^2)^{{\vert
F\vert\over 2}}}= \left( {\left(\underline \rho ((T^n
Q)^{-1})\right)^2\over 1+\left(\underline \rho ((T^n
Q)^{-1})\right)^2} \right)^{{\vert F\vert\over 2}}.
 \eeqn
where $\overline \rho (T^n Q)$ is the maximal characteristic root
of $T^n Q$ and $ \underline \rho ((T^n Q)^{-1})$ the minimal
characteristic root of $(T^n(Q))^{-1}$. Since $(T^n Q)^{-1}$ is
symmetric we have
$$
 \underline \rho ((T^n Q)^{-1})\ge \underline \rho (\im ((T^nQ)^{-1}))\ge
\alpha_1^n \underline \alpha^{-n} \underline \rho(\im Q^{-1}),
$$
using lemma (\ref{l.3.0.0})  in the last inequality. This is
enough to prove formula (\ref{f.3.8}).
 \ali
The proof of formula (\ref{f.3.9}) works similarly. Using
proposition (\ref{p.3.2}) and lemma (\ref{l.2.0})  we get:
 \beqn
\vert <{R^n(\exp \oeta Q\eta )\over \|R^n(\exp \oeta Q\eta
)\|},\Pi_{x\in F }\oeta_x\eta_x>\vert &=& {\vert \det(T^nQ)\vert
\over \|\exp \oeta T^n Q\eta \|}
\\
&\ge & {1\over (1+ \left(\overline \rho
((T^nQ)^{-1})\right)^2)^{{\cF\over 2}}}
\\
&=& \left( {\left(\underline \rho (T^n Q)\right)^2\over
1+\left(\underline \rho (T^n Q)\right)^2} \right)^{{\vert
F\vert\over 2}},
 \eeqn
and we conclude similarly using lemma (\ref{l.3.0.0}). $\Box$

\section{The continuous case}
We know that the Dirichlet forms $(a_\nn,\ddd^\pm_\nn)$ introduced in
section 1.2.2  satisfy the conditions i), ii), iii) of section  2.3
(cf \cite{Sabot1}).
For $\lambda\ge 0$, we denote by $a_{\nn, \lambda}$ the Dirichlet form defined by
$a_{\nn, \lambda}(f,g)=a_\nn (f,g)+\lambda \int fg dm_\nn$, and
by $A_{\nn,(\lambda)}$ the symmetric $F_\nn\times F_\nn$ matrix defined
by
$$<A_{\nn,(\lambda )}\cdot , \cdot >=(a_{\nn, \lambda})_{F_\nn}(\cdot ,\cdot).$$
We simply write $A_{(\lambda)}=A_{<0>,(\lambda)}$.
Using lemma (\ref{l.2.2}) we see that $A_{\nn,(\lambda)}$ can be extended to
a meromorphic function on $\BC$ with poles included
in the spectrum of $a_{\nn,\lambda}$ with
Dirichlet condition on $F_\nn$, i.e. in
the spectrum of $(a,\ddd^-)$ (indeed, $X_\nn\setminus F_\nn$ is the disjoint union of
$N^n$ copies of $\intX$).
\ali
We see that
\beq
\label{f.3.c.1}
A_{\nn,(\lambda)}=(A_{(\lambda)})_\nn,
\eeq
with $(A_{(\lambda)})_\nn$ defined by formula (\ref{f.1.6}) (where
$A$ is replaced by $A_{(\lambda)}$).
\begin{propos}
For all $\lambda\in \BC$ the following equality is satisfied (when the
terms are defined)
$$\alpha_1^{-1} T(A_{(\lambda)})=A_{(\gamma \lambda)}.$$
\end{propos}
\Rm:
\label{r.3.c.1}
This means that, at least locally, $A_{(\lambda)}$ is a holomorphic
curve invariant by the map $\alpha_1^{-1} T$.
We remark that $A_{(0)}$ is a fixed point of $\alpha_1^{-1} T$
(in general the existence of this fixed point is equivalent to
the existence of a self-similar diffusion
on the fractal, cf \cite{Sabot1}) and
that the direction  $({d\over d\lambda}A_{(\lambda)})_{\lambda=0}$ is an
instable direction of $\alpha_1^{-1} T$ since $\gamma>1$.
\ali
Proof:
For any $f$ in $\BR^{\partial F_{<1>}}\sim \BR^F$ we have:
\begin{eqnarray}
\nonumber
\alpha_1^{-1} T(A_{(\lambda)})(f)&=&
\alpha_1^{-1} (A_{<1>,(\lambda )})_{\partial F_{<1>}}(f)
\\
\nonumber
&=&
\alpha_1^{-1} ((a_{<1>,\lambda })_{F_{<1>}})_{\partial F_{<1>}} (f)
\\
\nonumber
&=&
\alpha_1^{-1} (a_{<1>,\lambda })_{\partial F_{<1>}} (f)
\\
\nonumber
&=&
(a_{\gamma \lambda })_F (f),
\end{eqnarray}
where the last relation comes from the scaling relation
$a_{<1>,\lambda}=\alpha_1 a_{\gamma \lambda}$.
$\Box$
\ali

Denote by $d_\nn^\pm$ the infinite dimensional determinant of
$(a_\nn, \ddd_\nn^\pm)$, as in section 2.3, and simply
$d^\pm=d_{<0>}^\pm$. The infinite dimensional determinant of
$a_\nn$ with Dirichlet boundary conditions on $F_\nn$ is equal, up
to a constant, to $(d^-)^{N^n}$ (indeed, $X_\nn\setminus F_\nn$ is
the disjoint union of $N^n$ copies of $\intX$). Hence, if we apply
lemma (\ref{l.2.3}) to $X_\nn$ and $F_\nn$ we see that \beq
\label{f.3.c.2} \det (A_{\nn,(\lambda)})=c_\nn^+
{d_\nn^+(\lambda)\over d^-(\lambda)^{N^n}},
\\
\label{f.3.c.3}
\det((A_{\nn,(\lambda)})_{|\intF_\nn})
=c_\nn^- {d_\nn^-(\lambda)\over d^-(\lambda)^{N^n}},
\eeq
for some constants $c^\pm_\nn$.

\begin{propos}
\label{p.3.c.1}
Formula (\ref{f.3.2.1}), (\ref{f.3.3.1}) and (\ref{f.3.4}) of
proposition (\ref{p.3.3})  are true in the continuous case
on the open ball $B(0,\vert \lambda^-_1\vert )$ with center 0
and radius $\vert \lambda_1^-\vert$ and when the function $\phi\;:\; B(0,\vert \lambda_1^-\vert)
\rightarrow \aaa$
is replaced by
\beq
\label{f.3.c.4}
\phi(\lambda)=\exp(\oeta A_{(\lambda)}\eta).
\eeq
\end{propos}
N.B.: Remind that $\lambda_1^-$ is the first Dirichlet eigenvalue
of $a$, and that $A_{(\lambda)}$ is well-defined on
$B(0,\vert\lambda^-_1\vert )$. \ali Proof: On $B(0,\vert
\lambda_1^-\vert )$ we have, using formula (\ref{f.3.c.2}), and
proposition (\ref{p.2.2}),
 \beqn
(\nu_\nn^+)_{|B(0,\vert\lambda_1^-\vert)}&=& ({1\over 2\pi}\Delta
\ln \vert d^+_\nn\vert)_{|B(0,\vert\lambda_1^-\vert)}
\\
&=& ({1\over 2\pi} \Delta \ln \vert \det A_{\nn,(\cdot )}\vert
+N^n \nu^-_{<0>})_{B(0,\vert\lambda_1^-\vert)}
\\&=&
({1\over 2\pi} \Delta \ln \vert <R^n\circ \phi(\lambda ),\Pi_{x\in
F} \oeta_x\eta_x> \vert)_{|B(0,\vert\lambda_1^-\vert)}.
 \eeqn
This proves formula (\ref{f.3.3.1}). The proof of formula
(\ref{f.3.2.1}) is similar. To prove formula (\ref{f.3.4})  we
first prove  that
\begin{eqnarray}\label{f.3.2003}
\dim E_{\nn,\lambda}^\ND=\dim
\ker^\ND(A_{\nn,(\lambda)})=\dim\ker^\ND((A_{(\lambda)})_\nn)
\end{eqnarray}
for $\lambda_1^-<\lambda\le 0$, where $E_{\nn,\lambda}^\ND$
denotes the vector space generated by the N-D eigenvalues of
$a_\nn$ with eigenvalues larger that $\lambda$. When this formula
is proved then formula (\ref{f.3.4}) is a direct consequence of
formula (\ref{f.3.2002}). Remark that $E_{\nn,\lambda}^\ND
\cap\{f\in \ddd_\nn,\; f_{|F_\nn}=0\}=\{0\}$ (otherwise, there
would be a Dirichlet eigenvalue of $(a,\ddd^-)$ with absolute
value smaller than $\vert \lambda^-_1\vert $). Hence, if $f$ is in
$E_{\nn,\lambda}^\ND$ then $f_{|F_\nn}$ is non-null and in
$\ker^\ND(A_{\nn,(\lambda)})$. Conversely, consider $g$ in
$\ker^\ND(A_{\nn,(\lambda)})$, its harmonic prolongation with
respect to $a_{\nn,\lambda}$ is well-defined and is in
$E_{\nn,\lambda}^\ND$. This means that the map $f\rightarrow
f_{|F_\nn}$ is a bijection from $E_{\nn, \lambda}^\ND$ to
$\ker^\ND(A_{\nn,(\lambda)})$, thus this proves the first equality
of (\ref{f.3.2003}). The second equality is given by relation
(\ref{f.3.c.1}). $\Box$
\begin{thm}
\label{t.3.c.1}
All formulas of theorem (\ref{t.3.1}) are true in the continuous case
on the ball $B(0,\vert \lambda_1^-\vert )$ and when $\phi$ is
given by (\ref{f.3.c.4}) as in proposition
(\ref{p.3.c.1}).
\end{thm}
Proof: It is similar to the lattice case, using lemma
(\ref{l.3.1}). We just have to check that $A_{(\lambda)}$ is in
$S_+$ for $\im(\lambda)<0$. This follows the classical relation
$$
\im (A_{(\lambda)}(f,f))= -\im (\lambda)\int \vert H_\lambda
f\vert^2dm,
$$
for $f\in \BR^F$ and $H_\lambda f$ the harmonic prolongation of
$f$ with respect to $a_\lambda$ (we remark from the explicit
expression of $H_\lambda f$ given in the proof of lemma 2.1 of
\cite{Sabot3} that $H_\lambda f$ admits an analytic prolongation
for $\im\lambda <0$). $\Box$

\chapter{Analysis of the psh function $G_{|\pi^{-1}(\lll)}$}
\setcounter{section}{0}
\section{The dichotomy theorem}
In this part we analyze the structure of the Green function $G$
restricted to the subvariety $\pi^{-1}(\lll)$, or equivalently the
current on $\lll$ with potential $G_{|\pi^{-1}(\lll)}$. This will
give some information on the measures $\mu^\ND$ and $\mu$. The
structure of this current is related to the dynamics of the
restriction to $\lll$ of the map induced by $R$ on $\ppp (\aaa)$
 (indeed as seen in section 3, the
subvariety $\pi^{-1}(\lll)$ is invariant by $R$). We first describe
precisely this map.
As in appendix C, we define the meromorphic map $g\;:\;\lll\rightarrow \lll$
and its iterates $g^n$
by their graph
$\Gamma_{g^n}\subset \lll\times \lll$ constructed as the
closure of the graph
\begin{eqnarray}
\label{f.4.1} \Gamma_{g^n}^0=\{(\pi(x),\pi(R^n(x))),\;\;\; x\in
\pi^{-1}(\lll)\setminus \{R^n(x)=0\} \}. \eeq We denote by
$\pi_1,\pi_2\;:\; \lll\times \lll\rightarrow \lll$ the projection
on the first and second coordinates. The set of indeterminacy
points of $g^n$, denoted $I_{g^n}$  is defined (cf appendix C) as
the set  of points where $\pi_1\;:\;\Gamma_{g^n}\rightarrow \lll$
is not a local biholomorphism. On $\lll\setminus I_{g^n}$ the map
is defined by $g^n(x)=\pi_2(\pi_1^{-1}(x))$. The codimension of
$I_{g^n}$ is at least 2 and it will be useful to describe the
structure of the set $I_{g^n}$ in terms of the map $R^n$. This can
be done locally: let $x$ be a point in $\lll$ and $U$ an open
subset of $\lll$ containing $x$, identified with a subset of
$\BC^{\dim \lll}$ by a local set of coordinates. If $s$ is a
section of the projection $\pi$ on the subset $U$ then we can find
holomorphic functions $f_1,\ldots ,f_k$ on $U$ and positive
integers $c_1,\ldots ,c_k$ such that
\begin{itemize}
\item
we can write
\beq
\label{f.4.3}
R^n\circ s =f_1^{c_1}\cdots f_k^{c_k} \tilde R_n,
\eeq
where $\tilde R_n$ is a holomorphic function from $U$ to $\aaa$ such that the set
$\{\tilde R_n=0\}$
 is at least of codimension 2.
\item
The analytic sets
\beq
\label{f.4.4}
Z_i=\{f_i=0\}
\eeq
are irreducible and
$f_i$ is a generator of the ideal
$\iii (Z_i)=\{f\hbox{ holomorphic on $U$},\; f=0 \hbox{ on $Z_i$}\}$.
\end{itemize}
Then the set of indeterminacy points $I_{g^n}\cap U$ is exactly
the set $\{\tilde R_n=0\}$ and on the set $\{\tilde R_n\ne 0\}$
the map $g^n$ is given by $g^n(x) =\pi(\tilde R_n(x))$. \ali This
gives us the opportunity to introduce the divisor associated with
the hypersurfaces of zeroes of the restriction of $R^n$ to $\lll$.
Precisely, we call divisor a formal sum $\sum_Z c_Z Z$ where the
sum runs over the set of irreducible subvarieties of codimension 1
of $\lll$ and the coefficients $c_Z$ are integers and null except
for a finite set of indices. We define $D_n=\sum c_Z Z$ as the
divisor on $\lll$ such that for any subset $U\subset \lll$ we have
\beq \label{f.4.4.1} (D_n)_{|U}= c_1 Z_1 +\cdots + c_k Z_k \eeq
where $Z_i$ and $c_i$ are defined by (\ref{f.4.3}) and
(\ref{f.4.4}) (the restriction to $U$ of an irreducible subvariety
$Z$ can be decomposed as a sum of irreducible hypervarieties of
$U$: this naturally allows us to represent $D_n\cap U$ as a
divisor on $U$ i.e. as a formal positive sum of irreducible
hypervarieties of $U$).

Since $R$ is of degree $N$ it is easy to see that \beq
\label{f.4.4.1.1}
D_{n+1}\ge N D_n
\eeq
for the natural ordering on divisors.

We denote by $S$ (resp. $S_n$) the closed positive $(1,1)$ current on
$\lll$ with potential $G_{|\pi^{-1}(\lll)}$ (resp. $(G_n)_{|\pi^{-1}(\lll)}$
where $G_n=\ln \| R^n\|$)
i.e. $S$ and $S_n$ are defined locally on an open subset $U$ of $\lll$ by
$$ S =d d^c G\circ s, \;\;\; S_n =dd^c \log \|R^n\circ s\|,
$$
if $s$ is a holomorphic section of the projection $\pi$ on $U$. By
definition the current $S_0$ is the restriction to $\lll$ of the
Fubini-Study form  defined on the projective space $\ppp (\aaa)$
and is therefore a K\"ahler form on $\lll$ (cf Appendix A.5). The
currents $S_n$ are well-defined for all $n$ since $G_n$ is not
equal to $-\infty$ on $\lll$ (indeed, we know that $R^n(\exp(\oeta
Q\eta))$ is non-null for $Q$ in $S_+$, cf lemma (\ref{l.3.0.0})).
Since ${1\over N^n} G_n$ converges pointwise in $\BR \cup
\{-\infty \}$ to $G$ we know that \beq \label{f.4.4.2} \lim_{n\to
\infty } {1\over N^n} S_n =S \eeq for the topology of current.

In Appendix C.2 we define the pull-back of a positive closed $(1,1)$ current by a
meromorphic map.
With this definition we have the following result.
\begin{propos}
\label{p.4.1}
For all integer $n$ we have
\beq
\label{f.4.5}
S_n=(g^n)^* S_0 + [D_n].
\eeq
\end{propos}
N.B.: For a divisor $D=\sum c_Z Z$ we denote by $[D]$ the current $[D]=\sum c_Z [Z]$ where
$[Z]$ is the current of integration on
the hypervariety $Z$ (cf appendix A).
\ali
Proof:
Let $U$ be an open subset of  $\lll$ identified with an open subset of
$\BC^{\dim \lll}$ thanks to a local chart.
If $s$ is a holomorphic section on $U$ of the projection $\pi$
then $R^n\circ s$ can be written (cf
formula (\ref{f.4.3}))
$$
R^n\circ s= f_1^{c_1}\cdots f_k^{c_k} \tilde R_n$$
and the map  $g^n$ is defined on the set $\{\tilde R_n\ne 0\}$ by
$$
g^n(x) =\pi (\tilde R_n (x)).
$$
The current $S_n$ is defined on $U$ by
\beq
\nonumber
S_n&=&dd^c G_n\circ s\\
\nonumber &=& dd^c \ln \|\tilde R_n\| +c_1dd^c \ln \vert f_1\vert
+\cdots + c_{k}dd^c\ln \vert f_{k}\vert, \eeq with notations of
formula (\ref{f.4.3}) and (\ref{f.4.4}). By Lelong-Poincar\'e
formula (cf appendix A) $dd^c \ln \vert f_i\vert =[Z_i]$ and we
will show that $dd^c \ln \| \tilde R_n\| =(g^n)^* S_0 $ on
$U\setminus \{\tilde R_n =0\}$, which implies the equality on all
$U$ since by Siu theorem (cf for example, \cite{Demailly1}), a
$(1,1)$ positive closed current cannot charge analytic subset of
codimension strictly larger than 1. Let $x_0$ be in $U\setminus
\{\tilde R_n=0\}$ and set $w_0=g(x_0)$. Let $r>0$ and $r_1>0$ be
such that $g(B(x_0,r_1))\subset  B(w_0,r)$. Let $\tilde s$  be a
holomorphic section of the projection $\pi$ on $B(w_0,r)$. We can
write $\tilde R_n(z)= j(z) \tilde s\circ g^n(z)$ on $B(x_0,r_1)$
where $j$ is a holomorphic function which does not take the value
0 on the set $\{\tilde R_n\ne 0\}$. This implies by definition
$dd^c \log \|\tilde R_n\| = (g^n)^* S_0$ on $B(x_0,r_1)\setminus
\{\tilde R_n=0\}$. $\Box$ \ali

Considering equation (\ref{f.4.4.1.1}) and  relation
(\ref{f.4.4.2}) we know that the limit of ${1\over N^n} [D_n]$
exists and that the limit of ${1\over N^n} (g^n)^* S_0$ exists.
The question is now to know whether these limits can be non null
at the same time. The following proposition answers the question.
\begin{propos}
\label{p.4.2}
If $D_n\ne 0$ for an integer $n$ then
$$\lim_{n\to\infty} {1\over N^n} (g^n)^* S_0 =0.$$
\end{propos}
The proof of this proposition relies on considerations on the cohomology classes of the
currents $S_n$ and is sent to section 4.2.
Actually, this result would be straightforward if $\lll$ was a projective space :
it would be obtained by simple considerations on the degree as in the proof of
theorem (\ref{at.2}) of Appendix B.
We deduce from the last proposition the following results.
\begin{thm}
\label{t.4.1}
i) If $D_n\ne 0$ for an integer $n$ then
$$ S=\lim_{n\to\infty}  {1\over N^n} [D_n]
$$
and $S$ is a countable sum of current of integration on hypersurfaces of $\lll$.
\ali
For both the continuous and the lattice case
(i.e. for any choice of $(A,b)$ in the discrete case)
we have:
\beq
\label{f.4.6}
\mu^\ND =\mu,
\eeq
in particular, for almost all blow-up the spectrum of the operator is
pure point and the eigenfunctions have compact support.

ii)
If $D_n=0$ for all $n$ then the map $g$ is algebraically stable (cf appendix C),
$$S=\lim_{n\to\infty} {1\over N^n} (g^n)^* S_0 $$
and $S$ is the Green current of $g$. In particular, the current $S$ does
not charge hypersurfaces and the support of
$S$ is included in the Julia set of $g$.

Moreover, in the lattice case for a generic choice of $(A,b)$ we have
$$\mu^\ND =0.$$
\end{thm}
\noindent N.B.: For a generic choice of $(A,b)$ means for any
$(A,b)$ in the complement of a proper  analytic subset. \ali \Rm:
The interesting information in this result is that a dichotomy
appears between situations where either the N-D eigenvalues
contribute for all the density of states or generically do not
exist. \ali \Rm: In proposition (\ref{p.4.4}) we will relate this
dichotomy theorem with an asymptotic degree associated with $g$.
\ali \ali Proof: i) Since ${1\over N^n} [D_n] $ is increasing it
is obvious that its limit will be a countable sum of currents of
integration. It is equal to $S$ by proposition (\ref{p.4.2}).

By proposition (\ref{p.3.3}), ii)  we know that $(\pi\circ\phi)^* [D_n]\le \mu_\nn^\ND $
($(\pi\circ \phi)^* [D_n]$ is the pull-back of the current $[D_n]$ as defined in
appendix A.6, i.e. on an open subset $U$ the current $[D_n]$ has potential
$\sum {c_i}\log\vert f_i\vert$
where the $c_i$'s and $f_i$'s come from formula (\ref{f.4.3}),
the pull-back is then defined
by $(\pi\circ\phi)^*[D_n]=dd^c \sum c_i \log \vert f_i\circ\pi\circ\phi\vert$).
There is not equality a priori since it may happen that the curve $\phi$ meets
some component of $\{R^n=0\}$ of codimension larger than 2 which do not appear in
$[D_n]$.
This implies that $(\pi\circ \phi)^* S \le \mu^\ND$, but $(\pi\circ \phi)^* S = \mu$ by
theorem (\ref{t.3.1}), so $\mu^\ND =\mu$.

ii)
We will see in the next section that $S_0$ satisfies an equation
in homology $N^n \{S_0\}=(g^n)^*(\{S_0\})$. Hence, $S$ is by
definition the Green current of $g$ as defined in appendix C.4 and
theorem (\ref{at.3}).
It only remains to prove that for a generic choice of $(A,b)$, $\mu^\ND =0$.
Suppose that $D_n=0$ for all $n$.
We want to prove that for a generic choice of $(A,b)$ the line $A+\lambda I_b$,
$\lambda \in \BC$, does not meet the set $\{R^n=0\}$.
This is equivalent to prove that for any choice of $A$ as in section 1.2.1,
and $D$ positive diagonal, $G$-invariant,
 with $\hbox{Trace} D=1$, in the complement of an analytic subset
of codimension at least 2,
the line $A+\lambda D$ does not meet the set
$R^n=0$. But the map
\beqn
j\;:\; \{A \;\hbox{ as in section 1.2.1}\}\times \{\hbox{$D$ diag. $\ge 0$,
$G$-inv.,  and $\hbox{Tr} D=1$}\}
\times \BC
&\rightarrow &\symG
\\
(A,D, \lambda) &\rightarrow & A+\lambda  D
\eeqn
is a local diffeomorphism for $\lambda\ne 0$.
Therefore the subset $j^{-1} ((R^n)^{-1}\{0\}) \cap \{\lambda \ne 0\}$ is
of codimension at least 2. The set $j^{-1}((R^n)^{-1}\{0\}) \cap \{\lambda = 0\}$
is also of codimension at least 2 since $R^n$ is not identically null on
$j(\{\lambda =0\})$.
This implies that the projection of $j^{-1}((R^n)^{-1}\{0\})$ on
the first 2 components $(A,D)$ is of codimension at least 1.
$\Box$

\section{Proof of Proposition (\ref{p.4.2}).
Structure of the variety $\lll$.}
\subsection{Notations and preliminary results}
We denote by $H^\11 (\lll)$ the $(1,1)$ Dolbeault cohomology group of
$\lll$ (cf Appendix C,
$H^\11$ is equal to the subspace of $H^2(\lll, \BC)$ generated by
the forms of type $(1,1)$).
If $\alpha$ is a $(1,1)$ closed form we denote by $\{\alpha \}$ its cohomology class.
As explained in the appendix the cohomology class of a current $\w$ can also be defined and
is denoted $\{\w\}$.
\ali
We remind from the appendix that if
$\alpha$ is a positive closed $(1,1)$ current then its
pull-back $(g^n)^*\alpha$ is well defined and
that if $\alpha$ is a smooth differential form
then
\begin{itemize}
\item
$(g^n)^* \alpha $ is in $L^1_\loc (\lll)$,
\item
$(g^n)^* \alpha $ is smooth on $\lll\setminus I_{g^n}$.
\end{itemize}
Moreover, the pull back $(g^n)^*$ induces a pull-back on cohomology group
$H^\11(\lll)$.
\ali
We first prove the following result.
\begin{lem}
\label{l.4.1}
For all integer $n$ then:
\beq
\label{f.4.9}
\{S_n\}=N^n \{S_0\}=(g^n)^* \{S_0\} +\{[D_n]\}.
\eeq
\end{lem}
Proof: The equality between $\{S_n\}$ and the last term
of the expression is immediately deduced from
proposition (\ref{p.4.1}).
We only have to prove that $\{S_n\}=N^n\{S_0\}$.
Remember that $S_n$ has potential
$G_n=\log\|R^n\|$ on $\pi^{-1}(\lll)$, and $G_n$
satisfies the following homogeneity relation:
$$G_n(\lambda x)= N^n\log\vert \lambda \vert +G_n(x).$$
This immediately implies that $\{S_n\}=N^n\{S_0\}$: indeed,
the function $u(z)=\log\|R^n(z)\|-N^n \log\|z\|$ is well defined on
$\lll$. Thus $S_n-N^n S_0=dd^c u$ and $\{dd^c u\}=0$ by definition. $\Box$

\subsection{Description of the structure of $\BL^G$}
To go further we need to describe the structure of the cohomology
group $\Hunl$. In Appendix E, we describe the topological
structure of $\BL^G$.  The main point is that $\BL^G$ is
isomorphic to the product ${\mathcal L}_0\times \cdots \times
{\mathcal L}_r$, where ${\mathcal L}_i$ is a smooth projective
variety with Betti numbers $b_1=\dim(H^1({\mathcal L}_i)=0$ and
$b_2=\dim(H^2({\mathcal L}_i)=1$. When $G$ is the trivial group
then $\BL^G=\BL^{\vert F\vert}$, and the situation is considerably
simpler. We treat this case separately, for convenience, and the
ready can restrict to this simpler situation, at a first reading.

 Let us be more precise. The structure of
$\BL^G$ depends on the decomposition of $\BR^F$ into real
irreducible representations of $G$. The space $\BR^F$ can be
decomposed into isotopic representation as
\begin{eqnarray}\label{f.4.10}
\BR^F=V_0\oplus\cdots \oplus V_r,
\end{eqnarray}
where each $V_i$ is an isotopic representation equal to the sum of
$n_i$ representations isomorphic to a single representation $W_i$:
we write $V_i=n_i W_i$ for simplicity. We denote by $V_i^\BC$ and
$W_i^\BC$ the complexifications of $V_i$ and $W_i$.  The
representation $W_i$ can be of one of the following  3 types
(mutually exclusive). If $W_i^\BC$ is $\BC$-irreducible we say
that $W_i$ is of type 2. Otherwise $W_i^\BC=U_i\oplus \overline
U_i$, where $U_i$ and its complex conjugate, $\overline U_i$, are
$\BC$-irreducible. If the character of  $U_i$ (and hence of
$\overline U_i$) is not real, then we say that $W_i$ is of type 1.
If the character of $U_i$ (and hence of $\overline U_i$) is real,
then we say that $W_i$ is of type 3 (cf Appendix E for details and
justification).

To state the result of appendix E, we need to introduce three
types of Grassmannians.  We denote by ${\Bbb G}^{n,2n}$ the
Grassmannian of $n$-dimensional subspaces of $\BC^{2n}$. The
Lagrangian Grassmannian $\BL^n$ has been defined in section 2.
Finally we define the orthogonal Grassmannian as follows: ${\Bbb
O}^n$ is the set of $n$-dimensional isotropic subspaces of
$\BC^{2n}$ for the non-degenerate symmetric bilinear form $(\cdot
,K_n \cdot)$, where
$$
K_n=\left( \begin{array}{cc} 0&\Id_n\\ \Id_n&0\end{array}\right).
$$
The set ${\Bbb O}^n$ is a smooth subvariety of ${\Bbb G}^{n,2n}$,
and has 2 connected components. We denote by ${\Bbb S}{\Bbb O}^n$
the connected component which contains the isotropic subspace
$\BC^n\oplus 0$.

These three types of Grassmannian are homogeneous spaces
associated respectively with the classical groups $Gl(2n,\BC)$,
$Sp(n,\BC)$ and $SO(2n,\BC)$ (cf Appendix E for details).

In appendix E, we proved that
$$
\BL^G\simeq {\mathcal L}_0\times \cdots \times {\mathcal L}_r,
$$
where
\begin{itemize}
\item
${\mathcal L}_i\simeq {\Bbb G}^{n_i, 2n_i}$ if $W_i$ is of type 1;
the dimension of ${\mathcal L}_i$ is $n_i^2$.
\item
${\mathcal L}_i\simeq \BL^{n_i}$ if $W_i$ is of type 2; the
dimension of ${\mathcal L}_i$ is $n_i(n_i+1)/2$.
\item
${\mathcal L}_i\simeq {\Bbb S}{\Bbb O}^{2n_i}$ if $W_i$ is of type
3; the dimension of ${\mathcal L}_i$ is $2n_i^2-n_i$.
\end{itemize}
\Rm: Since the subspaces $V_i^\BC$ are orthogonal for the
canonical symmetric bilinear form on $\BC^F$, the space
$\sym^G(\BC)$ is isomorphic to the product $\sym^G(V_0^\BC)\times
\cdots \times \sym^G(V_r)$. The tangent space to $\BL^G$ is
$\sym^G(\BC)$, and the dimension of each ${\mathcal L}_i$
corresponds to the dimension of $\sym^G(V_i^\BC)$. \ali
 \Rm: In particular, when all the irreducible representations of
$\BC^F$ are realizable over $\BR$, then $\BL^G$ is a product of
Lagrangian Grassmannians. This is the case in all the examples we
are going to consider.

The main point is that the first and second Betti numbers of
${\mathcal L}_i$ do not depend on the type of ${\mathcal L}_i$ and
are equal to
$$
b_1=\dim(H^1({\mathcal L}_i,\BC))=0, \;\;\; b_2=\dim(H^2({\mathcal
L}_i,\BC))=1
$$
(cf for example \cite{GriffithsH} for the case of ${\Bbb
G}^{n,2n}$, and \cite{SankaranV},  for the case of $\BL^n$ and
${\Bbb S}{\Bbb O}^n$). Since ${\mathcal L}_i$ is a smooth
projective variety, and hence a K\"ahler manifold, it implies that
$H^{1,0}({\mathcal L}_i)=H^{0,1}({\mathcal L}_i)=\{0\}$,
$H^{2,0}({\mathcal L}_i)=H^{0,2}({\mathcal L}_i)=\{0\}$ and
$H^{1,1}({\mathcal L}_i)\simeq\BC$ (cf for appendix A.5). Thus,
 \beq \label{f.4.14}
\Hunl=\Hun ({\mathcal L}_0)\oplus \cdots \oplus \Hun ({\mathcal
L}_r),
 \eeq
and $\Hunl$ is of dimension  $r+1$ (hence, we see that we are in
the situation described in appendix C.3).
 Each of these Grassmannians have a
canonical embedding in a projective space $\BP^k$: indeed, the
Grassmannian ${\Bbb G}^{n,2n}$ is naturally embedded in the
projective space ${\Bbb P}(\bigwedge^{n}\BC^{2n})$ by the
Pl\"ucker embedding, and $\BL^n$ and ${\Bbb S}{\Bbb O}^n$ are
subvarieties of ${\Bbb G}^{n,2n}$. We call canonical K\"ahler form
on ${\Bbb G}^{n,2n}$, $\BL^n$ and ${\Bbb S}{\Bbb O}^n$, the
restriction of the Fubiny-Study form on the projective space
${\Bbb P}(\bigwedge^{n}\BC^{2n})$, renormalized to be a generator
of the integral cohomology $H^{1,1}_\BZ$ (cf appendix A.5 and
C.3).  We denote by $\nu_i$ the canonical K\"ahler form on each
${\mathcal L}_i$. By abuse of notations we also denote by $\nu_i$
the pull-back of $\nu_i$ by the canonical projection $p_i\;:\;
\lll\rightarrow {\mathcal L}_i$. A natural basis for $\Hunl$ is
given by $(\{\nu_0\},\ldots ,\{\nu_r\})$  where $\{\nu_{i}\}$ is
the cohomology class of $\nu_{i}$. In this basis the pull-back
$(g^n)^*$ on $\Hunl$ is given by a $(r+1)\times (r+1)$ matrix
$d_n=(d_{n,i,j})_{0\le i,j\le r}$ defined by \beq \label{f.4.15.1}
(g^n)^*\{\nu_j\}=\sum_{i=0}^r  d_{n,i,j} \{\nu_i\}. \eeq We know,
cf appendix C proposition (\ref{ap.6}), that $(d_n)$ has
non-negative integer coefficients. As explained in the appendix
the matrix $d_n$ plays the same role as the degree in the case of
maps on projective spaces (i.e. if $\lll$ was a projective space
$\BP^k$ then $d_n$ would be scalar (since $\dim\Hun(\BP^k)=1$) and
would be the degree of the map $g^n$ as defined in appendix B).
Before going further we detail two particular cases where this
notion is easier to handle. \ali

{\bf The case $\lll=\BL^{\vert F\vert }$}. \ali This means that
$\BC^F$ is the sum of $\vert F\vert$ times the trivial
representation $W_0$. This happens if and only if $G$ acts
trivially on $F$. The cohomology group $\Hunl$ is then
1-dimensional and the map $(g^n)^*$ is scalar and represented by
the positive integer $d_n$. In this situation the proof of
proposition (\ref{p.4.2})
 is significantly simpler and we give it for the
convenience of the reader. Considering relation (\ref{f.4.9}) we
have
$$ N^n \{S_0\} =d_n\{S_0\} +\{[D_n]\},$$
thus $D_n\ne 0$ implies that $d_n<N^n$. The sequence $d_n$ being
submultiplicative  (cf proposition (\ref{ap.6})) this implies
$\lim_{n\to \infty}{1\over N^n} d_n=0$ and then $\lim_{n\to\infty
} {1\over N^n} (g^n)^* (S_0)=0$ since the total mass of the
current ${1\over N^n} (g^n)^* (S_0)$ goes to zero. $\Box$ \ali

{\bf The case $\lll\sim \BP^1\times \cdots \times \BP^1$ (r+1
times).}
 \ali
 This occurs  when $n_i=1$ for all $i$.
In \cite{Sabot2}, we gave a sufficient condition for this to
happen.
\begin{propos}
(\cite{Sabot2}, Th\'eor\`eme 3.2) \label{p.4.3} The two following
propositions are equivalent

i) For all $(x,y)$ in $F^2$ there exists $h\in G$ such that
$h\cdot x=y$ and $h\cdot y=x$.

ii) The representation $\BC^F$ can be decomposed into $r+1$
distinct $\BC$-irreducible representations and this decomposition
can be realized in $\BR$ (i.e. $n_i=1$ for all $i$ and the
representations $W_i$ are realizable in $\BR$).
\end{propos}
\Rm: It means that all the representations of $\BC^F$ are of type
2. Hence $\sym^G\simeq \BC^{r+1}$.
 \ali
\Rm : In particular this is true for nested fractals (cf examples,
section 1.1.4, and \cite{Lindstrom}). \ali \ali As explained in
the appendix, when $\lll=\BP^1\times \cdots \times \BP^1$ then
$g^n$ can be lifted to a polynomial map on $\BC^{2(r+1)}$ and the
matrix of degrees is equal to the degrees of the polynomials
involved in this map. In the example of the Sierpinski gasket we
will use this polynomial representation to compute the map $g$ and
analyze its dynamics.
\subsection{Proof of proposition (\ref{p.4.2}).}
The proof of the proposition rely on the following lemma.
\begin{lem}
\label{l.4.2} For any $n$ the matrix $d_n$ is primitive. More
precisely, for any $n$ and any $j=0,\ldots ,r$, $d_{n,j,0}>0$ and
$d_{n,0,j}>0$.
\end{lem}
N.B.: Primitive means that there exists a power with positive
coefficients. Here, we see that $(d_n)^2$ has positive
coefficients. \ali Suppose this lemma proved then the proof of
proposition (\ref{p.4.2}) runs very much like in the simpler case
$\lll=\BL^{\vert F\vert}$. Let us first prove that $\{S_0\}$ has
non-negative coordinates in the basis $(\{\nu_0\},\ldots
,\{\nu_r\})$. This is a direct consequence of the fact that $S_0$
is a K\"ahler form, as the restriction of the Fubini-Study form to
$\BL^G$ (cf appendix A.5). Indeed the real
$$
\int_{\BL^G} \nu_i \wedge S_0^{\dim\BL^G -1},
$$
is called the mass of $\nu_i$ and is a positive real when $S_0$ is
K\"ahler. Thus if $\{S_0\}= \sum_i c_i\{\nu_i\} $ then
$$
c_{i_0}={\int_{\BL^G} S_0^{\dim\BL^G}\over \int_{\BL^G}\nu_i\wedge
S_0^{\dim \BL^G -1} },
$$
and $\int_{\BL^G} S_0^{\dim \BL^G}$ is positive. (Actually, the
coefficients $c_i$ are even integral, since ${S_0}$ is an integral
class, cf \cite{LascouxB}, section 6.4.2, iii)).

 Then Equation (\ref{f.4.5})
reads as follows:
$$N^n\{S_0\}=d_n \{S_0\}+\{[D_n]\},
$$
where $\{S_0\}$ and $\{[D_n]\}$ are considered as vectors of
coordinates in the basis $(\{\nu_0\},\ldots \{\nu_r\})$. Denote by
$l_n$ the largest eigenvalue of the non-negative matrix $d_n$.
Suppose that $D_{n_0}\ne 0$, the primitivity of $d_{n_0}$
immediately implies that $l_{n_0}< N^{n_0}$. Indeed, $(d_{n_0})^2$
has positive coefficients, thus
$$
(N^{n_0}-d_{n_0})(d_{n_0}^2\{S_0\})
=(d_{n_0})^2((N^{n_0}-d_{n_0})\{S_0\})
$$
has positive coefficients which implies that $l_{n_0}<N^{n_0}$, cf
proof of theorem 1.1 of \cite{Seneta}). Thus, since  $d_n$  is
submultiplicative (cf proposition (\ref{ap.6})) we have
$\lim_{n\to \infty } {1\over N^n} (g^n)^* S_0 =0$. $\Box$ \ali

At any point $x\in \lll$ the tangent space $T_x(\lll)$ is
isomorphic to $\sym^G(V_0^\BC)\times \cdots \times
\sym^G(V_r^\BC)$. Therefore, at any point $x\in \lll\setminus
I_{g^n}$ where $g^n$ is smooth the differential $dg^n(x)$ of $g^n$
can be decomposed into blocks
 \beq \label{f.4.15}
B^n_{i,j}(x)\;:\; \sym^G(V_i^\BC)\rightarrow \sym^G(V_j^\BC),
\;\;\; 0\le i,j\le r.
 \eeq
We first prove the following lemma.
\begin{lem}
\label{l.4.3} The coefficient $d_{n,i,j}$ is positive if there
exists $x$ in $\lll\setminus I_{g^n}$ such that $B^n_{i,j}(x)\ne
0$.
\end{lem}
Proof: Let us first remark that when $\lll\sim\BP^1\times \cdots
\times \BP^1$ this lemma is trivial. Indeed, in this case the map
$g^n$ can be represented in homogeneous coordinates by \beq
\label{f.4.16} g([x_0:y_0],\ldots ,[x_r:y_r])= ([P^n_0:
Q^n_0],\ldots ,[P^n_r:Q^n_r]) \eeq where $P_i^n(x_0,y_0,\ldots
,x_r,y_r)$ and $Q_i^n(x_0,y_0,\ldots ,x_r,y_r)$ are homogeneous
polynomials of same degree in the variables $(x_j,y_j)$ (and $P_i$
and $Q_i$ are prime). As explained in the appendix, the degree of
$(P_i^n,Q_i^n)$ in the variables $(x_j,y_j)$ is equal to
$d_{n,i,j}$. Therefore $d_{n,i,j}>0$ means that $P^n_i$ and
$Q^n_i$ are not constant in $(x_j,y_j)$ and therefore that
$B^n_{i,j}$ is not the constant 0. \ali Let us now prove the lemma
in the general case. The volume of ${\mathcal L}_i$,
$\int_{{\mathcal L}_i} \nu_i^{\dim {\mathcal L}_i}$, is positive
and we set
 \beqn \label{f.4.17} C=\int_\lll \nu_0^{\dim{\mathcal
L}_0}\wedge \cdots \wedge \nu_r^{\dim{\mathcal L}_r} =
\int_{{\mathcal L}_0} \nu^{\dim{\mathcal L}_0}_0 \cdots
\int_{{\mathcal L}_r} \nu^{\dim{\mathcal L}_r}_r.
 \eeqn
Remark that if $\w$ is a closed current of bidegree $(1,1)$ and
$\alpha$ a smooth closed form of bidimension $(1,1)$,
 then  $\int_\lll \w\wedge \alpha $ depends only on the
cohomology class of $\w$. Therefore if $\w$ is a closed current of
bidegree $(1,1)$ with cohomology class $\{\w\}=c_0\{\nu_0\}+\cdots
c_r\{\nu_r\}$, then \beq \label{f.4.18} \int_\lll \w\wedge
\nu_0^{\dim{\mathcal L}_0}\wedge \cdots \wedge \nu^{\dim{\mathcal
L}_i -1}_i\wedge \cdots \wedge \nu_r^{\dim{\mathcal L}_r}= Cc_i.
\eeq Applying this formula to $(g^n)^*\nu_j$ we get \beq \nonumber
Cd_{n,i,j}=\int_\lll (g^n)^* \nu_j \wedge \nu_0^{\dim{\mathcal
L}_0}\wedge \cdots \wedge \nu^{\dim{\mathcal L}_i -1}_i\wedge
\cdots \wedge \nu_r^{\dim{\mathcal L}_r}. \eeq Since  $(g^n)^*
\nu_j$ is a positive closed $(1,1)$ current it does not charge the
analytic subset of codimension bigger than 1. Thus, we have:
 \beq
\label{f.4.19} Cd_{n,i,j}=\int_{\lll\setminus I_{g^n}} (g^n)^*
\nu_j \wedge \nu_0^{\dim {\mathcal L}_0}\wedge \cdots \wedge
\nu^{\dim{\mathcal L}_i -1}_i\wedge \cdots \wedge
\nu_r^{\dim{\mathcal L}_r}.
 \eeq
As explained in the appendix the current, $(g^n)^*\nu_j$ is smooth
on $\lll\setminus I(g^n)$ and we can write at any point $x\in
\lll\setminus I_{g^n}$ \beq \label{f.4.20} (g^n)^*\nu_j\wedge
\nu_0^{\dim {\mathcal L}_0}\wedge \cdots \wedge
\nu_i^{\dim{\mathcal L}_i -1}\wedge \cdots \wedge
\nu_r^{\dim{\mathcal L}_r} (x)= e^n_{i,j}(x) v(x), \eeq where
$e^n_{i,j}(x)$ is a smooth positive function on $\lll\setminus
I_{g^n}$ and $v$ the volume form $v=\nu_0^{\dim{\mathcal
L}_0}\wedge \cdots \wedge \nu_r^{\dim{\mathcal L}_r}$. To prove
that $d_{n,i,j}>0$ it is enough to prove that $e^n_{i,j}(x)>0$ for
at least one point $x$ in $\lll\setminus I_{g^n}$. Let
$x=(x_0,\ldots ,x_r)$ be in $\lll\setminus I_{g^n}$ and set
$g^n(x)=w=(w_0,\ldots ,w_r)$. We denote by $B_{{\mathcal
L}_i}(x,\epsilon)$ the ball of ${\mathcal L}_i$ with center $x$
and radius $\epsilon$.  We choose $\epsilon$ and $\tilde \epsilon$
such that $g^n( B_{{\mathcal L}_0}(x_0,\epsilon )\times \cdots
\times B_{{\mathcal L}_r}(x_r,\epsilon))\subset B_{{\mathcal
L}_0}(w_0,\tilde \epsilon )\times \cdots \times B_{{\mathcal
L}_r}(w_r,\tilde \epsilon)$, and holomorphic coordinates
$z_1^i,\ldots ,z^i_{\dim{\mathcal L}_i}$ and $\tilde z_1^i,\ldots
,\tilde z^i_{\dim{\mathcal L}_i}$ on $B_{{\mathcal L}_i}(x_i,
\epsilon)$ and $B_{{\mathcal L}_i}(w_i,\tilde \epsilon)$. If the
K\"ahler form $\nu_j$ has locally a psh potential $u_j$ on
$B_{{\mathcal L}_j}(w_j,\tilde \epsilon)$ then we can write
 \beq \label{f.4.21}
(g^n)^*\nu_j =\sum_{l,l'=0}^r \sum_{k=1}^{\dim{\mathcal
L}_l}\sum_{k'=1}^{\dim{\mathcal L}_{l'}} C_{k,k'}^{l,l'} idz_k^l
\wedge d\overline{z}_{k'}^{l'},
 \eeq
where
 \beq \nonumber
\left( C_{k,k'}^{l,l'}(z)\right)_{k\le \dim{\mathcal L}_l\atop
k'\le \dim{\mathcal L}_{l'}} &=& \left({\partial\over
\partial z_k^l}{\partial \over \partial \overline{z}_{k'}^{l'}}
(u_j\circ g^n)(z)\right)_{k\le \dim{\mathcal L}_l\atop k'\le
\dim{\mathcal L}_{l'}}
\\
 \label{f.correc.4.1}
&=& (B_{l,j}^n(z)) \left( {\partial^2 u_j\over \partial \tilde
z_m^j
\partial \overline{\tilde z}_{m'}^j}(g^n(z))\right)_{m,m'\le
\dim{\mathcal L}_j} \left(B^n_{l',j}(z)\right)^*,
 \eeq
where we wrote $B^n_{l,j}(z)$ for the matrix of the block
differential $B^n_{l,j}$ at the point $z$, for the local
coordinates $(z_k^l) $ and $(\tilde z_k^j)$. Since $\nu_l$ are
k\"ahler forms on ${\mathcal L}_l$ we can find $c>0$ such that
$$
\nu_l\ge c\sum_{k=1}^{\dim {\mathcal L}_l} idz_k^l\wedge
d\overline{z}_{k}^{l} ,
$$
on $B(x_l,\epsilon)$. This implies that
$$
(g^n)^* \nu_j \wedge \nu_0^{\dim{\mathcal L}_0}\wedge \cdots
\wedge \nu^{\dim{\mathcal L}_i -1}_i\wedge \cdots \wedge
\nu_r^{\dim{\mathcal L}_r} \ge  (\sum_k C^i_{k,k}) \Pi_{l=0}^r
\Pi_{k=1}^{\dim {\mathcal L}_l} c(idz_k^l\wedge
d\overline{z}_k^l).
$$
Thus $e^n_{i,j}(x)=0$ implies that $\sum_k C^{i,i}_{k,k}(x)=0$,
but the matrix $(C^{i,i}_{k,k'}(x))_{k,k'}$ is positive so this
implies that the matrix $(C^{i,i}_{k,k'}(x))_{k,k'}$ is null. From
formula (\ref{f.correc.4.1}), and since the matrix $\left(
{\partial u_j\over \partial \tilde z_m^j\partial \overline{\tilde
z}_{m'}^j}(g^n(x))\right)$ is of maximal rank, since $\nu_j$ is a
K\"ahler form,  we have $B^n_{i,j}(x)=0$. $\Box$
 \ali
 
Proof of lemma (\ref{l.4.2}).
We know that the map $Q\rightarrow \pi(\exp \oeta Q\eta )$ defines an embedding of
$\symG$ into $\lll$.
On $\symG$ the map $g$ is given by the map $T$ (cf proposition (\ref{p.3.2})).
Firstly, we compute the differential $dT$ at a point $Q$ where  $Q$ is
real and positive definite.
For $f$ in $\BR^F$ we denote by $H_Q(f)$ the harmonic prolongation of
$f$ with respect to $Q_{<1>}$.
We can easily see from formula (\ref{f.2.2})
that the differential $dT_Q$ of $T$ at the point $Q$
satisfies:
\beq
\label{f.4.22}
<dT_Q(Y)(f), f> =<Y_{<1>}(H_Q(f)), H_Q(f)>,
\eeq
for any $Y$ in $\symG$. At the point $Q=\Id$ the harmonic prolongation
$H_{\Id}(f)$ equals $\tilde f$ where:
$$
\tilde f=\left\{
\begin{array}{ll}
f& \hbox{ on $\partial F_{<1>}$}
\\
0& \hbox{ on $F_{<1>}\setminus \partial F_{<1>}$}
\end{array}
\right.
$$

We want to prove that $d_{n,j,0}>0$ and $d_{n,0,j}>0$. From lemma
(\ref{l.4.3}) it is enough to prove that $B_{0,j}(\Id)$ and
$B_{j,0}(\Id)$ are non-null. Let us prove that $B_{0,j}(\Id)\ne
0$. We choose a convenient $Y$ in $\symG (V_0)$: we choose
$Y=p_{V_0}$, the projection on the subspace $V_0$ for the
decomposition (\ref{f.4.10}). The projection on the component
$\symG (V_j)$ of $dT_{\Id}(p_{V_0})$ is just given by the
restriction to the subspace  $V_j$. To prove that $B_{0,j}(\Id)\ne
0$ it is enough to find $f$ in $V_j$ such that
$$<f, dT_{\Id}(p_{V_0})(f)>\; >0.$$
But \beqn <f, dT_{\Id}(p_{V_0})(f)>&=& \sum_{x\in F} \alpha_x^{-1}
<\tilde f_{|F_{<1>,x}}, p_{V_0} \tilde f_{F_{<1>,x}}>, \eeqn and
$\tilde f_{|F_{<1>,x}}(z)=\delta_x(z) f(x)$ where $\delta_x$ is
the Dirac function at the point $x$. Thus the last term equals,
since $p_{V_0}$ is an orthogonal projector,
$$
\sum_{x\in F} \alpha_x^{-1} \vert f(x)\vert^2 <p_{V_0}\delta_x, p_{V_0}\delta_x>.
$$
Since $W_0$ is the trivial representation of $G$, the subspace
$V_0$ is the subspace of functions invariant by $G$. Thus
$<p_{V_0}\delta_x,p_{V_0}\delta_x>\;>0$ for all $x$ in $F$.
(Indeed, the function ${1\over \vert G\vert} \sum_{g\in G}
\delta_{g\cdot x}$ is non null and contained in $V_0$. This
implies that $\delta_x$ cannot be in the orthogonal complement of
$V_0$.) This implies that the block $0,j$ of the matrix $dT_{\Id}$
is non-null, i.e. that $B_{0,j}(\Id)$ is non-null. \ali To prove
that $B_{j,0}(\Id)$ is non-null we proceed similarly. We consider
$Y=p_{V_j}$ and $f=1 \in V_0$. We have: \beqn
<1,dT_{\Id}(p_{V_j})(1)>= \sum_{x\in F} \alpha_x^{-1} <\delta_x,
p_{V_j}(\delta_x)>. \eeqn But $<p_{V_j}(\delta_x),\delta_x>$
cannot be null for all $x$ since $(\delta_x)_{x\in F}$ generates
the space $\BR^F$. $\Box$
\section{Asymptotic degree of $g^n$.}
We denote by $l_n$ the maximal eigenvalue of $d_n$.
Using the submultiplicativity of $d_n$ we set
\beq
\label{f.3.47}
d_\infty =\lim_{n\to \infty}
 (l_n)^{1\over n}.
\eeq
N.B.: In the publication, we made a mistake and wrote
$d_\infty ={1\over n}\log l_n$, which must obviously be 
$\log d_\infty$.
 \ali
Considering formula (\ref{f.4.9}) we see that $d_\infty\le N$.
If we denote by $\|d_n\|$ the $L_\infty$ norm of $d_n$, i.e.
$\|d_n\|=\sup_i \sum_{j=0}^r (d_n)_{i,j}$, we can easily check that
\beq
\label{f.3.47.1}
d_\infty=\lim_{n\to\infty} \|d_n\|^{1\over n}.
\eeq
(Indeed, classically $\|d_n\|\ge l_n$ and for all $n_0$ we can find
a constant $K>0$ such that for all $p\ge 0$ and $k=0\ldots n-1$,
$\|d_{n_0p+k}\|\le \|(d_{n_0})^pd_k\|\le K(l_{n_0})^p$,
which immediately gives the other inequality $\lim \|d_n\|^{1\over n}
\le d_\infty$).
\begin{propos}
\label{p.4.4}
i) We are in the case i) of theorem (\ref{t.4.1}) if and only if
$d_\infty <N$.

ii)In the lattice case for all choice of $(A,b)$
\beq
\label{f.3.48}
\log d_\infty \ge \limsup_{n\to\infty}
 {1\over n} \log \vert \nu_\nn^\pm-\nu_\nn^\ND\vert,
\eeq
where $\vert \nu_\nn^\pm-\nu_\nn^\ND\vert$ is the total mass of the measure.
Furthermore, for a generic choice of $(A,b)$ we have the equality
\beq
\label{f.3.49}
\log d_\infty =\lim_{n\to\infty} {1\over n}
\log \vert \nu_\nn^\pm-\nu_\nn^\ND\vert.
\eeq
\end{propos}
Proof:
i) Indeed, we remarked in the proof of
proposition (\ref{p.4.2})  that $D_n\ne \emptyset$ is equivalent
to $l_n<N^n$.
\ali
ii)
Let $V^\pm$ be the analytic subsets defined by
\beqn
V^-=\{X\in\pi^{-1}(\lll),\; <X,1>=0\},
\\
V^+=\{X\in \pi^{-1}(\lll),\; <X,\Pi_{x\in F}\oeta_x\eta_x>=0\}.
\eeqn Proceeding exactly as in proposition (\ref{p.4.1}) we know
that $dd^c \ln\vert <R^n,1>\vert$ is a potential for
$(g^n)^*[V^-]+[D_n]$ (and idem for $<R^n,\Pi_{x\in
F}\oeta_x\eta_x>$ and $V^+$). Hence, by proposition (\ref{p.3.3})
we have \beq \label{f.3.50} \nu_\nn^\pm =(\pi\circ \phi)^*((g^n)^*
[V^\pm] +[D_n]), \eeq where $(\pi\circ \phi)^*$ is the pull-back
of $\pi\circ \phi$ as defined in appendix A.6. By proposition
(\ref{p.3.3}) we know that the N-D spectrum corresponds to the
zeroes of $R^n$ and we deduce
 \beq \label{f.3.51}
\nu_\nn^\pm-\nu_\nn^\ND \le (\pi\circ \phi)^* ((g^n)^* [V^\pm]),
 \eeq
with equality when $\phi(\lambda)$ does not meet the indeterminacy
points of $g^n$ (which is the case for a generic choice of
$(A,b)$, as seen in the proof of theorem (\ref{t.4.1})). The map
$\phi$ can be extended into a holomorphic function on $\BP^1$ by
setting $\phi(\infty)=\pi(\exp(\oeta I_b\eta))$. Remark that
$\phi(\infty)$ does not belong to $V^+$ or $V^-$, so that formula
(\ref{f.3.50}) and (\ref{f.3.51}) remain valid when $(\pi\circ
\phi)^*$ is the pull-back of the extension of $\pi\circ \phi$ to
$\BP^1$. In $\BP^1$ the cohomology class of a positive
$(1,1)$-current (i.e. a positive measure) corresponds to its total
mass, more precisely, we have $\{\nu\}=\vert \nu\vert
\{\delta_0\}$ where  $\nu$ is a positive measure and $\vert
\nu\vert$ its integral on $\BP^1$. One can easily check that the
coordinates of the linear map $(\pi\circ
\phi)^*:H^{1,1}(\lll)\rightarrow H^{1,1}(\BP^1)$ in the basis
$(\{\nu_0\},\ldots ,\{\nu_r\})$ and $\{\delta_0\}$ are positive as
well as the coordinates of $\{[V^\pm]\}$ in the basis
$(\{\nu_0\},\ldots ,\{\nu_r\})$. Considering equation
(\ref{f.3.51}) in homology we get
$$\vert \nu_\nn^\pm-\nu_\nn^\ND\vert \le (\pi\circ \phi)^* (d_n\{[V^\pm]\}),$$
with equality for a generic choice of $(A,b)$.
Thus we have
$$\vert \nu_\nn^\pm-\nu_\nn^\ND\vert \le \|(\pi\circ \phi)^*\| \;\|d_n\|\;
\|\{[V^\pm]\}\|
$$
where $\|\pi\circ \phi)^*\|$ and  $\|d_n\|$ are the $L^\infty$ matrix
norm and $\|\{[V^\pm]\}\|$ the $L^\infty$ norm of the vector
$\{[V^\pm]\}$.
Thus (\ref{f.3.48}) easily follows form (\ref{f.3.47.1}).
On the other hand, since $\{[V^\pm]\}$ and $(\pi\circ \phi)^*$ have positive
coordinates we can find $K>0$ such that $\vert \nu_\nn^\pm-\nu_\nn^\ND\vert
\ge K\|d_n\|$ when equality in (\ref{f.3.51}) is satisfied.
Thus, for a generic choice of $(A,b)$,
equality (\ref{f.3.49}) is satisfied.
$\Box$

\section{Regularity of the density of states}
We first state a conjecture and show how it can be related to results
on Lelong numbers of the Green current.
\begin{conjecture}
The measure $\mu-\mu^\ND$ is continuous, i.e. it does not charge
any point.
\end{conjecture}
We introduce the notion of Lelong numbers of a psh function. Let
$u$ be a psh function in a neighborhood $U\subset \BC^n$ of 0.
Then the Lelong number of $u$ at 0 is defined  by
$$\nu(u,0)=\max\{c\ge 0\; \hbox{s.t.} \; u(z)\le c\log \|z\| +O(1)\}.
$$
The Lelong number of a positive closed $(1,1)$ current $T$ is defined by
$\nu(T,p)=\nu(u,p)$ for any local psh potential $u$ of $T$, $T=dd^c u$.
On $\BC$ it is easy to see that the Lelong number at $p$ of a
positive closed $(1,1)$ current (i.e. a positive measure) is the mass of
the point $p$.

Coming back to our situation, we suppose that we are in the
lattice case and that $d_\infty=N$ (otherwise $\mu-\mu^\ND=0$). We
remark that the Lelong number of $dd^c \ln \| R^n \circ \phi \|$
at $\lambda$ is the order of vanishing  $\ord (R^n\circ
\phi,\lambda)$. By proposition (\ref{p.2.3}), ii), we know that
this order of vanishing is equal to $\nu_\nn^\ND(\{\lambda\})$.
Hence, we see that the conjecture is equivalent to
$$
\lim_{n\to \infty} {1\over N^n} \nu(dd^c \ln \| R^n\circ \phi\|, \lambda)=
\nu(dd^c G\circ \phi,\lambda).
$$
Since the current $S= dd^c G$ is the limit of the sequence of
currents ${1\over N^n} S_n$ with potential $ {1\over N^n} dd^c
\ln\| R^n\|$ the question is equivalent to know whether the limit
of the Lelong numbers of the restriction of the currents $S_n$ to
a curve is equal to the Lelong numbers of the restriction of the
limit $S$. There is, at the present time, only one result on
Lelong numbers of Green currents: it says, cf \cite{Favre1}
theorem 2.4.6, that the Lelong numbers of the Green current are
null on the complement of the indeterminacy points of the
iterates: with our notations it means that the Lelong number of
$S$ is null when the Lelong numbers of the $S_n$ are null.
Unfortunately, it cannot be applied easily to our case. Firstly,
because it does not state that the Lelong numbers of $S$ is the
limit of the Lelong numbers of ${1\over N^n} S_n$ at any point,
and secondly, because we need information on the restriction of
the Green current of $S$ to the curve $\phi(\lambda)$, which can
be strictly bigger than the Lelong number of $S$. Nevertheless, it
could certainly be possible, with a little more work, to prove a
result in the generic case, but we prefer to leave the general
result as a conjecture. \ali

The following result gives a criterion for the regularity in a much
stronger sense.
\begin{propos}
\label{p.4.5}
When $d_\infty=N$, the integrated density of states
$\mu(\lambda)=\int_\lambda^0 d\mu$ is locally h\"older continuous on the
set of $\lambda$'s such that
there exist open subsets $U\subset \BC $ and $V\subset \lll$ such that
$\lambda\in U$,   $\cup_{n=1}^\infty  I_{g^n}\subset V$
and $g^n\circ \phi(U)\cap V=\emptyset$ for all
$n$.
\end{propos}
\noindent
N.B.:
By locally H\"older continuous we mean that for any
relatively compact open set $U\subset {\Bbb C}$, we can find
$\alpha_0>0$ and $C>0$ such that $\vert \mu(\lambda)-\mu(\lambda')\vert
\le C\vert \lambda -\lambda'\vert ^{\alpha_0}$ on $U$.
\ali
Proof:
By proposition VI.3.9 of \cite{CarmonaL}
we know that the H\"older regularity of $\mu(\lambda)$ is
equivalent to the H\"older regularity of $G\circ \phi(\lambda)$.
Thus the result is a direct adaptation of
Theorem 7.1 of \cite{Sibony1}
(cf appendix, theorem (\ref{at.1}) iii))
which states that the Green function is locally H\"older continuous in the
set of normal points: a careful reading of the proof shows that
it can be straightforwardly modified to prove the H\"older
regularity of $G\circ\phi$ under the milder conditions
of proposition (\ref{p.4.5}). $\Box$

\section{Some related rational maps.}
As explained in the introduction, the map $g$ defined on the
projective variety $\lll$ is from the theoretical point of view
the best-suited to our problem, but from the computational point
of view it is not easy to handle. In this section we introduce
related rational maps defined on projective spaces. The map $\hat
g$ we introduce is the map considered in the initial work of
Rammal in the case of the Sierpinski gasket, cf \cite{Rammal}, as
well as in previous work of the author, cf \cite{Sabot3}.

The map $T$ is 1-homogeneous, hence it can be written
\beq
\label{f.4.24}
T={\hat R\over p},
\eeq
where $\hat R:\symG\rightarrow \symG$ is a homogeneous polynomial
map and $p$ a homogeneous polynomial, prime with $\hat R$.
We define $\tilde R:\symG\times \BC \rightarrow \symG\times \BC$
by
$$\tilde R(Q,z)=(\hat R(Q), p(Q)z).
$$
The map $\tilde R$ has no common factor and induces a rational map
$\tilde g$ on $\BP^{\dim \symG }$. We denote by $\tilde d_n$ the
degree of the iterate $\tilde g^n$ (cf appendix B).
\begin{lem}
The map $\tilde g$ is birationally equivalent to the map $g$,
i.e. there exists a rational map $h:\BP^{\dim \symG}\rightarrow \lll$ and
two analytic subsets $E\subset \lll$ and
$\tilde E\subset \BP^{\dim \symG}$ of
codimension at least 1,
such that $h$ is a biholomorphism from $\BP^{\dim \symG}\setminus \tilde E$
onto $\lll\setminus E$ and such that
$$g=h\circ \tilde g\circ h^{-1}.$$
\end{lem}
Proof:
This is clear from formula ((\ref{f.3.1}) since $h:\symG\rightarrow \lll$ given by
$h(Q)=\pi(\exp(\oeta Q\eta))$ is
biholomorphic from $\symG$ onto $\pi \{X\in \pi^{-1}(\lll),\; <X,1>=0\}$.
Hence, the map $g$ and $\tilde g$ extends the map $T$ to 2 different
compactifications of $\symG$. $\Box$

Using the 1-homogeneity of $T$, one can introduce a rational map on
$\BP^{\dim \symG-1}$ which contains most of the information on $\tilde g$.
Precisely, for all $n$ one can write
$$\hat R^n =h_n \hat R_n,$$
for a homogeneous polynomial map $\hat R_n$ with no common factor
and a homogeneous polynomial $h_n$. This induces a map $\hat g^n$
on $\BP^{\dim \symG-1}$ with degree $\hat d_n=\degree (\hat R_n)$.
Set
$$ \hat d_\infty= \lim_{n\to \infty }\hat d_n^{1\over n},
\;\;\;  \tilde d_\infty= \lim_{n\to \infty }\tilde d_n^{1\over n},
$$
which are well-defined thanks to the subadditivity of $\ln \hat
d_n$ and $\ln \tilde d_n$.
 \ali \Rm: As we shall see in the
examples, the map $\hat g$ is the renormalization map that was
previously considered, for example, in \cite{Rammal},
\cite{FukuShima2}, \cite{Sabot3}. The map $\hat g$ is
1-dimensional when the space $\symG$ is of dimension 2. Hence, the
property of spectral decimation is related to the fact that $\dim
\symG =2$.
\begin{propos}
\label{p.4.6}
We have the equality $d_\infty=\tilde d_\infty=\hat d_\infty$.
\end{propos}
\noindent Proof: To prove that $\tilde d_\infty =d_\infty$ we
proceed as in \cite{DFavre}: we can write in homology $(g^n)^*=h_*
(\tilde g^n)^* h^*$ and $(\tilde g^n)^*=h^*(g^n)^* h_*$. If $\|
\|$ denotes the $L^\infty$ norm then we have $ \| d_n\|\le \tilde
d_n \|h_*\|\|h^*\|,$ from which we get $d_\infty\le \tilde
d_\infty$ (N.B: note that $\tilde d_n$ is a scalar, i.e. $\tilde
d_n=\| \tilde d_n\| )$. Conversely, $\tilde d_n\le
\|h^*\|\|h_*\|\|d_n\|$, from which we deduce $\tilde d_\infty\le
d_\infty$.

To prove that $\hat d_\infty =\tilde d_\infty$ we write
\beqn
\tilde R^n (Q,z)&=&
(\hat R^n(Q), p\circ \hat R^{n-1}(Q)\cdots p (Q) z)
\\
&=&
(h_n \hat R_n, \left( \Pi_{j=0}^{n-1} h_j^{\degree (p)}(Q) \right)
p\circ \hat R_{n-1}(Q)\cdots p (Q)z).
\eeqn
But $h_{n-1}^{\degree (p)+1}$ divides $h_n$
(since $\degree (\hat R)=\degree (p)+1$),
thus $\Pi_{j=0}^{n-1} h_j^{\degree (p)}$
divides $h_n$ and
$\tilde d_n \le \degree (p) (\hat d_{n-1}+\cdots \hat d_1+1)+1$.
It immediately follows that $\hat d_\infty =\tilde d_\infty$.
$\Box$

Remark that from the proof we see that $\tilde R$ is  algebraically stable
if and only if $\degree (h_n)=0$ for all $n$ and that in this case
$\tilde G(Q,z)=\hat G(Q)$ if $\tilde G$ and $\hat G$ are the Green functions
of respectively $\tilde R$ and $\hat R$.
But this result is not really useful since in general
the maps $\hat g$ and $\tilde g$ are not algebraically stable.
To convince the reader of this fact we explicitly compute the degrees
$\tilde d_n$ when $d_\infty =N$.
Set $p_\nn^-(Q)=\det ((Q_\nn)_{|F_\nn\setminus \partial F_\nn})$.
We claim that when $d_\infty =N$
(i.e. when we are in the case ii) of theorem (\ref{t.4.1})) then
$$\tilde R^n (Q,z)= (p_\nn^-(Q) T^n(Q), p_\nn^- (Q)z).$$
Indeed, we know from formula (\ref{f.3.1}) that $p_\nn^-$ simplifies
the singularities of $T^n$, and we know from the same formula that
$p_\nn^-$ and $p_\nn^- T^n$ cannot have a common factor
because otherwise it would imply that $R^n$ is null on  a hypersurface
of $\lll$, which is impossible since we are in case ii) of theorem (\ref{t.4.1}).
Thus,
\beqn
\tilde d_n &=& \degree (p_\nn^- )+1
\\&=& \vert F_\nn\setminus \partial F_\nn \vert +1
= {N^n-1\over N-1} \vert F_{<1>}\setminus \partial F_{<1>}\vert
+1. \eeqn So, we remark that $\tilde d_n$ grows like $aN^n+b$ for
some $a$ and $b$. In particular, we see that $\tilde d_n$ cannot
be equal to $d_\infty^n=N^n$ except in the case where $N-1=\vert
F_{<1>}\setminus\partial F_{<1>}\vert$. In general it is not the
case. However, in the example of the interval we have $N=2$ and
$\vert F_{<1>}\setminus
\partial F_{<1>}\vert=1$ so there is a priori no incompatibility
between $d_\infty =N$ and the algebraic stability of $\hat g$ and
$\tilde g$. As we shall see, in this particular case $\tilde R$
and $\hat R$ are indeed algebraically stable with degree 2 and one
can express the density of states in terms of the Green function
$\hat G$ (this was shown directly in \cite{Sabot3}). But this is
an exceptional situation: indeed, it is not difficult to see that
the equality $N-1=\vert F_{<1>}\setminus \partial F_{<1>}\vert$
can occur only when the lattice is based on the unit interval of
$\BR$, i.e. when we are, up to isomorphism, in the following
situation: the self-similar set $X$ is the interval $[0,1]$ and
$\Psi_1([0,1]), \ldots ,\Psi_N([0,1])$ are $N$ subintervals of
$[0,1]$.

\noindent\Rm: From the formula we derived for $\tilde d_n$, we see
that the map $\tilde g$ is not algebraically stable (and none of
its iterates is) in the case $N=d_\infty$, at the exception of the
unit interval (indeed, otherwise $\tilde d_n$ would be equal to
$N^n$ after a certain level). This means that there is no natural
way to define a non degenerate Green current associated with
$\tilde R$ and $\tilde g$. This clearly confirms the essential
role played by the map $g$ defined on the algebraic subvariety
$\BL^G$. It is quite interesting to note that this is very
coherent with the general philosophy that seems to emerge in the
study of iteration of rational map which is roughly that when a
map is not algebraically stable then one must seek for a
birational transformation that makes it algebraically stable (cf
\cite{DFavre}, theorem 0.1, where a result in this direction is
proved for a particular case). In our case, the birational
transformation has some "physical" meaning, since the complex
Lagrangian Grassmanian is known to be a natural compactification
of the space of complex symmetric matrices (cf for example,
\cite{Collin1}).

\chapter{Examples}
\setcounter{section}{0}
\section{The Sierpinski Gasket}
{\it The maps $T$ and $g$} \ali For the Sierpinski Gasket the
group of symmetries is $G=S_3$ and $\BC^F$ can be decomposed into
a sum of 2 irreducible representations $\BC^F=W_0\oplus W_1$,
where $W_0$ is the subspace of constant functions and $W_1$ its
orthogonal complement (for the usual scalar product on $\BC^F$).
Hence, any $Q$ in $\symG$ can be written \beq \label{f.5.1}
Q=u_0p_{|W_0} +u_1p_{|W_1}, \eeq where $(u_0,u_1)$ are in $\BC^2$
and $p_{|W_0}$ and $p_{|W_1}$ are the orthogonal projections on
$W_0$ and $W_1$ respectively. We denote by $\Qu01$ the element
(\ref{f.5.1}) and we have $\symG\sim \BC^2$. \ali As in section
4.2.2,
 $\lll\sim \BP^1\times \BP^1$.
A point in $\BP^1\times \BP^1$ will be represented
in homogeneous coordinates by
$$([u_0:v_0],[u_1:v_1]).$$
We denote by
$\pi\times \pi :(\BC^*)^2\times (\BC^*)^2
\rightarrow \BP^1\times \BP^1$ the canonical
projection.
The space $\symG$ is embedded in $\BP^1\times \BP^1$ by the injection
$\Qu01\rightarrow ([u_0:1],[u_1:1])$.

An easy computation shows that with the isomorphism $\symG\sim
\BC^2$ the map $T$ is given by \beq \label{f.5.2} T(u_0,u_1)=
3\left( {u_0u_1\over 2u_0+u_1},{u_1(u_0+u_1)\over
5u_1+u_0}\right). \eeq Thus, in homogeneous coordinates the maps
$g$ is given by
 \beq \label{f.5.3}
\nonumber &&g\left( [u_0:v_0],[u_1:v_1]\right)
\\
&=&\left([3u_0u_1:2u_0v_1+u_1v_0], [3u_1(u_0v_1+u_1v_0):
5u_1v_0v_1+u_0v_1^2] \right). \eeq The matrix of degrees is \beq
\label{f.5.4} d_1= \left(
\begin{array}{cc}
1&1
\\
1&2
\end{array}
\right).
\eeq
\ali
{\it The map $\hat g$ on $\BP^1$}
\ali
As shown in section 4.4 the 1-homogeneity of $T$ naturally induces
a map on $\BP^1$.
Indeed, if we set $z={u_0\over u_1}$ and $\tilde z={\tilde u_0\over \tilde u_1}$
where $(\tilde u_0,\tilde u_1)=T(u_0,u_1)$ then we have
$\tilde z=\hat g (z)$ where
\beq
\label{f.5.5}
\hat g(z)={z(z+5)\over (2z+1)(z+1)}.
\eeq
In homogeneous coordinates in $\BP^1$, we see that $\hat g$ is given by
\beq
\label{f.5.6}
\hat g (
[z_0:z_1])
=[z_0(5z_1+z_0):(2z_0+z_1)(z_0+z_1)]
\eeq
(i.e., in (\ref{f.5.5}), $\hat g$ is given in coordinates $[z,1]$).
More formally, if we denote by $\hat s$
the rational map $\hat s :\BP^1\times \BP^1\rightarrow \BP^1$
given by
$$\hat s([u_0:v_0],[u_1:v_1])=[u_0v_1:u_1v_0]$$
then the following diagram is commutative
$$
\begin{CD}
{\BP^1\times\BP^1} @>\mathrm{g}>> {\BP^1\times\BP^1}\\
@VV{\mathrm{\hat s}}V @VV{\mathrm{\hat s}}V\\
{\BP^1}@>\mathrm{\hat g}>> {\BP^1}
\end{CD}
$$
We set $C_z=\hat s^{-1} (\{z\})$ for $z$ in $\BP^1$.
We see that
$$g(C_z)\subset C_{\hat g (z)}.$$
\Rm: At this point we already know, by proposition (\ref{p.4.6}),
that the asymptotic degree $d_\infty$ is equal to the degree of
$\hat g$, i.e  is equal to 2. So we know that we are in case ii)
of theorem (\ref{t.4.1}), thus that we have $\mu^\ND=\mu$ and that
for almost all blow-up the spectrum is pure point with compactly
supported eigenfunctions. \ali \ali {\it The map $R$} \ali At this
point we are not yet in a position to describe the current $S_n$.
Indeed, we computed the map $g$ on $\lll$ but not the map $R$ on
$\pi^{-1}(\lll)$ (and to describe the sequence of currents $S_n$
we need to compute the map $g$ and the hypersurface $D_1$ of
zeroes of $R$). In general, it is not easy to compute the map $R$
on $\aaa$, since $\aaa$ has quite large dimension. To overcome
this difficulty we will lift the restriction of $R$ to
$\pi^{-1}(\lll)$ to a polynomial map on $\BC^2\times \BC^2$. We
first describe more precisely the isomorphism $\lll\sim
\BP^1\times \BP^1$. Let $\psi_0$ be an orthonormal vector of $W_0$
and $\psi_1$, $\psi_1'$ be an orthonormal basis of $W_1$. We
consider the homogeneous polynomial map $s:\BC^2\times \BC^2
\rightarrow \aaa$ given by
$$s\left(
(u_0,v_0),(u_1,v_1) \right)
=(v_0 +u_0\opsi_0\psi_0)
(v_1 +u_1\opsi_1\psi_1)(v_1 +u_1\opsi_1'\psi_1').
$$
We see that
$$
s\left(
(u_0,1),(u_1,1) \right)=
\exp(u_0\opsi_0\psi_0+u_1(\opsi_1\psi_1+\opsi_1'\psi_1'))=
\exp\oeta \Qu01 \eta,
$$
and that $s$ is $(1,2)$ homogeneous, i.e. that
$$s\left(\beta
(u_0,v_0),\beta' (u_1,v_1) \right)
=
\beta(\beta')^2 s\left(
(u_0,v_0),(u_1,v_1) \right).
$$
Hence, the map $s$ takes values in $\pi^{-1}(\lll)$ and the isomorphism
$\lll \sim \BP^1\times \BP^1$ is represented by the commutation of the
following diagram
$$
\begin{CD}
{(\BC^2)^*\times(\BC^2)^*} @>\mathrm{\pi\times\pi}>> {\BP^1\times\BP^1}\\
@VV{\mathrm{\hat s}}V @VV{\mathrm{\sim}}V\\
{\pi^{-1}(\lll)}@>\mathrm{\pi}>> {\lll}
\end{CD}
$$
Remind from formula (\ref{f.3.1}) that
$$R(\exp(\oeta Q\eta ))=\det ((Q_{<1>})_{|F_{<1>}\setminus \partial F_{<1>}})
\exp(\oeta TQ\eta).
$$
An easy computation gives
$$\det((Q_{<1>}^{u_0,u_1})_{|F_{<1>}\setminus \partial F_{<1>}})=
4 (2u_0+u_1)(u_0+5u_1)^2.
$$
By homogeneity, we get
$$ R(s((u_0,v_0),(u_1,v_1)))=(v_0v_1^2)^3 R(s(({u_0\over v_0},1),({u_1\over v_1},1)).
$$
Hence, we see that
$$ <R(s\left(
(u_0,v_0),(u_1,v_1) \right)),1>= 4
v_1^3(2u_0v_1+u_1v_0)(u_0v_1+5u_1v_0)^2,$$ and that the value of
$R(s\left( (u_0,v_0),(u_1,v_1) \right))$ on the monomials of
degree 1 of $\aaa$ is \beqn &12 u_0u_1v_1^3(u_0v_1+5u_1v_0)^2
\left(\oeta p_{|W_0}\eta\right)
&\\
&+12 u_1v_1^2(u_0v_1+v_1u_0)(2u_0v_1+u_1v_0)(u_0v_1+5u_1v_0) \left(\oeta p_{|W_1}\eta\right).
&\eeqn
Thus, if we denote by $R:\BC^2\times \BC^2 \rightarrow \BC^2\times \BC^2$ the
homogeneous polynomial map given by (we adopt the same notation for this map and
for the map on $\aaa$, since they are different representations of the same map)
\beq
\nonu
&&
R\left(
(u_0,v_0),(u_1,v_1) \right)
\\
&=&
\label{f.5.8}
\left(
(3u_0u_1v_1, 2u_0v_1^2+v_0u_1v_1),
(6u_1(u_0v_1+u_1v_0), 2(5u_1v_0v_1+u_0v_1^2))
\right),
\eeq
then we see that the following diagram commutes
$$
\begin{CD}
{\BC^2\times \BC^2} @>\mathrm{R}>> {\BC^2\times \BC^2}\\
@VV{\mathrm{s}}V @VV{\mathrm{s}}V\\
{\pi^{-1}(\lll)\cup \{0\} }@>\mathrm{R}>> {\pi^{-1}(\lll)\cup\{0\}}
\end{CD}
$$
(Indeed, by the previous computation we see that
$R(s((u_0,v_0),(u_1,v_1)))$ and $s(R((u_0,v_0),(u_1,v_1)))$
coincide on the unit 1 of $\aaa$ and on the monomials of degree 1
of $\aaa$. Since they are also elements of
$\pi^{-1}(\lll)\cup\{0\}$ they are equal.) \ali \Rm: This means
that we are able to lift the map $R$ on $\pi^{-1}(\lll)\cup\{0\}$
to a polynomial map on $\BC^2\times \BC^2$. This will be useful
for computation. It is clear that it would be possible to do the
same thing when $\lll$ is a product of $\BP^1$, i.e. for example,
in the case of nested fractals.

Hence, we see that in $\BP^1\times \BP^1$
the analytic set $D_1$ is
given by $D_1=\pi\times \pi\{v_1=0\}=\BP^1\times [1,0]$.
Thanks to (\ref{f.5.8}) we will be able to describe $S_n$ and $S$.
Indeed $\|s( (u_0,v_0),(u_1,v_1))\|^2=(u_0^2+v_0^2)(u_1^2+v_1^2)^2$,
thus $S_0$ is the current with potential
$$\ln \|(u_0,v_0)\|+2\ln \|(u_1,v_1)\|$$
on $\BC^2\times \BC^2$
(i.e. this means that we have $(\pi\times \pi)^* S_0 =dd^c \ln \|(u_0,v_0)\|+2dd^c \ln \|(u_1,v_1)\|$).
In particular, we remark that $\{S_0\}=\left(
\begin{array}{c}
1\\2\end{array}
\right)$
for the canonical basis of $H^{1,1}(\BP^1\times \BP^1)$.
Since $\{[D_1]\}=\left(\begin{array}{c} 0\\1\end{array}\right)$
we see that equation (\ref{f.4.5}) is indeed verified for $n=1$.
The current $S_n$ will be defined by its potential on $\BC^2\times \BC^2$,
$\ln\|(R^n)_0\|+2\ln\|(R^n)_1\|$ where $(R^n)_0$ and $(R^n)_1$ are the
coordinates of $R^n$ on the first and second components of $\BC^2\times \BC^2$.
\ali
\ali{\it The Green current $S$}
\ali
We are now in a position to describe the iterates $R^n$ and the currents $S_n$.
We set $z_0=u_0v_1$, $z_1=u_1v_0$.
We remark that equation (\ref{f.5.8}) can be rewritten
$$R\left( (u_0,v_0),(u_1,v_1)\right)=
\left( (u_1P_1(z_0,z_1), v_1P_2(z_0,z_1)), (u_1P_3(z_0,z_1),v_1P_4(z_0,z_1))
\right),
$$
where
\beqn
P_1(z_0,z_1)= 3 z_0,
&
\;\;\;
&
P_2(z_0,z_1)=2z_0+z_1,
\\
P_3(z_0,z_1)=6 (z_0+z_1),
&\;\;\;
&
P_4(z_0,z_1)=2(z_0+5z_1).
\eeqn
We define $\hat R$ by
$$\hat R (z_0,z_1)= (P_1(z_0,z_1)P_4(z_0,z_1), P_2(z_0,z_1)P_3(z_0,z_1)).
$$
Note that $\hat R$ is a lift on $\BC^2$ of $\hat g$.
For all $k\ge 0$ we set
\beqn
P_{3,k}= P_3 P_3\circ \hat R \cdots P_3\circ \hat R^k,
\\
P_{4,k}= P_4 P_4\circ \hat R \cdots P_4\circ \hat R^k.
\eeqn
An easy computation shows
\beqn
R^n\left( (u_0,v_0),(u_1,v_1)\right)
&=&
(u_1v_1)^{3^{n-2}}
\Pi_{k=0}^{n-3} (u_1v_1 P_{3,k}P_{4,k})^{3^{n-k-3}}
\\
&&
\left(
(P_{3,n-2} P_1\circ \hat R^{n-1}, P_{4,n-2} P_2\circ \hat R^{n-1}),
(P_{3,n-1},P_{4,n-1})\right).
\eeqn
We remark that $P_3(z_0,z_1)$ and $P_4(z_0,z_1)$ are null respectively on
$C_{-1}$ and $C_{-5}$, thus we have
\beqn
[D_n]&\ge&
3
{3^{n-1}-1\over 2}\left( [u_1=0]+[v_1=0]\right)
\\
&& \;\;\; +\sum_{k=0}^{n-3} 3{3^{n-k-2}-1\over 2}
\left( \sum_{z, \hat g^k z=-1} [C_z]+
\sum_{z, \hat g^k z=-5} [C_z] \right)
.
\eeqn
Since $S=\lim_{n\to \infty} {1\over 3^n} S_n$ we see that
\beq
\label{f.5.10}
S\ge
\demi([u_1=0]+[v_1=0]) +
\sum_{k=0}^\infty  {3^{-k-1}\over 2}
\left( \sum_{z, \hat g^k z=-1} [C_z]+
\sum_{z, \hat g^k z=-5} [C_z] \right).
\eeq
To prove that there is actually equality in the last formula it is
enough to check that there is equality in homology. It is easy if we remark
that $\{S\}=\{S_0\}=\left(\begin{array}{c} 1\\2\end{array}\right)$ and
$\{[C_z]\}=\left(\begin{array}{c} 1\\1\end{array}\right)$.
Finally we sum-up our results in the next theorem
\begin{thm}
\label{t.5.1}
For the Sierpinski gasket we are in the case ii) of theorem (\ref{t.4.1}).
The asymptotic degree is 2 and the Green current $S$ is given by the right hand
term of (\ref{f.5.10}).
\end{thm}
{\it Description of $\mu=\mu^\ND$ in the discrete case} \ali It is
useful to make the change of variable $v={3z\over 1-z}={3u_0\over
u_1-u_0}$ in  (\ref{f.5.5}). With this change of variable $\hat g$
is conjugated to the polynomial \beq \label{f.5.11} \hat p(v)=
v(5+2v). \eeq \Rm: This polynomial is, up to a change of variable
the polynomial that appears in the initial work of Rammal,
\cite{Rammal}, and in subsequent works, \cite{FukuShima1},
\cite{Teplyaev}. \ali \ali With the choice made in section 1.2.3
we see that the coordinates of $A$ are $(0,3)$ in $\BC^2\sim
\symG$. We take for $b$ the uniform measure on $F$. In $\lll\sim
\BP^1\times \BP^1$, $\pi(\phi(\lambda))$ has homogeneous
coordinates $([\lambda:1],[3+\lambda:1])$, which means that in
coordinates $v$ it corresponds to $v(\lambda)={3\lambda\over
(3+\lambda)-\lambda}=\lambda$. Applying theorem (\ref{t.3.1}) and
theorem (\ref{t.5.1}) we get
\begin{thm}
\label{t.5.2}
For the Sierpinski gasket in the discrete case we have
$$\mu=\mu^\ND=
\demi\delta_{-3}+\sum_{k=0}^\infty  {3^{-k-1}\over 2}
\left( \sum_{\lambda, \hat p^k \lambda={-3\over 2}} \delta_\lambda +
\sum_{\lambda, \hat p^k \lambda ={-5\over 2}} \delta_\lambda \right).
$$
\end{thm}
One can remark that $[-{5\over 2},0]$ is backward invariant by
$\hat p$, hence the Julia set $\jjj$ of $\hat p$ is included
in $[-{5\over 2},0]$ and it is not difficult to see that
$\jjj$ is a Cantor subset of $[-{5\over 2},0]$.
\ali
Since $\hat p(-{5\over 2})=0$ we remark that $-{5\over 2}$
is in the Julia set of $\hat p$, hence the Dirac masses
obtained by preimages of $-{5\over 2}$ are accumulating
points of $\mu$.
Iterating $\hat p$ we see that $-{3\over 2}$ is in the Fatou
set of $\hat p$ and hence the
Dirac masses at the preimages of $-{3\over 2}$ are isolated in
$\mu$.
The point $-3$ is in the complement of the Julia set and hence is an
isolated mass in $\mu$.
\section{The unit interval}
We will show that for this particular case, the density of states
can be expressed thanks to a rational map simpler than the map
$g$, namely we express the density of states in terms of the Green
current of $\hat g$, defined on $\BP^2$, which has been introduced
in section 4.5. This relates the present work with our previous
work \cite{Sabot3}. But firstly, we  illustrate some of the
notions we introduced in the text by an explicit computation of
the map $R$ on this example.

In this case, $F=\{0,1\}$ and $G$ is the trivial group so
$\symG=\symF$ can be identified with $\BC^3$:
$Q$ in $\symF$ is represented by the point of coordinates
$(a,d,q)$ if
$$Q=\left(
\begin{array}{cc}
a&q\\q&d
\end{array}
\right)
.$$
In the Grassmann algebra generated by $\{\oeta_0,\eta_0,\oeta_1,\eta_1\}$
we have:
\beq
\label{f.5.20}
\exp(\oeta Q\eta)=1+a\oeta_0\eta_0+d\oeta_1\eta_1+q(\oeta_0\eta_1+\oeta_1\eta_0)
+(ad-q^2)\oeta_0\eta_0\oeta_1\eta_1.
\eeq
Since $G$ is the trivial group $\lll=\BL^2$  and from the last formula
we easily deduce that
\beq
\label{f.5.20.1}
\BL^2 =\pi\{Z1+a\oeta_0\eta_0+d\oeta_1\eta_1+q(\oeta_0\eta_1+\oeta_1\eta_0)
+D\oeta_0\eta_0\oeta_1\eta_1,\; ad-q^2=DZ\}.
\eeq
We set $\delta ={\alpha \over 1-\alpha}$.
An easy computation shows that
\beq
\nonumber
&& R(Z1+a\oeta_0\eta_0+d\oeta_1\eta_1+q(\oeta_0\eta_1+\oeta_1\eta_0)
+D\oeta_0\eta_0\oeta_1\eta_1)
\\
\label{f.5.20.2}
&=&\tilde Z1+\tilde a\oeta_0\eta_0+\tilde d\oeta_1\eta_1+
\tilde q(\oeta_0\eta_1+\oeta_1\eta_0)
+\tilde D\oeta_0\eta_0\oeta_1\eta_1,
\eeq
where
\beqn
&\tilde Z=\delta (a +\delta^{-1} d) Z,\;\;
\tilde D= \delta^2 (a +\delta^{-1}d) D
,&\\
&\tilde a =\delta (a^2+\delta^{-1} DZ),\;\;
\tilde d =\delta ( d^2+\delta DZ)
,&\\
&\tilde q=-\delta q^2&
\eeqn
On $\pi^{-1}(\lll)$, considering formula (\ref{f.5.20.1}),
we see that $\tilde a$ and $\tilde d$ are
also equal to
$$\tilde a=\delta (a(a+\delta^{-1}d) -
\delta^{-1} q^2),\;\;
\tilde d =\delta (\delta d( a+ \delta^{-1} d)-\delta  q^2).
$$
It is then clear that \beqn \pi\{R^n=0\}\cap \BL^2 = \BL^2\cap
\cup_{k=1}^n \pi\{\alpha^ka+(1-\alpha)^kd=0,\;\; q=0\}. \eeqn
(hence the zeroes of $R^n$ are of codimension 2 in $\BL^2$, so
$D_n=\emptyset$ for all $n$ and we are in case ii) of theorem
(\ref{t.4.1})). \ali In the lattice case we can easily describe
the function $\phi(\lambda)$. Indeed, $A$ is the usual discrete
Laplace operator with coordinates $(a=1,d=1,q=-1)$. If $b$ is the
measure that gives weights $m_0$ and $m_1$ to the points 0 and 1
then we see that $\phi(\lambda)=\exp(\oeta (A+\lambda I_b)\eta)$
is a point of the form (\ref{f.5.20}) with $a=1+\lambda m_0$,
$d=1+\lambda m_1$ and $q=-1$. In particular we see that
$\phi(\lambda)$ does not meet the zeroes of $R^n$ so that
$\mu^\ND=0$ (actually, it is very easy to see directly that there
cannot be any Neumann-Dirichlet eigenfunction for this
1-dimensional Laplace operator). \ali \ali {\it The maps $T$ and
$\hat g$} \ali For this example the easiest way to describe the
density of states is to consider the simpler map $\hat g$
introduced in section 4.5. A simple computation or an application
of formula (\ref{f.5.20.2}) gives in coordinates $(a,d,q)$:
$$T(a,d,q)={1\over a+\delta^{-1} d}(a(a+\delta^{-1}d) -
\delta^{-1} q^2, \delta d( a+ \delta^{-1} d)-\delta  q^2, -q^2).
$$
Set $p (Q)=\det ((Q_{<1>})_{|F_{<1>}\setminus \partial F_{<1>}})=
\delta(a+\delta^{-1}d)$.
As in section 4.5. we define the map $\hat R:\BC^3\rightarrow \BC^3$
obtained from $T$ by simplifying the singularities:
\beqn
\hat R (Q) &=& p(Q) TQ
\\
&=& \delta (a(a+\delta^{-1}d) -
\delta^{-1} q^2, \delta d( a+ \delta^{-1} d)-\delta  q^2, -q^2).
\eeqn
We denote by $\hat g$ the rational map on $\BP^2$ induced by $\hat R$.
The map $\hat g$ has a simple indeterminacy point $[1,-\delta,0]$
and we easily see that $\hat g$ is algebraically stable
(cf \cite{Sabot3}, for more details).
We denote by $\hat G$ the Green function
$\hat G=\lim_{n\to \infty} {1\over 2^n}\log\|\hat R^n\|$.
In the lattice case we set $\hat \phi(\lambda)=A-\lambda I_b$ and in the continuous
case $\hat \phi(\lambda)=A_{(\lambda)}$.
\begin{thm}
\label{t.5.3}
In the lattice case
$$\mu={1\over 2\pi} \Delta \hat G\circ \hat \phi.$$
In the continuous case the same formula holds on the ball
$B(0,\vert \lambda^-_1\vert )$.
\end{thm}
\Rm: This  result was proved directly in theorem 3.1 of
\cite{Sabot3}. As shown in \cite{Sabot3}, this formula is related
to the classical Thouless formula, and we can relate the Green
function $\hat G\circ \hat \phi$ with the Lyapounov exponent of
the propagator of the PDE associated with our second order
differential operator. In \cite{Sabot3} we used this formula to
prove that the density of states is continuous and supported by a
Cantor subset of $\BR$ for $\alpha\ne \demi$ (for $\alpha=\demi$
we are in the situation of the classical Laplacian and everything
is well-known). We also proved the H\"older regularity of $\mu$
for some values of the parameter $\alpha$. In \cite{Sabot7} we go
further and describe the spectral type of the operator, depending
on $\alpha$ and on the blow-up $\w$.
 \ali
Proof: Thanks to  the expressions of $R$ and $\hat R$ we can
easily deduce that $\hat G(Q)\le G(\exp\oeta Q\eta )$ for any $Q$
in $\symF$. On the other hand for $Q$ in $S_+$ we know that (cf
lemma (\ref{l.3.1}))
$$G(\exp\oeta Q\eta )=\lim_{n\to\infty} {1\over N^n }\log\vert <R^n(\exp\oeta Q\eta ),1>\vert,
$$
and by direct computation we see that $<R^n(\exp\oeta Q\eta ),1>=
\Pi_{k=0}^{n-1} p\circ \hat R^k (Q).$
Hence we see that for any $Q$ in $S_+$ we have $\hat G(Q)\ge G(\exp\oeta Q\eta )$.
The equality $ \hat G\circ \hat \phi(\lambda)=G\circ \phi(\lambda)$ is satisfied
for $\lambda $ in $\BC\setminus \BR$, so in $L^1_{loc}$.
Theorem  (\ref{t.5.3}) is proved. $\Box$
\section{Nested fractals}
For nested fractals we shall prove that we are in the case ii) of
theorem (\ref{t.4.1}). (The existence of N-D eigenvalues was
proved initially in \cite{BarlowK}. The fact that $\mu^\ND=\mu$
was proved directly in \cite{Sabot4}). Denote by $W_0,\ldots ,W_r,
W_{r+1}, \ldots ,W_{r'}$ the list of irreducible representations
of the group $G$, and assume that the representations $W_0,\ldots
,W_r$ are contained in $\BR^F$. In \cite{Sabot4}, proposition 2.3,
we proved that $r'>r$, i.e. that there exists at least one
irreducible representation which is not contained in $\BR^F$. The
space $\BR^{F_\nn}$ can be decomposed in
$$\BR^{F_\nn}=V_{\nn,0}\oplus \cdots \oplus V_{\nn,r'},
$$
where $V_{\nn,j}$ is the isotopic representation associated with
$W_j$. It is easy to check that for $n$ large enough,
$V_{\nn,j}\ne \emptyset$ for all $j$ (indeed, for $n$ large
enough, there is at least one point $x$ in $F_\nn$ such that
$g\rightarrow g\cdot x$ is injective; thus $\BR^{F_\nn}$ contains
the representation $\BR^G$ and since the representation $\BR^G$
contains at least once each irreducible representation, cf
\cite{Serre}, we know that $V_{\nn,j}\ne \emptyset$ for all $j$).
For $Q$ in $\symG$ the operator $Q_\nn$ can be decomposed in
blocks on $V_{\nn,0},\ldots ,V_{\nn,r'}$. Consider, for example,
$(Q_{\nn})_{|V_{\nn,r'}}$: clearly, we have $\ker
((Q_{\nn})_{|V_{\nn,r'}})\subset \ker^\ND(Q_\nn)$ since
$V_{\nn,r'}\subset \ddd_\nn^-$. Hence $R^n(\exp\oeta Q\eta )$ is
vanishing on $\{\det ((Q_{\nn})_{|V_{\nn,r'}})=0\}$, which is an
analytic set of codimension 1 in $\symG$. This implies that
$D_n\ne \emptyset$ and that we are in case ii) of theorem
(\ref{t.4.1}).

Remark that this implies that $d_\infty<N$ but that we have no more
information on the value of $d_\infty$.

\chapter{Remarks, questions and conjecture.}
\setcounter{section}{0} The main open problem is to understand the
almost sure Lebesgue decomposition of the operator $H_\infi$. Let
us be more precise. Consider a typical blow-up $\w$, in particular
for which $\partial F_\infi=\emptyset$. The Hilbert space
$\ddd_\infi$ can be decomposed into three parts $\hhh_{ac},
\hhh_{sc}, \hhh_{pp}$ such that the restriction of $H_\infi$ to
these subspaces is respectively absolutely continuous, singular
continuous or purely punctual (i.e. this means that the spectral
measure of any function in these subspaces is resp. absolutely
continuous, singular continuous or purely punctual). The spectrum
of $H_\infi$ and the spectrum of the restriction of $H_\infi$  to
$\hhh_{ac}, \hhh_{sc}, \hhh_{pp}$ are respectively denoted by
$\Sigma(\w), \Sigma_{ac}(\w), \Sigma_{sc}(\w), \Sigma_{pp}(\w)$
(they a priori depend on $\w$). Proposition
 2 of \cite{Sabot6}
states that these sets are almost surely constant in $\w$, i.e.
equal to deterministic sets $\Sigma, \Sigma_{ac}, \Sigma_{sc},
\Sigma_{pp}$ for almost all blow-up $\w$. \ali One can be more
precise and split the Hilbert space $\hhh_{pp}$ into two parts:
the first ,that we denote $\hhh_{ND}$, is the subspace generated
by the Neumann-Dirichlet eigenfunctions, i.e. by the eigenfunctions
of $H_\infi$ with compact support, the second, $\tilde \hhh_{pp}$,
is defined as its orthogonal complement in $\hhh_{pp}$. We denote
by $\Sigma_{\ND} $ and $\tilde\Sigma_{pp}$ the spectrum of the
restriction of $H_\infi$ to resp. $\hhh_\ND$ and
$\tilde\hhh_{pp}$. It is clear that $\Sigma^\ND=\supp \mu^\ND$,
and we proved in proposition 2 of \cite{Sabot6} that $\tilde
\Sigma_{pp}(\w)$ is also determined almost surely in $\w$. We also
know that $\Sigma=\supp\mu$ from proposition 1 of \cite{Sabot6}.
In this text we showed that $\mu$ and $\mu^\ND$ can be computed
from the Green function of the renormalization map $R$ and the
order of vanishing of $R$. The natural question is then whether it
is possible to characterize the other parts of the Lebesgue
decomposition of the spectrum
$\Sigma_{ac},\Sigma_{sc},\Sigma_{pp},\tilde\Sigma_{pp}$ in terms
of the renormalization map $R$. We can even ask more: in
\cite{Sabot6}, we introduced several measures $\mu^{ac},
\mu^{sc},\mu^{pp},\tilde\mu^{pp}$ which split $\mu$ in several
parts corresponding to the different components of the spectrum by
$\Sigma_\cdot =\supp \mu^\cdot$. The question is whether it is
possible to compute these measures from some characteristics of
the map $R$, as we computed $\mu$ and $\mu^\ND$ from its Green
function and the asymptotic multiplicities of its zeroes. In
particular, it would be interesting to understand at which
condition there may be a pure point component $\tilde \Sigma_{pp}$
not created by the Neumann-Dirichlet eigenfunctions (i.e. at which
conditions there are $L^2$ eigenfunctions which are not generated
by compactly supported eigenfunctions). But all these questions
seem to be very difficult.

Let us now make few remarks and a conjecture. When $N>d_\infty$ we
know that the spectrum is pure point with compactly supported
eigenfunctions, i.e. that
$\Sigma_{ac}=\Sigma_{sc}=\tilde\Sigma_{pp}=\emptyset$. Consider
now the case where $N=d_\infty$. We showed in this text that the
N-D spectrum is related to the zeroes of $R$, i.e. to the
indeterminacy points of $g$. Considering that eigenfunctions of
$H_\infi$ which are not with compact support are in some sense
approximated  by functions with compact support which are not far
from being eigenfunctions we can ask the following very imprecise
question: does the existence of a pure point component in the
spectrum different from the N-D component, i.e.
$\tilde\Sigma_{pp}\ne \emptyset $ imply that the iterates of $g$
on $\phi(\lambda)$ approach the indeterminacy points, in a sense
to be made precise. This leads to propose the following
conjecture.
\begin{conjecture}
Consider the case $d_\infty=N$. Assume that the
condition of proposition \ref{p.4.5} is satisfied for all
$\lambda\in\BR$, i.e. that for any $\lambda\in \BR$
there exists two open subsets $U\subset \BC$ and
$V\subset \BL^G$ such that $\lambda\in U$,
$\cup_{n=1}^\infty I_{g^n}\subset V$ and
$g^n(\phi(U))\cap V=\emptyset$.
Is it true  that in this case $\Sigma_{pp}=\emptyset$?
\end{conjecture}
\noindent
\Rm:
It is already known from theorem \ref{t.3.1} that
under these conditions $\Sigma_{\ND}=\emptyset$.
\ali

One of the main problem is the lack of examples where computations
are possible. There are very few examples where the spectral type
of the operator can be analyzed. The case of the Sierpinski gasket
is now well understood and computations in this example are easy.
In this case $N>d_\infty$ and so the Neumann-Dirichlet
eigenfunctions are complete. The case of nested fractal is also
understood, in this case also $N>d_\infty$.
 \ali We
present the spectral analysis of the self-similar Sturm-Liouville
operator on the real line in \cite{Sabot7}. In this case we can
prove that $\Sigma_{pp}=\emptyset$.

But we have no example where $\tilde \Sigma_{pp}\ne\emptyset$,
i.e. where there is a pure point component not induced by the N-D
spectrum. We think it could be interesting to understand the
following situation (at least by numerical computations), which
could be a candidate for $\tilde \Sigma_{pp}\ne\emptyset$.
Consider the Sierpinski gasket and take $G=\{\Id\}$, the trivial
group as group of symmetries of the picture. In this case the
renormalization map $T$ is defined on the bigger space $\Sym_F$ of
symmetric matrices on $F=\{1,2,3\}$. The subvariety $\BL^G$ is the
Lagrangian Grassmanian $\BL^3$. There is then no reason that
$N>d_\infty$. We would bet, on the contrary, that $N=d_\infty$ in
this case. The question is the following: what happens if we take
for the initial operator $A_{<0>}$ and initial measure $b_{<0>}$ a
small perturbation  of the usual discrete Laplace operator and of
the uniform measure? Does the spectrum remains pure point or not?

\vfill\break

\renewcommand{\thesection}{\Alph{section}}
\renewcommand{\thesubsection}{\Alph{section}.\arabic{subsection}}
\renewcommand{\theremarques}{A.\arabic{remarques}}
\def\Rm{\refstepcounter{remarques}{\bf Remark \theremarques} }
\renewcommand{\thethm}{A.\arabic{thm}}
\renewcommand{\thepropos}{A.\arabic{propos}}
\renewcommand{\theproposdefin}{A.\arabic{proposdefin}}
\renewcommand{\thelem}{A.\arabic{lem}}
\renewcommand{\thecorrol}{A.\arabic{corrol}}
\renewcommand{\thedefin}{A.\arabic{defin}}
\setcounter{subsection}{0}
\chapter*{Appendix}
\setcounter{section}{0}
For the convenience of the reader we review several notions of
pluricomplex analysis and pluricomplex dynamics. Our point of view
is very partial and only motivated by the notions we need in the
main text. The reader is strongly advised to refer to the relevant
literature for a better understanding. In appendix A1 and A2 we
introduce the notion of plurisubharmonic functions and of currents
on complex manifolds. Most of the material is taken from the
appendix of
 review texts  of Fornaess and
Sibony (cf \cite{Sibony1}, \cite{FSibony2} and related literature
as, for example, \cite{LascouxB}, \cite{GriffithsH},
\cite{Hormander}). In  B1 we give a short introduction to basic
notions that appear in relation with iteration of rational maps of
the projective spaces. All the material comes from \cite{Sibony1},
\cite{FSibony2}. In B2 we introduce some notions on iteration of
meromorphic maps of compact complex manifolds. The material comes
from works of Favre, Diller-Favre and Guedj-Favre (cf
\cite{Favre1}, cf \cite{DFavre}, \cite{FavreG}).
\section{Plurisubharmonic functions and positive currents.}
\subsection{Plurisubharmonic functions}
Let $\W$ be a domain of $\BC^m$. A function
$f\;:\; \W\rightarrow \BR\cup \{-\infty\}$
is said to be plurisubharmonic (resp. pluriharmonic) if
\begin{itemize}
\item
$f$ is upper-semi continuous,
\item
$f$ is not constant equal to $-\infty $,
\item
the restriction of $f$ to any complex line is subharmonic (resp. harmonic).
\end{itemize}
Plurisubharmonic (psh for short) functions are in $L^1_\loc$ and can be characterized
by the following property: a function $v$ in $L^1_\loc(\W)$ is almost surely
equal to a psh function if and only if for all vectors $w\in \BC^m$
\beq
\label{af.0}
\sum_{j,k} {\partial^2 v\over \partial  z_j \overline z_k} w_j \overline w_k
\ge 0
\eeq
in the sense of distributions, i.e. the first term is a positive
measure.
\ali
\ali
{\bf Examples and basic properties}
\ali
(1) If $f$ is a holomorphic function in $\W$ then
$\log \vert f\vert$ is a psh function. Moreover
$\log\vert f\vert$ is pluriharmonic on $\W\setminus \{f=0\}$.
\ali
(2) The function $u=\log \| z \| $ is psh in $\BC^m$.
\ali
(3) If $g\;:\; \W \rightarrow \W'$ is a holomorphic map
between an open subset of $\BC^m$ to an open subset of
$\BC^{m'}$  and if $u$ is psh in $\W'$ then $u\circ g$ is
psh or equal to $-\infty$ (note that if $g$ has generic maximal rank $m'$ then
$u\circ g$ is not identically $-\infty$).
In particular the notion of psh functions can be
extended to complex manifolds.
\begin{propos}
\label{ap.1}
Let $v_j$ be a sequence of psh functions in a domain $\W$ of
$\BC^m$. Suppose that $v_j$ is uniformly bounded from above in any compact
subset of $\W$, then

(i) either $v_j$ converges to $-\infty $ on compacts or
there exists a subsequence $v_{j_k}$ which is convergent
in $L^1_\loc$ to a subharmonic function.

(ii) If $v$ is subharmonic and $v_j\to v$ in $L^1_\loc$ then for any compact
$K\subset \W$ and any continuous function $f$:
$$\limsup_{j\to \infty }\sup_K (v_j -f )\le \sup_{K} (v-f).
$$
\end{propos}
N.B.: This result is a corollary of the same statement for subharmonic functions on $\BR^n$ (cf \cite{Sibony1}),
as psh functions on $\BC^m$ are subharmonic on $\BR^{2m}$.
\subsection{Currents}
We denote by $D_\pq (\W)$ the space of $C^\infty$ differential
forms with compact support on $\W$ and of bidegree $\pq$,
i.e. of the type
$$ \phi=\sum_{\vert I\vert=p , \vert J\vert=q}
\phi_{I,J} dz_I\wedge d\overline z_J,$$
where $I=(i_1,\ldots ,i_p)$, $J=(j_1,\ldots ,j_q)$ and
 $dz_I=dz_{i_1}\wedge \cdots \wedge dz_{i_p}$,
$d\overline z_J= d\overline z_{j_1}\wedge \cdots \wedge d\overline
z_{j_q}$. We denote by $D_\pq'(\W)$ the space of currents of
bidimension $(p,q)$, i.e. the space of continuous linear forms on
$D_\pq(\W)$. One can also talk about currents of bidegree
$(m-p,m-q)$ since a current $S$ can be represented as a
$(m-p,m-q)$ differential form with distributional coefficients:
$$S=\sum_{\vert I'\vert=m-p , \vert J'\vert=m-q}
S_{I',J'} dz_{I'}\wedge d\overline z_{J'}.$$
The differential $d$ is defined by duality by
$$<dS, \phi>=(-1)^{p+q+1} <S,d\phi>,$$
or equivalently on coefficients by
$$dS=\sum_j\sum_{I',J'} ({\partial S_{I',J'}\over \partial z_j}dz_j+
{\partial S_{I',J'}\over \partial \overline z_j}d \overline z_j)
\wedge dz_{I'}\wedge d\overline z_{J'}
.$$
It can be decomposed in $d=\partial + \overline \partial$ where
$$\partial S=
\sum_j\sum_{I',J'} {\partial S_{I',J'}\over \partial z_j}dz_j
\wedge dz_{I'}\wedge d\overline z_{J'}
,\;\;
\overline \partial S=
\sum_j\sum_{I',J'} {\partial S_{I',J'}\over \partial \overline z_j}d\overline z_j
\wedge dz_{I'}\wedge d\overline z_{J'}.$$
We write $d^c={i\over 2\pi} (\partial -\overline \partial )$ so that
we have $d d^c={i\over \pi} \partial \overline \partial$.
\subsection{Positive currents}
 A current of bidimension $(p,p)$ (or bidegree $(m-p,m-p)$)
is positive if $<S,\phi>\ge 0$ for all $\phi$ of the form
$$\phi=i\alpha_1\wedge \overline \alpha_1\wedge \cdots \wedge i\alpha_p\wedge
\overline \alpha_p,$$
with $\alpha_i\in D_{(1,0)}(\W)$.
A current $S$ of bidegree $(1,1)$ can be written
\beq
\label{af.0.1}
S=\sum_{j,k} S_{j,k} i dz_j\wedge d \overline z_k,
\eeq
and is positive if for any $w$ in $\BC^m$
the distribution
$$ \sum_{j,k} S_{j,k} w_j\overline w_k$$
is a positive measure.
\ali
With formula (\ref{af.0}) we see that a function $u$ in $L^1_\loc (\W)$  is almost
surely equal to a psh function if and only if $d d^c u$ is a positive
current.
We have the following converse result.
\begin{propos}
\label{ap.2}
Let $S$ be a $(1,1)$ positive closed current on an open ball
of $\BC^m$ then there exists a psh function $u$ such that
$S=dd^c u$.
We say that $u$ is a potential of $S$.
\end{propos}
{\bf Examples and basic properties.}
\ali
(i)
Let $Z$ be an analytic subset of $\W$ with pure dimension $p$.
Let $Reg (Z) $ be the subset  of regular points of $Z$
(i.e the subset of points where $Z$ is locally
a complex manifold of dimension $p$).
We define $[Z]$ as the current of bidimension $(p,p)$ defined by
$$<[Z], \phi> =\int_{Reg(Z)} \phi, \;\;\; \forall \phi\in D_{(p,p)}(\W).$$
This current is positive of bidimension $(p,p)$ and in fact closed,
as shown by Lelong.
\ali
(ii)
Let $f$ be a holomorphic function on $\W$.
We call divisor a formal sum of irreducible analytic hypersurfaces of $\W$.
The divisor of $f$ is the divisor $\sum m_i Z_i $, where the $Z_i$'s are
irreducible analytic hypersurfaces, such that $f$ can be written
$$f=g\Pi_i f_i^{m_i},
$$
for a holomorphic function $g$ which does not take the value 0,
and holomorphic functions $f_i$ such that $Z_i=\{f_i=0\}$ and
$f_i$ is a generator of the ideal $V(Z_i)=\{ f\hbox{ holomorphic
on } \W,\; f(z)=0 \hbox{ on $Z_i$}\}$ (cf for example,
\cite{GriffithsH} section 2.1). Then the Lelong-Poincar\'e
equation states \beq \label{af.1} d d^c \log \vert f\vert =\sum_i
m_i [Z_i]. \eeq (iii) All these definitions can be extended to
complex manifolds using a local chart.

\subsection{Currents of bidegree $(1,1)$ on $\BP^k$.}
Let $\BP^k$ be the complex projective space of dimension $k$
and $\pi\;:\; \BC^{k+1}\setminus \{0\} \rightarrow \BP^k$ the canonical projection.
Let $\ppp$ be the convex cone of psh functions $u$ on $\BC^{k+1}$
such that for a
real $c>0$
\beq
\label{af.2.1}
 u(\lambda z)=c \ln \vert \lambda \vert +u(z), \;\;\; \lambda \in \BC, \; z \in \BC^{k+1}.
\eeq
To any $u$ in $\ppp$ we can associate a positive closed current
of bidegree $(1,1)$ on $\BP^k$: let $s$ be a holomorphic
section of $\pi$ on an open subset $U\subset \BP^k$, then $dd^c (u\circ s)$
defines a positive closed current $S$ of bidegree $(1,1)$ on $U$.
If $s'$ is another section of $\pi$ then $s'=j\cdot s$ for a holomorphic
function $j$ which does not take the value 0. Hence $dd^c (u\circ s)=dd^c (u\circ s')$
so $S$ does not depend on the particular  section and defines a positive
closed current  on all $\BP^k$.
We denote by $L\;:\; \ppp \rightarrow D_{k-1,k-1}'$ the
operator defined by $L(u)=S$.
\begin{propos}
(\cite{Sibony1}, theorem A.5.1)
For any positive closed current $S$ of bidegree $(1,1)$ on $\BP^k$ there exists
a unique (up to an additive constant) function $u\in \ppp$ such that $L(u)=S$.
The function $u$ is called a potential of $S$.
\end{propos}
For example, if $P$ is an irreducible
homogeneous polynomial of degree $d$ on $\BC^{k+1}$ then
$$\log \vert P(\lambda z)\vert =d\log \vert \lambda \vert +\log \vert P(z)\vert,
$$
and by the Lelong-Poincar\'e formula (\ref{af.1})
the current $[P=0]$ has potential $\log\vert P\vert $.
\subsection{The Fubini-Study form, K\"ahler forms}
Consider on $\BP^k$ the closed positive form $\w$ of bidegree $(1,1)$ with
potential $\log\|z\|$ on $\BC^{k+1}$.
We take homogeneous coordinates
$$[z_0:z_1:\cdots :z_k]$$
on $\BP^k$ (i.e. the point $[z_0:\cdots :z_k]$ represents the
image by $\pi$ of the point $(z_0,\ldots ,z_k)$ of $\BC^{k+1}$).
The space $\BC^k$ is identified with $ \BP^k\setminus
\pi(\{z_0=0\})$ taking for coordinates $w_i={z_i\over z_0}$. On
$\BC^k$ the form $\w$ is given by (cf for example,
\cite{GriffithsH}, page 30):
$$\w= {i\over 2\pi}\left( {\sum dw_i\wedge d\overline w_i\over
1+w_i\overline w_i}-{(\sum \overline w_i dw_i)\wedge (\sum w_i d\overline w_i)
\over (1+\sum w_i\overline w_i)^2}\right).
$$
The form $\w$ is called the Fubini-Study form on $\BP^k$ and has
the following properties: it is smooth, closed, and at any point
the coefficients $(S_{j,k})$ defined by equation (\ref{af.0.1})
defines a positive definite matrix. In general, on a complex
manifold a $(1,1)$ form with these properties is called a K\"ahler
form and a complex manifold with such a form is called a K\"ahler
manifold. We do not want to enter into the details of this notion
(cf for example, \cite{GriffithsH} or \cite{LascouxB}) but we just
want to point out that if $X$ is a smooth analytic subvariety of
$\BP^k$ then the restriction of the Fubini-Study form $\w$ to $X$
defines a K\"ahler form on $X$ (which is canonical for the
embedding $X\subset \BP^k$). The volume of $X$ defined as
$$\int_X (\w_{X})^{\dim X}$$
is finite and actually equal to the degree of $X$.
\ali

If $S$ is a positive closed current on $\BP^k$ of bidegree $(p,p)$ on
$\BP^k$ then the total mass of $S$ is defined as
$$\|S\|=\int_{\BP^k} S\wedge \w^{k-p}.$$
If $S$ is of bidegree $(1,1)$ and has potential $u$
then $\|S\|=c$ where $c$ is the homogeneity constant
appearing in formula (\ref{af.2.1}).
Indeed, the function $v=u-c\log\|\cdot \|$ is well defined on
$\BP^k$, hence $\|S\|=c\int w^k +\int dd^c v\wedge \w^{k-1} =c \int \w^k$
since $\w^{k-1}$ is closed, and $\int \w^k=1$.
Finally, we mention that if $(S_n)$ is a sequence of positive currents
such that the total mass converges to 0 then $S_n$ converges to 0
in the sense of currents (cf \cite{Sibony1}).
\ali
In the same way, if $S$ is a current on $X$ (or on any compact
K\"ahler manifold) then $\|S\|$ is defined by
$\int_{X}S\wedge (\w_{|X})^{\dim X -p}$.
\subsection{Pull-back, push-forward of a current}
Let $f\;:\; \W \rightarrow \W'$ be a holomorphic map
between open subsets of $\BC^{m}$ and $\BC^{m'}$.
The pull-back $f^*\alpha $ of a smooth form $\alpha$ in $D_\pq ( \W')$ is well-defined
as an element of $D_\pq (\W)$.
Thus we can define the push-forward $f_* S$ of a current $S$ of bidimension
$(p,q)$ by duality, i.e.
$$<f_* S, \phi >=< S, f^*\phi >, \;\;\; \forall \phi \in D_\pq (\W').
$$

For some particular class of maps we can define the push-forward
of differential forms and then the pull-back of currents (cf
\cite{Sibony1}, A3) but we do not want to enter into details since
we will only consider the case of $(1,1)$ positive closed currents
for which the situation is simpler. \ali Suppose that the map $f$
is dominating, i.e. that its differential is generically
surjective. Let $S$ be a positive closed current of bidegree
$(1,1)$ on $ \W'$. Let $z_0$ be in $\W$ and set $w_0=f(z_0)$. For
$r>0$ we can write $S=dd^c u$ for a psh function $u$ on
$B(w_0,r)$. Choose $r_1$ such that $f(B(z_0,r_1))\subset
B(w_0,r)$. The function $u\circ f$ is psh (indeed, it is not equal
to $-\infty$ since $f$ is dominating). This definition does not
depend on the choice of the potential $u$ since if $u_1$ and $u_2$
are 2 potential of $S$ then $u_1-u_2$, and hence $(u_1-u_2)\circ
f$ are pluriharmonic. Then the pull-back $f^* S$ is defined
locally by $f^* S=dd^c u\circ f$. In \cite{Sibony1} it is proved
that the pull-back is continuous on the set of positive closed
currents. Remark that when $f$ is not dominating the pull-back can
be defined similarly, as soon as $f(\W)$ is not included in the
set where the potential of $S$ is $-\infty$ (when $f$ is not
dominating the pull-back is a priori not continuous). We will see
in next sections that actually the pull-back can be defined for
meromorphic maps on compact complex manifolds.

\section{Dynamics of rational maps on the projective space $\BP^k$}
\subsection{Definitions, indeterminacy points}
Let $\BP^k$ be the complex projective space of dimension $k$ and
$\pi\;:\; \BC^{k+1}\setminus \{0\} \rightarrow \BP^k$ the canonical projection.
A point $z$ in $\BP^k$ can be represented in homogeneous coordinates
by
$$z=[z_0:\cdots :z_k],
$$
where $[z_0:\cdots :z_k]$ denotes the point $\pi(z_0,\ldots ,z_k)$.
Consider now a homogeneous polynomial map $R\;:\; \BC^{k+1}\rightarrow \BC^{k+1}$
of degree $d$, i.e. $R=(R_0,\ldots ,R_k)$ where the $R_i$'s are
homogeneous polynomials of degree $d$ in the variables
$(z_0,\ldots ,z_k)$. Suppose that the $R_i$ have no common factor.
It is natural to associate a map $f$ on the projective space $\BP^k$
such that the following diagram commutes:
$$
\begin{CD}
{\BC^{k+1}} @>\mathrm{R}>> {\BC^{k+1}}\\
@VV{\pi}V @VV{\pi}V\\
{\BP^k}@>\mathrm{f}>> {\BP^k}
\end{CD}
$$
The map $f$ can be defined only on the set $\BP^k\setminus I$
where $I=\pi\{z,\; R(z)=0\}$.
The set $I$ is called the set of indeterminacy points of $f$ and is an analytic subset of
codimension at least 2 (indeed if all the $R_i$'s vanish on an analytic
hypersurface then a polynomial can be factorized in $R$).
Usually one writes $f$ in homogeneous coordinates as
$$f=[R_0,\ldots, R_k].$$
The function $f$ is a rational map of $\BP^k$ (a precise meaning
to this notion is given at the beginning of appendix C) and the
polynomial map $R$, which is called the lift of $f$ to
$\BC^{k+1}$, exists and is unique (up to a multiplicative
constant) for any rational map of $\BP^k$. The degree of $f$ is
defined as the degree of $R$, i.e. $d$. The map $f$ is holomorphic
on $\BP^k$ if and only if $I=\emptyset$. The map $f$ is said to be
dominating if its differential is generically surjective. If $f$
is dominating and  if $S$ is a positive closed current of bidegree
$(1,1)$ with potential $u$ on $\BC^{k+1}$ we define the pull-back
$f^* S$ as the current with potential $u\circ R$. \ali

At a point $p\in I$ the image of $f$ can be defined as a subset of $\BP^k$:
let $\bbb_p$ be the subset of
$\BP^k$ defined by
$$\bbb_p=\cap_{\epsilon>0} \overline{ f(B(p,\epsilon)\setminus I)}
$$
where $B(p,\epsilon)$ is the ball of center $p$ and radius
$\epsilon$. Then $\bbb_p$ is an analytic subset of $\BP^k$ called
the blow-up of $f$ at $p$ (as we will see in appendix C, the
blow-up can also be defined using the graph of $f$). On the other
hand  an irreducible subvariety of dimension $p$ can be sent into
a subvariety of dimension strictly smaller. These phenomenons are
new compared to the situation of 1-dimensional complex dynamics.
This results in the fact that the degree of the iterates of $f$ do
not necessarily grow like $d^n$. Let us explain this clearly:
suppose, for example, that a hypersurface $V$ is sent by $f$ on a
point of indeterminacy. Then the map $R^2$ is null on $V$ and a
hence a polynomial can be factorized in $R^2$. The degree of $f^2$
is then smaller than $\degree (f)^2$ since $f^2$ is lifted to a
polynomial map with degree smaller than that of $R^2$. In general,
we can always write
$$ R^n=h_n R_n,$$
where $h_n$ is a homogeneous polynomial and $R_n$ a homogeneous
polynomial map, with no common factor, of degree
$d_n=d^n-\deg(h_n)$. The map $R_n$ is a reduced lift of $f^n$,
thus  the degree of $f^n$ is $d_n$. We say that there is
decreaseness in the degree if $d_n<d^n$.
\begin{propos}
(\cite{Sibony1}, proposition 4.3)
\label{ap.0}
Let $f$ and $g$ be rational maps of $\BP^k$ of degree $d$ and $d'$.
The degree of $f\circ g$ is smaller than $dd'$ and equal to
$dd'$ if and only if there does not exist
any hypersurface $V$ such that $g(V\setminus I_g)\subset I_f$.
\end{propos}
\begin{proposdefin}
\label{ad.1}
The map $f$ is algebraically stable if there does not exist an integer
$n$ and a hypersurface $V$ such that
$$f^n(V\setminus I)\subset I.
$$
If $f$ is algebraically stable the degree of $f^n$ is $d^n$, but in
general we only have the inequality
$$d_{n+m}\le d_n d_m.$$
We define the dynamical degree as the limit
$$d_\infty=\lim_{n\to\infty} {1\over n} \log d_n.
$$
The map $f$ is algebraically stable if and only if $d_\infty=d$.
\end{proposdefin}
Proof: the proof of this result is clear from proposition (\ref{ap.0}).
Indeed $d_n=d^n$ for all $n$ if and only if $f$ is algebraically stable
and if $d_{n}<d^{n}$ for a $n>0$ then clearly the limit $d_\infty$
(which exists by subadditivity) is smaller than $d$.
$\Box$

\subsection{Green function and Green current. The algebraically stable case.}
Let $f$ be a rational map, algebraically stable, with degree $d$.
We denote by $I_n$ the set
of indeterminacy points of $f^n$, we have $I_n\subset I_m$ for $n\le m$, and
we set $I_\infty =\cup_{n=1}^\infty I_n$.
We recall the following definitions.
\begin{itemize}
\item
A point $p$ in $\BP^k\setminus I_\infty$ is in the  Fatou set of
$f$ if there exists a neighborhood $U$ of  $p$ such that
$f^n_{|U}$ is equicontinuous.
\item
The Julia set is the complement of the Fatou set.
\item
A point $p$ is normal if there exist  neighborhoods $U$ of $p$ and
$V$ of $I$ such that $f^n (U)\cap V=\emptyset $ for all $n$. We
denote by $N$ the set of normal points.
\end{itemize}
Denote by $G_n\;:\; \BC^{k+1}\rightarrow \BR\cup \{-\infty\}$
the psh function
\beq
\label{af.2}
G_n (z)= \log\| R^n (z)\|,\;\;\; z\in \BC^{k+1}.
\eeq
Remark that $G_0$ is by definition the potential of the
Fubini-Study form
$\w$ on $\BP^k$ and that $G_n$ is the potential of the pull-back
$S_n=(f^n)^*\w$.
\begin{thm}
\label{at.1}
(\cite{Sibony1}, Th\'eor\`eme 6.1, 6.5, 7.1)
Let $f$ be a rational map, algebraically stable, with degree $d$.
The sequence ${1\over d^n}G_n$ converges pointwise and in $L^1_\loc$ to
a psh function $G$ satisfying:
\beqn
\left\{
\begin{array}{l}
G(\lambda z)=\log\vert \lambda\vert +G(z)\\
G(R(z))=dG(z)
\end{array}
\right.
\end{eqnarray*}
The function $G$ is called the Green function of $f$.
The current $S$ with potential $G$ is called the
Green current of $f$. It is the limit of the sequence ${1\over d^n} (f^n)^*\w$
and satisfies
$$ f^* S=dS.
$$
\indent
i) The current $S$ does not charge hypersurfaces.

ii) The support of $S$ is contained in the Julia set of $f$. The
set $N\cap (\BP^k \setminus \supp S)$ is contained in the Fatou
set (in particular, when $N=\BP^k$, the Julia set equals the
support of $S$).

iii) The Green function is H\"older continuous in the set of normal points
$\pi^{-1} (N)$.
\end{thm}
\noindent \Rm: The question of whether the sequence of currents
${1\over d^n} (f^n)^*S_0$ converges to $S$ when we start from a
particular current $S_0$ is not easy in general. In particular, it
is interesting to consider the preimages of the current $[V]$ of
integration on a hypersurface $V$.  In this case the limit
${1\over d^n} (f^n)^* [V]$ represents the asymptotic repartitions
of the preimage of $V$. Generically, the limit is $S$ (cf
\cite{Sibony1}), but for a particular $V$ the problem to know
whether the limit is $S$ is a priori not easy. This is more or
less the problem we encounter in a particular case to prove
theorem (\ref{t.3.1}). There are general results for this problem
in the case of birational maps on compact K\"ahler manifold, cf
\cite{DFavre}.

\subsection{The non algebraically stable case}
When $f$ is not algebraically stable then we can write for $n$ large enough
$$R^n=h_n R_n,$$
where $h_n$ is a homogeneous polynomial of positive degree, $R_n$
a homogeneous polynomial map with non common factor. We can still
define $G_n$ by equation (\ref{af.2}) and $S_n$ as the current
with potential $G_n$. Remark that in this case $S_n$ is not equal
to the pull-back $(f^n)^* \w$ since the map $R^n$ is not reduced
(i.e. its components have common factors). Precisely, we have
$$ S_n=(f^n)^* \w +[h_n=0].
$$
Indeed, we see that
$$ dd^c \log \|R^n \|=dd^c \log \|R_n \| +dd^c \log\vert h_n\vert.$$
The first term equals $(f^n)^* \w$ since $R_n$ is a reduced lift for $f^n$
and $\log\vert h_n\vert$ is a potential of $[h_n=0]$ (counting multiplicities,
i.e. $[h_n=0]$ stands for the current of integration on the divisor of $h_n$)
by the Lelong-Poincar\'e equation.
Remark also that $h_{n+1}$ divides $h_n^d$  so that we have $[h_{n+1}=0]\ge d[h_n=0]$
(counting multiplicities).
This implies that ${1\over d^n} [h_n=0]$ converges towards a current with support contained in
$\cup_n \{h_n=0\}$.
On the other hand, we remark that the total mass of $(f^n)^*\w$ is equal to the
degree of $R_n$, i.e. $d_n$, thus ${1\over d^n} (f^n)^* \w$ converges to 0.
This sums-up in the following result.
\begin{thm}
(cf \cite{Sibony1}, Th\'eor\`eme 9.1)
\label{at.2}
When $f$ is not algebraically stable then
\beq\label{af.11}
S=\lim_{n\to\infty}
{1\over d^n} [h_n=0]
\eeq
and the current $S$ is supported by a countable union of
hypersurfaces.
\end{thm}
\Rm:
We see that in this case the current $S$ does not contain
much information about the dynamics of $f$, but just about the distribution of the
hypersurfaces going to indeterminacy points.
\section{Iteration of meromorphic maps on compact complex manifolds}
\subsection{Definitions, indeterminacy points}
Let $X$ and $Y$ be compact complex manifolds.
Denote by $\pi_1\;:\; X\times Y\rightarrow X$
and $\pi_2\;:\; X\times Y\rightarrow Y$
the projection on the first and second components of
$X\times Y$.
A meromorphic function $f\;:\; X\rightarrow Y$ is defined by its graph
$\Gamma_f \subset X\times Y$, an irreducible subvariety of $X\times Y$
for which the first projection is a proper modification,
i.e. such that there exists a proper subvariety $V\subset X$
such that $\pi_1$ is a biholomorphism from $\Gamma_f\setminus \pi_1^{-1} (V)$
to $X\setminus V$.
We denote by $I_f\subset X$ the set of points of indeterminacy,
i.e. the set of points where $\pi_1$ has no
local inverse.
The set $I_f$ is an analytic subset of codimension at least
2. Of course, at a point $x\in X\setminus I_f$ the image
by $f$ is defined by $f(x)=\pi_2(\pi_1^{-1} (x))$.
When $x$ is an indeterminacy point the image
by $f$ is an analytic set defined by $f(x)=\pi_2(\pi_1^{-1} (\{x\}))$.
The map $f$ is holomorphic on $X\setminus I_f$.
The map is said to be dominating if $\pi_2$ is surjective.

If $g\,:\, Y\rightarrow Z$ is another meromorphic map
then the graph $ \Gamma_{g\circ f}$ is defined as the closure:
$$\Gamma_{g\circ f}=\overline {\{ (x, g(f(x))),\;\;\; x\in X\setminus I_f,\;
f(x)\in Y\setminus I_g\}}
$$
One can also consider the graph
$$\Gamma_g\circ \Gamma_f =\{ (x,z),\;\;\; \exists y\in Y \hbox{ such that
$(x,y)\in \Gamma_f$ and $(y,z)\in \Gamma_g$}\}.
$$
\begin{propos}
\label{ap.4}
(cf \cite{DFavre}, proposition 1.5)
The equality $\Gamma_{g\circ f}= \Gamma_g\circ \Gamma_f$ is satisfied if
and only if $\Gamma_g\circ \Gamma_f$ is irreducible and this is
true if and only if there is no hypersurface $V\subset X$ such that
$f(V)\subset I_g$.
\end{propos}
\begin{defin}
\label{ad.4}
A dominating meromorphic map $f\;:\; X\rightarrow X $ is said to be
analytically stable if $\Gamma_f\circ \Gamma_{f^n} =\Gamma_{f^{n+1}}$,
i.e. there does not exist an hypersurface $V\subset X$ and
an integer $n$ such that $f^n(V) \subset I_f$.
\end{defin}
\Rm:
When $X$ is a smooth algebraic projective  variety (i.e. a smooth irreducible analytic subset
of a projective space) the terminology meromorphic map and analytically stable
are replaced by rational map and algebraically stable (due to the algebraicity of
the manifold).
In the case of a rational map on the projective space defined
as in appendix B the reader can check that the definitions are
consistent.

\subsection{Action on cohomology groups.}
We suppose now that $X$ is a smooth algebraic variety, i.e. that it
can be embedded as a smooth irreducible analytic subset of a projective space
(actually, we could only suppose that $X$ is a K\"ahler manifold).
The counterpart of the degree in the case of projective
spaces will come from the action of $f$
on the cohomology groups of $X$.
We first need to introduce some notions and notations.
We denote by $H^r(X,\BC)$ the De-Rahm $r$-cohomology
group of $X$, i.e. the quotient of closed $r$-differential
forms by exact $r$-differential forms.
The cohomology class of a current can also be defined since there is
identification between $H^r(X,\BC)$ and the quotient of
closed currents of degree $r$ by exact currents of degree $r$.
We denote by $\{\alpha\}$ the class of a differential form (or current) $\alpha$.
For us the Dolbeault cohomology group $H^\pq (X)$ will be the subspace
of $H^{p+q}(X,\BC)$ of classes of differential
forms of bidegree $(p,q)$ (in general  the Dolbeault
cohomology groups are not defined in this way but in the
case of K\"ahler manifolds there is identification).

Let $f\;:\; X\rightarrow X$ be a rational map.
Using a desingularization of the graph $\Gamma_f$
(we do not want to enter into the definition of this
notion here) it is  possible  to consider the pull-back
of $\pi_2 $ in the sense of forms and the push-forward
of $\pi_1$ in the sense of currents. This allows us to
define the pull-back $f^*\alpha =\pi_{1*}(\pi_{2}^*\alpha) $ for all
smooth form $\alpha \in D_{p,q} (X)$ as
a current on $X$.
The push-forward $f_*$ can be defined similarly.
We do not want to enter into details of the construction  here
(cf \cite{DFavre} or \cite{Favre1}), but we just
want to point out that $f^*$ and $f_*$ have the following
properties
\begin{itemize}
\item
$f^*\alpha $ is smooth on $X\setminus I_f$,
\item
$f^*\alpha $ is in $L^1_\loc$,
\item
$f^*$ and $f_*$ commutes with $d$.
\end{itemize}
By the third property we see that the operators $f^*$ and $f^*$
induce linear operators on the Dolbeault cohomology groups
$H^\pq(X)$ by
$$ f^*\{\alpha \}=\{f^*\alpha\}, \;\;\;\forall \alpha\in D_{p,q}(X).
$$

In general,  it is not possible to define the pull-back $f^*$
on the space of currents in a continuous way (cf
\cite{Favre1}) but when $S$ is a closed positive $(1,1)$
current we can define $f^* S =\pi_{1*}\pi_2^{*} S $
where the pull-back  $\pi_2^* S$ is defined
as in appendix A.6 using locally a psh potential
for $S$ and the push forward is defined in the sense
of current.
The pull-back  $f^*$ has the following properties
(cf \cite{Favre1}, proposition 2.2.8):
\begin{itemize}
\item
$f^*$ is continuous on the cone of  closed positive $(1,1)$ currents.
\item
For any hypersurface $V\subset X$
$$f^*[V]=[f^{-1}(V)].$$
\item
$f^*\{S\}=\{ f^* S\}$ for any closed positive $(1,1)$ current $S$.
\end{itemize}
\subsection{The matrix of degrees.}
Suppose now that \beq \label{af.6} X=X_1\times \cdots \times X_r.
\eeq where the $X_i$'s are smooth algebraic varieties, simply
connected, and such that $\dim_\BC H^{1,1} (X_i)=1$ for all $i$ in
$1,\cdots ,r$. Since $X$ is K\"ahler and simply connected we have
$H^{1,0}(X_i)=H^{0,1}(X_i)=0$ and thus $H^{1,1}(X)=
H^{1,1}(X_1)\oplus \cdots \oplus H^{1,1}(X_r)$ is of dimension
$r$. Consider $\nu_i$ the K\"ahler form on $X_i$ obtained by
restriction of the Fubini-Study form to $X_i$ for an embedding of
$X_i$ in a projective space and renormalized to be a generator
of the $\BZ$-cohomology. By abuse of notations we also denote
by $\nu_i$ the form on $X$ obtained as the pull-back of the form
$\nu_i$ on $X_i$  by the canonical projection on the $i$-th
coordinate of the right term of (\ref{af.6}). The family
$(\{\nu_1\},\ldots ,\{\nu_r\})$ gives a natural basis of
$H^{1,1}(X)$. In this basis the linear operator $(f^n)^*$ on
$H^{1,1}(X)$ is represented by  a matrix $d_n=(d_{n,i,j})$ defined
by
$$ (f^n)^*\{\nu_j\}=\sum_i d_{n,i,j}\{\nu_i\}.
$$
\begin{propos}
\label{ap.6}
i) The matrix $d_n$ has non-negative integer coefficients.

ii) The sequence of matrices $(d_n)$ is submultiplicative, i.e.
$d_{n+m}\le d_n d_m$ with equality for all $n,m$ if and
only if $f$ is algebraically stable.
\end{propos}
Proof: i) Consider the group of integral classes $H^{1,1}_\BZ
(X)=H^{1,1}(X)\cap H^2 (X,\BZ)$ where $H^2(X,\BZ)$ is the
cohomology group with values in $\BZ$ (actually a class $\alpha$
is integral if its integral along any 1-simplex is an integer).
By hypothesis, the element $\{\nu_i\}$ is a generator of the group
$H^{1,1}_\BZ(X_i)$.
Therefore  $(\{\nu_1\},\ldots ,\{\nu_r\})$ generates
$H^{1,1}_\BZ(X)$ and the coefficients of $d_n$ are integers since
the  linear operator $f^*$ leaves invariant the lattice
$H^{1,1}(X)\cap H^2(X,\BZ)$ (cf \cite{DFavre}, proposition 1.11).
The positivity of coefficients comes from the following fact: let
$H^{1,1}_{psef}$ be the cone of classes $\{S\}$ generated by
closed positive currents. In our case it is easy to see that
$H^{1,1}_{psef}$ is the cone of elements with positive coordinates
in the basis $(\{\nu_1\},\ldots ,\{\nu_r\})$: indeed, suppose that
$\{S\}\in H^{1,1}_{psef}$ then $<S,\nu_{i_1}\wedge \cdots \wedge
\nu_{i_{n-1}}>\ge 0$ for all choice $(i_1,\ldots ,i_{n-1})$. Thus,
if $\{S\}=\sum_{i=1}^{n}s_i \{\nu_i\}$, we can choose $(i_1,\ldots
,i_{n-1})$ such that $<S,\nu_{i_1}\wedge \cdots \wedge
\nu_{i_{n-1}}>=s_i$. Finally, we conclude the proof using the fact
that the cone $H^{1,1}_{psef}$ is left invariant by $f^*$, cf
\cite{DFavre}, proposition 1.11. \ali ii) This is a direct
application of proposition 1.13 of \cite{DFavre}. \ali \Rm: In our
case a considerable simplification comes from the fact that we
assumed (\ref{af.6}). This implies that the cones $H^{1,1}_{psef}$
and $H^{1,1}_{nef}$ considered in \cite{DFavre} are equal and
coincide with the cone $\BR_+^r$ of classes which have positive
coordinates in the basis $(\{\nu_1\},\ldots ,\{\nu_r\})$.
\subsection{Green currents}
We take from \cite{Favre1}, \cite{Sibony1} the following result.
\begin{thm}
\label{at.3}
Let $f$ be a dominating meromorphic map, algebraically stable.
Let $\alpha$ be a smooth closed positive form of bidegree
$(1,1)$ such that the cohomology class satisfies
$f^* \{\alpha \}=\rho \{\alpha \}$ for a positive real $\rho>1$.

(i) The sequence of currents $\rho^{-n} (f^n)^* \alpha $ converges towards
a positive closed $(1,1)$ current $S$ such that
\beq
\label{f.a.8}
f^* S=\rho S.
\eeq
The current $S$ depends only on the cohomology class of $\alpha$, i.e.
if $\alpha'$ is a smooth differential form cohomologous to $\alpha$
then the limit is the same.

(ii)
The support of $ S$ is included in the Julia set of $f$.
\end{thm}
\Rm: When the matrix $d_1$ is primitive (i.e. when $d_1$ admits a power with strictly positive coefficients)
there exists a unique (up to a multiplicative constant)
class with positive coordinates which satisfies equation (\ref{f.a.8}).
Thus, the current $S$ defined in this way is unique (up to a constant)
and it is natural to call it the Green current of $f$.

\subsection{Examples}
{\bf The projective spaces} \ali Let $X=\BP^k$. Since $\dim_\BC
H^{1,1} (\BP^k)=1$ the matrix $d_n$ is scalar and actually equals
the degree of the map $f$ as defined in appendix B. Indeed, if
$\w$ is the Fubini-study form on $\BP^k$ then the potential
$\log\|R_n\|$ of $(f^n)^* \w$ has the following homogeneity
$$\log\|R_n (\lambda z)\| =\deg (f^n) \log \vert \lambda \vert + \log\|R_n(z)\|.
$$
This implies that $u(z)=\log \| R_n(z) \| - \deg (f^n )\log\|z\|$
is defined globally on $\BP^k$ and thus that $\deg (f^n) \w - (f^n)^* \w=dd^c u$.
Thus $d_n\{\w\}=\{(f^n)^*\w \}=\deg (f^n) \{ \w\}$.
\ali
\ali
{\bf$ X=\BP^1\times \cdots \times \BP^1$ (r times).}
\ali
A point $z$ in $X$ can be represented in homogeneous coordinates by
\beq
\label{af.7}
z=([x_1: y_1],\ldots ,[x_r:y_r]).
\eeq
A map $f\;:\; X \rightarrow X$ can be represented in homogeneous coordinates
by
\beqn
f=([P_1: Q_1], \ldots ,[P_r: Q_r]),
\eeqn
where $P_j(x_1,y_1,\ldots ,x_r,y_r)$ and $Q_j(x_1,y_1,\ldots ,x_r,y_r)$
are polynomials homogeneous in the variables $(x_i,y_i)$ with same
degree and no common factor, i.e. we have
$$ P_j (x_1, y_1, \ldots ,\lambda x_i, \lambda y_i, \ldots , x_r, y_r)=
\lambda^{d_{i,j}} P_j(x_1,y_1,\ldots ,x_r,y_r)
$$
(and idem for $Q_j$) if the degree of $(P_j,Q_j)$ in the variables
$(x_i,y_i)$ is $d_{i,j}$.
It happens that the degrees $(d_{i,j})$ coincide with the matrix
of degrees of $f$ defined previously (cf \cite{Guedj1}).

\section{A different proof of lemma \ref{l.3.1}}

We present here a different proof of lemma \ref{l.3.1}. This proof
does not involve any direct estimate but just the fact that $S_+$
is $T$-invariant and general properties of holomorphic maps on
$S_+$.

On $S_+$ we consider the hyperbolic metric given by (cf
\cite{Siegel}, \cite{Terras})
$$
ds^2=\hbox{trace} (V^{-1} dQ V^{-1} \overline{dQ}), \;\;\; \hbox{
for $Q=U+iV\in S_+$}.
$$
It is well known that $S_+$ can be identified with $sp(n,
\BR)\backslash U(n)$, and that $ds^2$ is invariant under the
action of the symplectic group $sp(2n,\BR)$ (cf \cite{Siegel},
page 3, cf \cite{Terras}). The geodesic distance induced by $ds^2$
on $S_+$ is given by (cf \cite{Siegel})
$$
(d_{S_+}(Q,Q_1))^2= \sum_{i=1}^{\vert F\vert} (\log^2({1
+r_k^\demi\over 1-r_k^\demi})),
$$
where $0\le r_1\le \cdots \le r_{\vert F\vert}<1$ are the
characteristic roots of the cross ratio $R(Q,Q_1)$ of $Q,Q_1$
given by
$$
R(Q,Q_1)=(Q-Q_1)(Q-\overline Q_1)^{-1}(\overline Q-\overline
Q_1)(\overline Q-Q_1)^{-1}
$$
(i.e. this means that the $r_k$'s are the eigenvalues of
$\sqrt{R(Q,Q_1)R(Q,Q_1)^*}$. These values $r_k$ satisfy $0\le
r_k\le 1$, cf \cite{Siegel}, page 16).

The upper-half plane $S_+$ can be mapped holomorphically onto the
matrix ball
$$
\{\eee \hbox{ complex sym, } I-\eee\overline \eee >0\}
$$
by the Cayley transform
\begin{eqnarray}\label{f.A.d.3}
Q\rightarrow \eee=(Q-i\Id)(Q+i\Id)^{-1}.
\end{eqnarray}
Inverting this relation we get
\begin{eqnarray}\label{f.A.d.4}
Q=i(\Id+\eee)(\Id-\eee)^{-1}.
\end{eqnarray}
With these notations it is clear that the cross ratio $R(Q, i\Id)$
is given by
\begin{eqnarray}
\label{f.A.d.5} R(Q, i\Id)=\eee \overline \eee.
\end{eqnarray}

The key property we are going to use is the following
generalization of the Schwartz-Pick lemma (cf \cite{Koranyi}). Let
$f$ be a holomorphic map from $S_+$ to $S_+$, then for any $x,y$
in $S_+$
\begin{eqnarray}
\label{f.A.d.7} d_{S_+}(f(x),f(y))\le \sqrt{\vert
F\vert}d_{S_+}(x,y),
\end{eqnarray}
($\vert F\vert$ is the rank of the symmetric space $S_+$. The
exact value of the constant does not matter, the important point
is that it does not depend on the function $f$).

Then we prove the following lemma instead of lemma \ref{l.3.0.0}.
\renewcommand{\thelem}{\ref{l.3.0.0}-bis}
 \begin{lem} \label{l.A.d.1}
 i) The map $T$ is holomorphic on $S_+$ (resp. on
$S_+^G$) and $S_+$ (resp. $S_+^G$) is $T$-invariant. Moreover, for
any $Q$ in $S_+$ and $n$ we have
 \beq
\label{f.A.d.8}
 d_{S_+}(i\Id , T^n Q)\le \sqrt{\vert F\vert} (d_{S_+}(i\Id ,Q)+nd_{S_+}(i\Id, T(i\Id))).
\end{eqnarray}

ii) For any $Q$ in $S_+$ we have the following inequality
$$
\det(\Id +\overline Q Q)\le 2^{\vert F\vert}\exp( 2\vert F\vert
d_{S_+}(i\Id, Q)).
$$
\end{lem}
Proof:
 The fact that $S_+$ is $T$-invariant is proved in lemma
\ref{l.3.0.0}. Using (\ref{f.A.d.7}) for each of the iterates
$T^n$ (indeed, each $T^n$ is a holomorphic map from $S_+$ to
$S_+$) we get
\begin{eqnarray*}
d_{S_+}(i\Id, T^nQ)&\le & d_{S_+}(i\Id,
T^n(i\Id))+d_{S_+}(T^n(i\Id), T^nQ)
\\
&\le& \sum_{k=0}^{n-1} d_{S_+}(T^k(i\Id), T^{k+1} (i\Id))+
\sqrt{\vert F\vert} d_{S_+}(i\Id, Q)
\\
&\le & \sqrt{\vert F\vert}(nd_{S_+}(i\Id, T(i\Id))+d_{S_+}(i\Id,
Q)).
\end{eqnarray*}
ii) Let $\eee$ be the image of $Q$ by the Cayley transform
(\ref{f.A.d.3}). Let $\overline \rho$ be the largest eigenvalue of
$\eee \overline \eee$. We have $\overline \rho<1$ and we deduce
from (\ref{f.A.d.5}) that
\begin{eqnarray*}
d_{S_+}(i\Id, Q)\ge \log\left({1+\overline \rho^\demi\over
1-\overline \rho^\demi}\right)
\end{eqnarray*}
Considering now relation (\ref{f.A.d.4}) we deduce
$$
Q\overline Q\le \left( {1+\overline \rho^\demi\over 1-\overline
\rho^\demi}\right)^2,
$$
which immediately implies
$$
\det(\Id+ Q \overline Q)\le 2^{\vert F\vert} \exp(2\vert F\vert
d_{S_+}(Q,i\Id)).\; \Box
$$

\ali

 Proof of lemma (\ref{l.3.1}). This will be a consequence
of the estimates we proved in lemma (\ref{l.A.d.1}). We first
remark that the terms of the sequences in formulas (\ref{f.3.8})
and (\ref{f.3.9}) are non-positive. By proposition (\ref{p.3.2}),
lemma (\ref{l.2.0}) and lemma (\ref{l.A.d.1}) we have:
 \beqn
\vert <{R^n(\exp \oeta Q\eta )\over \|R^n(\exp \oeta Q\eta )\|}
,1>\vert&=& \vert <{\exp \oeta T^n Q\eta \over \|\exp \oeta T^n
Q\eta \|} ,1>\vert
\\
&=& {1\over  \|\exp \oeta T^n Q\eta \|} \\
&=& \left({1\over \det(\Id +T^nQ \overline{T^n Q})}\right)^{\demi}
\\
&\ge& 2^{-\demi \vert F\vert }\exp(-\vert F\vert d_{S_+}(T^n
Q,i\Id))
\\
&\ge & 2^{-\demi \vert F\vert} \exp(-\vert F\vert \sqrt{\vert
F\vert }(d_{S^+}(Q,i\Id) +n d_{S^+}(i\Id, T(i\Id)))).
 \eeqn
This immediately implies equality (\ref{f.3.8}).
 \ali The proof of
formula (\ref{f.3.9}) works similarly. Remark first that for any
$Q$ in $S_+$, $d_{S^+}(i\Id, Q)=d_{S^+}(i\Id,- Q^{-1})$ (indeed,
$Q\rightarrow -Q^{-1}$ is an isometrie of $S_+$ fixing $i\Id$).
Using proposition (\ref{p.3.2}) and lemma (\ref{l.2.0}) we get:
 \beqn \vert <{R^n(\exp \oeta Q\eta )\over \|R^n(\exp \oeta Q\eta
)\|},\Pi_{x\in F }\oeta_x\eta_x>\vert &=& {\vert \det(T^nQ)\vert
\over \|\exp \oeta T^n Q\eta \|}
\\
&=& \left({1\over \det (\Id+ (T^nQ)^{-1}\overline
{(T^nQ)^{-1}})}\right)^\demi
 \eeqn
and we conclude similarly using lemma (\ref{l.A.d.1}). $\Box$

 \ali
\Rm: We proved in theorem \ref{t.3.1} the convergence of the
counting measure for the Neumann and Dirichlet boundary condition.
The technic presented in this appendix allows, with a little extra
effort, to prove the convergence of the counting measure for any
boundary condition. Indeed, for each $n$ let $B_n$ be a real
symmetric operator on $\BR^{F_\nn}$ supported on $\partial F_\nn$
(i.e. $B_n f=0$ if $f_{|\partial F_\nn}=0$) (we can thus also
consider $B_n$ as an operator on $\BR^F$ by identifying $\partial
F_\nn$ with $F$). We consider the boundary condition induced by
$B_n$, i.e. we denote by $\nu_\nn^{B_n}$ the counting measure of
the eigenvalues of $A_\nn-B_n$. Proceeding as for the Neumann and
Dirichlet boundary condition, we can prove that $\nu_\nn^{B_n}$ is
given by
$$
\nu_\nn^{B_n} ={1\over 2\pi} \Delta \ln \vert <\exp(-\oeta
B_n\eta)\cdot R^n(\phi(\lambda)), \prod_{x\in
F}\oeta_x\eta_x>\vert.
$$
We can as well replace the last expression by
$$
{1\over 2\pi} \Delta \ln \vert <{\exp(-\oeta B_n\eta) \over
\|\exp(\oeta B_n\eta)\|} R^n(\phi(\lambda)), \prod_{x\in
F}\oeta_x\eta_x>\vert.
$$
By the same strategy as before we see that the convergence of
$\nu_\nn^{B_n}$ would be implied by the following convergence for
$Q$ in $S_+$
$$
\lim_{n\to \infty}{1\over N^n} \ln \vert <{\exp(-\oeta B_n\eta)
\over \|\exp(-\oeta B_n\eta)\|} {R^n(\exp\oeta Q\eta)\over \|
R^n(\exp\oeta Q\eta) \|}, \prod_{x\in F}\oeta_x\eta_x>\vert=0.
$$
We see that as before the term in the logarithm is bounded from
above by 1. The term inside the logarithm is equal to
\begin{eqnarray}
\nonumber &&\vert <{\exp\oeta (T^nQ -B_n)\eta \over \|\exp(-\oeta
B_n\eta)\| \|\exp\oeta T^n Q\eta  \|}, \prod_{x\in
F}\oeta_x\eta_x>\vert
\\
\label{f.A.d.12} &=& {\vert \det(T^nQ-B_n)\vert \over \left( \det
(\Id +\overline{T^nQ}T^nQ)\det(\Id+ B_n^2)\right)^\demi}
\end{eqnarray}
The key point is that the change of boundary condition can be
viewed as a "rotation" in $S_+$ (by a rotation in $S_+$ we mean
the image by the Cayley transform of a rotation in the unit ball).
It is actually easier to map everything in the unit ball since
rotations have a simple expression. Let us first derive some
simple formula. Let $Q_1$ and $Q_2$ and $\eee_{Q_1}$ and
$\eee_{Q_2}$ there images by the Cayley transform. Then a direct
computation gives
\begin{eqnarray*}
&&{\vert \det(Q_1)\vert^2\over \det(\Id +\overline Q_1 Q_1)}=
{\vert
\det(\Id+\eee_{Q_1})\vert^2\over\det(2(\Id+\overline\eee_{Q_1}\eee_{Q_1}))},
\\
&& {\vert \det(Q_1-Q_2)\vert^2\over \det(\Id +\overline Q_1
Q_1)\det(\Id +\overline Q_2 Q_2)}= {\vert
\det(2(\eee_{Q_1}-\eee_{Q_2}))\vert^2\over
\det(2(\Id+\overline\eee_{Q_1}\eee_{Q_1}))\det(2(\Id+\overline\eee_{Q_2}\eee_{Q_2}))}.
\end{eqnarray*}
In particular in the last expression if $Q_2=B$ is a real matrix
then $\eee_B$ is unitary and we get
\begin{eqnarray*}
{\vert \det(Q_1-B)\vert^2\over \det(\Id +\overline Q_1
Q_1)\det(\Id +B^2)}= {\vert \det(\eee_{Q_1}-\eee_{B})\vert^2\over
\det(2(\Id+\overline\eee_{Q_1}\eee_{Q_1}))}.
\end{eqnarray*}
If we apply this to (\ref{f.A.d.12}) we get
$$
{\vert \det(\eee_{T^n Q}-\eee_{B_n})\vert \over
\det(2(\Id+\overline\eee_{T^nQ }\eee_{T^n Q}))^\demi }.
$$
The isometries of the unit ball that fixes the point 0 (the
rotations) are exactly the maps $\eee\rightarrow U^t \eee U$ for
$U$ unitary (cf \cite{Siegel}, page 11). If we take any $U_n$ that
sends $\eee_{B_n}$ to $-\Id$ and if we denote by $\tau_n(T^n Q)$
the point of $S_+$ such that $\eee_{\tau_n(T^nQ)}=U_n^t
\eee_{T^nQ}U_n$, then we have $d_{S_+}(i\Id, \tau(T^n
Q))=d_{S_+}(i\Id, T^n Q)$ and the last expression equals
\begin{eqnarray*}
&&{\vert \det(\Id + \eee_{\tau(T^n Q)})\vert \over
\det(2(\Id+\overline\eee_{\tau(T^nQ) }\eee_{\tau(T^n Q)}))^\demi}.
\\
&=& {\vert \det(\tau_n(T^nQ))\vert\over \left(\det(\Id
+\overline{\tau_n(T^nQ)}\tau_n(T^nQ))\right)^\demi}.
\end{eqnarray*}
Then we can apply as previously the estimates of lemma
\ref{l.A.d.1} to conclude.

\section{$G$-Lagrangian Grassmannian}
Let us first recall the classification of complex irreducible
representations, and the Frobenius-Schur Theorem. Let $G$ be a
finite group and $U$ an irreducible representation of $G$ over
$\BC$. We denote be $\chi$ its character. The representation $U$
is said to be
\begin{itemize}
\item
of type 1 if the character $\chi$ is not real.
\item
of type 2 if the character of $\chi$ is real, and the
representation $U$ is realizable over $\BR$ (i.e., $U$ can be
realized as the complexification of an irreducible representation
of $G$ over $\BR$).
\item
of type 3 if the character $\chi$ is real, but $U$ is not
realizable over $\BR$.
\end{itemize}
Let us now recall a consequence of the Frobenius-Schur theorem (cf
\cite{Serre}, proposition 38).
\begin{propos}\label{p.FrobSchur}
i) If $G$ does not have a non-zero invariant bilinear form on $U$,
then $U$ is of type 1.

ii) If such a form exists, it is unique up to homothety, is
non-degenerate, and is either symmetric or skew-symmetric. If it
is symmetric, then $U$ is of type 2, and if it is skew-symmetric,
then $U$ is of type 3.
\end{propos}

Let us now consider an irreducible representation $W$ of $G$ over
$\BR$, and denote by $W^\BC$ its complexification. Then, there are
three possible cases (mutually exclusive). If $W^\BC$ is
irreducible in $\BC$, then $W^\BC$ is of type 2, and by extension
we say that $W$ is of type 2. If $W^\BC$ is not irreducible in
$\BC$, then $W=U\oplus \overline U$, where $U$ and $\overline U$
(the complex conjugate of $U$) are irreducible in $\BC$. In this
case $U$ is necessarily of type 1 or 3. If $U$ is of type 1, then
so is $\overline U$, and $U$ and $\overline U$ are not isomorphic.
By extension, in this case, we say that $W$ is of type 1. If $U$
is of type 3, then so is $\overline U$, and $U$ and $\overline U$
are isomorphic. By extension, in this case, we say that $W$ is of
type 3.
 \ali

Let us now introduce some definitions. Let $n$ be an integer. We
consider $\BC^{2n}=\BC^n\oplus \BC^n$. We denote by $( , )$ the
canonical symmetric bilinear form on $\BC^{2n}$, $(X,Y)=X^t Y$,
and by $<,>$ the canonical hermitian scalar product, $<X,Y>=(X,
\overline Y)$. We denote $\Id_n$ the $n\times n$ identity matrix
and by $J_n$ the $2n\times 2n$ antisymmetric matrix defined by
\begin{eqnarray}\label{f.correc.1}
\left( \begin{array}{cc} 0& -\Id_n
\\
\Id_n & O
\end{array}\right).
\end{eqnarray}

We define three types of Grassmannian on $\BC^{2n}$. We first
denote by ${\Bbb G}^{n,2n}$, the Grassmannian of $n$ dimensional
subspaces of $\BC^{2n}$. The group $Gl(2n, \BC)$ acts transitively
on ${\Bbb G}^{n,2n}$, and ${\Bbb G}^{n,2n}$ is isomorphic to the
homogeneous space $Gl(2n, \BC)/P_n$, where
$$
P_n=\{\left( \begin{array}{cc} A& B \\ C& D\end{array}\right)\in
Gl(2n,\BC), \;\;\; C=0\}.
$$
In particular, the tangent space of ${\Bbb G}^{n,2n}$ is
isomorphic to $M_n(\BC)$.
 \ali
We recall that we defined ${\Bbb L}^n$ as the set of Lagrangian
subspaces of $\BC^{2n}$, i.e. as the set of $n$-dimensional
isotropic subspaces for the canonical symplectic form $\w(X,Y)=
(X, J_n Y)$. We denote by $Sp(n,\BC)$ the linear symplectic group,
i.e. the group of complex $2n\times 2n$ matrices $S$, such that
$S^t J_n S=J_n$. The group $Sp(n,\BC)$ acts transitively on the
set of Lagrangian subspaces of $\BC^{2n}$, and ${\Bbb L}^n$ is
isomorphic to the homogeneous spaces $Sp(n,\BC)/P_n$, where $P_n$
is the parabolic subgroup
$$
P_n=\{\left( \begin{array}{cc} A& B \\ C& D\end{array}\right)\in
Sp(n,\BC), \;\;\; C=0\}.
$$
In particular, the tangent space of ${\Bbb L}^n$ is isomorphic to
$\sym_n(\BC)$, the space of complex symmetric $n\times n$
matrices.
 \ali
We consider now the non degenerate symmetric bilinear form $(X,K_n
Y)$ on $\BC^{2n}$, where $K_n$ is the symmetric matrix
\begin{eqnarray}\label{f.correc.Kn}
\left( \begin{array}{cc} 0& \Id_n
\\
\Id_n & O
\end{array}\right).
\end{eqnarray}
We denote by ${\Bbb O}^n$ the set of maximal isotropic subspaces,
for the symmetric bilinear form $(\cdot, K_n \cdot)$, i.e., the
set of $n$-dimensional isotropic subspaces of $\BC^{2n}$. The
group $O(\BC^{2n}, K_n)$ of linear transformation which preserve
the bilinear form $(\cdot, K_n\cdot)$ is isomorphic to the
classical orthogonal group $O(2n,\BC)$ (indeed, $(\cdot ,
K_n\cdot)$ is symmetric and non degenerate) and acts transitively
on ${\Bbb O}^n$. Hence, ${\Bbb O}^n$ is isomorphic to $O(\BC^{2n},
K_n)/P_n$, where $P_n$ is the parabolic subgroup which leaves
invariant the isotropic subspace $\BC^n\oplus 0$,
\begin{eqnarray}\label{f.correc.PO}
P_n=\{\left( \begin{array}{cc} A& B \\ C& D\end{array}\right)\in
O(2n,\BC), \;\;\; C=0\}.
\end{eqnarray}
In particular, the tangent space of ${\Bbb O}^n$ is isomorphic to
the space of $n\times n$ complex skew symmetric matrices. We
denote by $SO(\BC^{2n}, K_n)$ the subgroup of $O(\BC^{2n},K_n)$ of
element of determinant 1. Since $P_n\subset SO(\BC^{2n},K_n)$,
${\Bbb O}^n$ has two connected components, $SO(\BC^{2n},K_n)/P_n$
and $\ksi SO(\BC^{2n},K_n)/\ksi P_n$, where $\ksi$ is any element
of $O(\BC^{2n},K_n)$ with determinant -1. We denote by ${\Bbb
S}{\Bbb O}^n=SO(\BC^{2n},K_n)/P_n$, the connected component which
contains the $\Id\cdot P_n$.
 \ali

We suppose now that $k$ is an integer and that $\BR^k$ is an
orthogonal representation of a finite group $G$, i.e. we consider
$G$ as a finite subgroup of the orthogonal group $O(k,\BR)$. The
vector space $\BC^k$ is a complex representation of $G$, and we
consider the diagonal action of $G$ on $\BC^{2k}=\BC^k\oplus
\BC^k$, i.e. $G$ is the subgroup of $O(2k)$ of elements of the
form
$$
\left(\begin{array}{cc} g & 0 \\ 0 & g \end{array}\right), \;\;\;
\forall g\in G.
$$
We consider the Lagrangian Grassmannian $\BL^k\simeq
Sp(k,\BC)/P_k$. We denote by $\hat \BL^G\subset \BL^k$ the
subvariety of $G$-invariant Lagrangian subspaces. We know that
$\sym_k(\BC)$, the space of $k\times k$ symmetric complex
matrices, is embedded in $\BL^k$ as follows
\begin{eqnarray*}
\sym_k(\BC)&\rightarrow &\BL^k
\\
Q&\rightarrow& L_Q=\vect\{e_i+\sum_{j=1}^k Q_{i,j}e'_j\},
\end{eqnarray*}
where $\{e_1, \ldots ,e_k,e'_1, \ldots ,e'_k\}$ is the canonical
basis of $\BC^{2k}$. We denote by $\sym^G(\BC)$ the subspace of
$\sym_k(\BC)$ of $Q$ which commutes with $G$. It is clear that the
Lagrangian subspace $L_Q$ is invariant under $G$ if and only if
$Q\in\sym^G(\BC)$. We denote by $\BL^G$ the closure in $\BL^k$ of
$\sym^G(\BC)$. It is clear that $\BL^G\subset \hat \BL^G$.
 \ali

We suppose that $\BR^k$ is decomposed into isotopic
representations as follows,
\begin{eqnarray}\label{f.correc.decomp}
\BR^k=V_0\oplus \cdots \oplus V_r,
\end{eqnarray}
where $V_i$ is an isotopic representation equal to the direct sum
of $n_i$ representations isomorphic to a single $\BR$-irreducible
representation $W_i$ (and the $W_i$ are not isomorphic). This
means that $V_i=W_{i,1}\oplus\cdots \oplus W_{i,n_i}$, where the
$W_{i,j}$ are isomorphic to $W_i$ and we can choose the $W_{i,j}$
orthogonal. For simplicity, we write sometimes $V_i=n_iW_i$.

\begin{thm}
The subvariety $\BL^G$ is the connected component of $\hat \BL^G$
which contains the Lagrangian subspace $\BC^k\oplus 0$, and is
isomorphic to
$$
{\mathcal L}_0\times \cdots \times {\mathcal L}_r,
$$
where
\begin{itemize}
\item
${\mathcal L}_i\simeq {\Bbb G}^{n_i, 2n_i}$ if $W_i$ is of type 1;
the dimension of ${\mathcal L}_i$ is $n_i^2$.
\item
${\mathcal L}_i\simeq \BL^{n_i}$ if $W_i$ is of type 2; the
dimension of ${\mathcal L}_i$ is $n_i(n_i+1)/2$.
\item
${\mathcal L}_i\simeq {\Bbb S}{\Bbb O}^{2n_i}$ if $W_i$ is of type
3; the dimension of ${\mathcal L}_i$ is $2n_i^2-n_i$.
\end{itemize}
\end{thm}
Proof: Let us first introduce some notations.  If $M=(M_{i,j})$ is
a $n\times n$ matrix and $N$ a $m\times m$ matrix, we denote by
$M\otimes N$, the $mn\times mn$ matrix of the tensor product,
given by blocks by
$$
\left( \begin{array}{ccc} &&\\ & M_{i,j}N & \\
&&\end{array}\right)_{i,j=1\cdots n}.
$$

If $U$ is a subspace of $\BC^k$, then we will denote by $U$ and
$U'$ the subspaces of $\BC^{2k}$, equal respectively to the copy
of $U$ on the first and the second component of $\BC^k\oplus
\BC^k$, so that we have $U'=J_kU$. Similarly, if $f$ is a vector
of $\BC^k$, we denote by $f$ and $f'$, respectively the copy of
$f$ on the first and second component of $\BC^k\oplus \BC^k$, so
that $f'=Jf$. We denote by $V_i^\BC$ the complexification of
$V_i$. Hence, with the previous notations we have
\begin{eqnarray}\label{f.correct.decompcomp}
\BC^{2k}=V_0^\BC\oplus \cdots \oplus V_r^\BC\oplus
(V_0^\BC)'\oplus \cdots \oplus (V_r^\BC)'.
\end{eqnarray}

Let us first prove that $\BL^G$ is a connected component of $\hat
\BL^G$. The space $\sym_k(\BC)$ is embedded in $\BL^k$ and, with
this embedding,
$$
\sym_k(\BC)=\BL^k\setminus \{L\in \BL^k, \; L\cap (0\oplus
\BC^k)\neq \{0\}\}.
$$
As we already remarked, we know that a Lagrangian subspace of the
type $L_Q$ is in $\hat \BL^G$ if and only if $Q$ is in
$\sym^G(\BC)$. Otherwise stated, it means that
$$
\sym^G(\BC)=\hat \BL^G \setminus \{L \in \hat \BL^G, \; L\cap
(0\oplus \BC^k)\neq \{0\}\}.
$$
Since $\sym^G(\BC)$ is connected, it means that $\BL^G$ is a
connected component of $\hat \BL^G$. We set
$$Sp^G(\BC)=\{S\in Sp(k, \BC), \;\; g S=Sg, \; \forall g\in G\},
$$
and $P^G=P_k\cap Sp^G$. Let us now prove that  $\BL^G$ is
isomorphic to the connected component of $Sp^G/P^G$ which contains
$\Id\cdot P^G$. Indeed, $Sp^G/P^G$ is isomorphic to the subset of
$\BL^k\simeq Sp(k,\BC)/P_k$ of Lagrangian subspaces $L$ such that
there exists $S$ in $Sp^G(\BC)$, such that
$$
S(\BC^k\oplus 0)=L.
$$
Hence $Sp^G/P^G\subset \hat \BL^G$. But for $Q$ in $\sym^G(\BC)$,
$$
S=\left(\begin{array}{cc} \Id_k &0 \\ Q & \Id_k\end{array}\right),
$$
is in $Sp^G(\BC)$ and $S(\BC^k\oplus 0)=L_Q$. Hence,
$\sym^G(\BC)\subset Sp^G/P^G$, and thus  $\BL^G\subset Sp^G/P^G$,
since $Sp^G/P^G$ is compact. This implies that $\BL^G$ is a
connected component of $Sp^G(\BC)/P^G$.
 \ali
\Rm: It is not true in general that $\hat \BL^G\simeq Sp^G/P^G$.
Actually, we have the inclusions $\BL^G\subset Sp^G/P^G\subset
\hat \BL^G$, and each of these inclusions can be strict. We will
try to clarify this point in a subsequent work.

Let us first prove that we can reduce the problem to the case
where $\BR^k$ contains a unique type of irreducible
representation, i.e. to the case where $r=0$.  Since the subspaces
$V_i$, $V_i'$ are real orthogonal, the subspaces $V_i^\BC\oplus
(V_i^\BC)'$ are $\w$-orthogonal and the restriction of the
symplectic form $\w$ to $V_i^\BC\oplus (V_i^\BC)'$ is non
degenerated, hence is a symplectic form. This implies that
$$
Sp^G(\BC)\simeq Sp^G(V_0^\BC\oplus (V_0^\BC)')\times \cdots \times
Sp^G(V_r^\BC\oplus (V_r^\BC)'),$$
 and
$$
P^G\simeq P^G(V_0^\BC\oplus (V_0^\BC)')\times \cdots \times
P^G(V_r^\BC\oplus (V_r^\BC)'),
$$
where $Sp^G(V_i^\BC\oplus (V_i^\BC)')$ is the group of
$G$-invariant symplectic transformation on $V_i^\BC\oplus
(V_i^\BC)'$, and $P^G(V_i^\BC\oplus (V_i^\BC)')$ the subgroup
which leaves invariant $V_i^\BC$. Hence, $\BL^G\simeq {\mathcal
L}_0\times \cdots \times {\mathcal L}_r$, where ${\mathcal L}_i$
is the connected component of $Sp^G(V_i^\BC\oplus
(V_i^\BC)')/P^G(V_i^\BC\oplus (V_i^\BC)')$, which contains
$\Id\cdot P^G(V_i^\BC\oplus (V_i^\BC)')$.

Hence, we suppose now that $\BR^k$ contains a unique type of
irreducible representations, i.e. that $\BR^k=nW=W_1\oplus \cdots
\oplus W_n$. This means that we have the decomposition
\begin{eqnarray}\label{f.correc.decompnW}
\BC^{2k}=W_1^\BC\oplus \cdots \oplus W_n^\BC\oplus
(W_1^\BC)'\oplus \cdots\oplus (W_n^\BC)'.
\end{eqnarray}
 \ali

{\it If $W$ if of type 2.}
 \ali
This is the simplest situation. In this case  $W_j^\BC$ and
$(W_j^\BC)'$ are irreducible over $\BC$. Let us set $p=\dim W$,
and choose real orthonormal basis $(g_{1,j}, \ldots ,g_{p,j})$ of
$W_j$, which realize the isomorphism $W_j\simeq W_{j'}$. We denote
by $(g_{i,j}')$ the corresponding basis of $W_j'$. By Schur lemma,
we know that, in this base, any element of
$\hbox{End}^G(\BC^k\oplus \BC^k)$ (where $\hbox{End}^G $ is the
space of endomorphism commuting with $G$), is of the form
$$
M\otimes \Id_p,
$$
where $M\in M_{2n}(\BC)$. Since the change of base is orthogonal,
the matrix of the symplectic form $\w$ remains equal to
$$
J_k= J_n\otimes \Id_p.
$$
This implies that, by this change of bases, $Sp^G(\BC)$ becomes
the space of matrix of the type $M\otimes \Id_p$ such that
$$(M\otimes \Id_p)^t (J_n\otimes \Id_p) (M\otimes \Id_p)= J_n\otimes
\Id_p,
$$
which is equivalent to $M^t J_n M=J_n$. Hence $Sp^G(\BC)\simeq
Sp(n,\BC)$. Similarly, $P^G$ is isomorphic to $P_n$ and
$\BL^G\simeq \BL^n$, since $Sp(n,\BC)/P_n$ is connected.
 \ali

{\it If $W$ is of type 1.}
 \ali
In this case, $W^\BC=U\oplus \overline U$, and $U$ and $\overline
U$ are $\BC$-irreducible, not isomorphic, and orthogonal for the
hermitian scalar product $<,>$. Let $p=\dim W$, and choose an
orthonormal basis $(g_1, \ldots ,g_p)$ of $W$ (for the hermitian
scalar product). The family $(\overline g_1, \ldots ,\overline
g_p)$ is an orthonormal basis of $\overline U$.

By isomorphism, we have the corresponding decomposition
$W_j^\BC=U_j\oplus \overline U_j$, $(W_j^\BC)'=U_j'\oplus
\overline U_j'$, and the corresponding basis $(g_{1,j}, \ldots
,g_{p,j})$, $(\overline g_{1,j}, \ldots ,\overline g_{p,j})$, and
$(g'_{1,j}, \ldots ,g'_{p,j})$, $(\overline g'_{1,j}, \ldots
,\overline g'_{p,j})$. We rewrite now the decomposition of
$\BC^{2k}$ as
$$
\BC^{2k}=U_1\oplus \cdots \oplus U_n\oplus U_1'\oplus \cdots
\oplus U_n'\oplus \overline U_1\oplus \cdots \oplus \overline
U_n\oplus \overline U_1' \oplus \cdots \oplus \overline U_n',
$$
and we denote by ${\mathcal B}$ the corresponding basis (i.e. we
endow each component with the basis $(g_{i,j})$, $(g'_{i,j})$,
$\cdots$, we just described).

By Schur lemma, in this basis, any element of
$\hbox{End}^G(\BC^k\oplus \BC^k)$ has the form
\begin{eqnarray}\label{f.correc.3}
Z=\left( \begin{array}{cc}Z_1\otimes \Id_p & 0
\\
0& Z_2\otimes \Id_p\end{array}\right),
\end{eqnarray}
where $Z_1$ and $Z_2$ are $2n\times 2n$ complex matrices. Let us
now compute the matrix of the symplectic form $\w$ in this new
basis (the change of basis, from the canonical basis of $\BC^{2k}$
to ${\mathcal B}$ is unitary, but not orthogonal).
 Since $\w(X,Y)=(X,JY)=<X, \overline{JY}>$, we see that $\w$
is null on all term except on the  $U_j\times \overline U'_{j}$,
$U'_j\times \overline U_{j}$, and the symmetric terms.  On
$U_j\times \overline U_j'$, we have
$$
\w(g_{i,j}, \overline g'_{i',j})=-<g_{i,j},
g_{i',j}>=-\delta_{i,i'}.
$$
On $U'_{j}\times \overline U_{j}$, we have $\w( g'_{i,j},
\overline g_{i',j})=-\overline{\w(g_{i',j},
g'_{i,j})}=\delta_{i,i'}$. Hence, the matrix of $\w$ in the base
${\mathcal B}$ is
$$
\hat J=\left( \begin{array}{cc} 0& J_n\otimes \Id_p
\\
J_n\otimes \Id_p& 0
\end{array}\right).
$$
Hence, by this change of basis, $Sp^G(\BC)$ is isomorphic to the
group of matrices $Z$ of the form (\ref{f.correc.3}), which
satisfies $Z^t \hat J Z=\hat J$. Hence $Sp^G(\BC)$ is isomorphic
to
\begin{eqnarray}\label{f.correc.5}
\{ Z=\left( \begin{array}{cc}Z_1 & 0
\\
0& Z_2
\end{array}\right), \;\; Z_1, Z_2 \in M_{2n}(\BC), \;\;
Z_2^t J_n Z_1 =J_n\}.
\end{eqnarray}
Similarly, $P^G$ is isomorphic to the subgroup of element $Z$ of
(\ref{f.correc.5}), such that $Z_1$ and $Z_2$ are of the form
$$
\left( \begin{array}{cc} *&*\\0&*\end{array}\right).
$$
Since $Z_2$ is determined by $Z_1$ in (\ref{f.correc.5}) by
$Z_2=-J_n(Z_1^t)^{-1}$, we see that $Sp^G(\BC)/P^G$ is isomorphic
to
\begin{eqnarray*}
Gl(2n, \BC)/ \left\{Z\in Gl(2n, \BC), \;\; Z \hbox{ of the form }
\left( \begin{array}{cc} *&*\\0&*\end{array}\right)\right\}
&\simeq & {\Bbb G}^{n,2n}.
\end{eqnarray*}
 \ali

{\it If $W$ is of type 3.}
 \ali
In this case $W=U\oplus \overline U$, where $U$ and $\overline U$
are irreducible and isomorphic. Let us first remark that $U$ and
$\overline U$ are necessarily orthogonal for the hermitian scalar
product $<,>$. Indeed, by proposition \ref{p.FrobSchur}, we know
that the symmetric scalar product $(,)$ is null on $U$, thus
$<x,y>=(x,\overline y)=0$ for $x\in U$ and $y\in \overline U$. Let
us now describe an explicit isomorphism between $U$ and $\overline
U$. By proposition \ref{p.FrobSchur}, we know that there exists a
non degenerate $G$-invariant skew symmetric bilinear form $B$ on
$U$. Thus, for all $x$ in $U$, there exists $\phi(x)\in U$ such
that
$$
B(x,y)=\overline{<\phi(x),y>}, \;\;\; \forall y\in U.
$$
The map $\phi$ is antilinear (i.e. $\phi(\lambda x+y)= \overline
\lambda \phi(x)+\phi(y)$) and bijective, since $B$ is non
degenerate. Thus, $\psi=\overline \phi:U\rightarrow \overline U$,
is an isomorphism, commuting with the action of $G$. Since $B$ is
skew-symmetric and non degenerate, $\dim U=2p$, and there exists a
symplectic basis $(g_1, \ldots ,g_{2p})$ of $U$, i.e. a basis such
that
$$
B(g_i,g_j)= \delta_{i,j+p}-\delta_{i,j-p},
$$
(i.e., the matrix of $B$ in this basis is $J_p$). We set
$f_i=\psi(g_i)$. The family $(f_1, \ldots,f_{2p})$ is a basis of
$\overline U$, which realizes the isomorphism $U\simeq \overline
U$.

Let us come back to $\BC^{2k}$. In each term  $W_j^\BC$ (resp.
$(W_j^\BC)'$) of the decomposition (\ref{f.correc.decompnW}) we
make the corresponding decomposition $W_j^\BC=U_j\oplus \overline
U_j$ (resp. $(W_j^\BC)'=U_j'\oplus \overline U_j'$), and we define
the corresponding basis $(g_{1,j},\ldots , g_{2p,j})$ and
$(f_{1,j},\ldots ,f_{2p,j})$ (resp. $(g_{1,j}', \ldots
,g_{2p,j}')$ and $(f_{1,j}', \ldots ,f'_{2p,j})$).

We rewrite the decomposition of $\BC^{2k}$ in
$$
\BC^{2k}=U_1\oplus \cdots \oplus U_n\oplus \overline U_1\oplus
\cdots \oplus \overline U_n\oplus  U_1'\oplus \cdots \oplus
U'_n\oplus \overline U_1' \oplus \cdots \oplus \overline U_n',
$$
and we denote by $\mathcal B$ the corresponding basis (i.e., we
endow each component by the basis $(g_{1,j}, \ldots ,g_{2p,j})$,
$(f_{1,j},\ldots ,f_{2p,j})$, $\ldots$, we just described). By
Schur lemma, the matrix of any element of $\hbox{End}^G(\BC^{2k})$
in the base $\mathcal B$, is of the type $M\otimes \Id_{2p}$,
where $M\in M_{4n}(\BC)$.

 Let us
now compute the matrix of $\w$ in the basis $\mathcal B$. Clearly,
since $\w(X,Y)=(X,\overline{JY})$, we see that $\w$ is null on all
component except of $\BC^{2k}\times \BC^{2k}$ except on components
of the type $U_j\times \overline U_j'$, $\overline U_j\times
U_j'$, and the symmetric components. On $U_j\times \overline
U_j'$, we have
\begin{eqnarray*}
\w(g_{i,j}, f_{i,j'}')&=& -<g_{i,j}, \overline f_{i',j}>
\\
&=& -\overline{ < \phi(g_{i',j}), g_{i,j}>}
\\
&=& -B(g_{i',j}, g_{i,j})= \delta_{i,i'+p}-\delta_{i,i'-p}.
\end{eqnarray*}
Similarly, $\w(f_{i,j}, g'_{i',j})=
-(\delta_{i,i'+p}-\delta_{i,i'-p})$. Thus, the matrix of $\w$ in
$\mathcal B$ is
$$
\hat J=\left(\begin{array}{cc} 0& -J_n\otimes J_p
\\ J_n\otimes J_p & 0\end{array}\right)= \hat K\otimes J_p,
$$
where $\hat K$ is the symmetric matrix
$$
\hat K=\left(\begin{array}{cc} 0& -J_n
\\ J_n & 0\end{array}\right).
$$
Hence, we see that $Sp^G(\BC)$ is isomorphic to the set of
matrices of the form $M\otimes \Id_{2p}$ such that
$$
(M\otimes \Id_{2p})^t (\hat K\otimes J_p)(M\otimes \Id_{2p})= \hat
K\otimes J_p.
$$
Since the first term is equal to $(M^t \hat K M)\otimes J_p$, we
see that the condition becomes
$$
M^t \hat K M=\hat K.
$$
By the orthogonal change of base given by
$$
O=\left( \begin{array}{cc} \Id_{2n}&0 \\ 0& J_n\end{array}\right),
$$
we see that $\hat K$ becomes
$$
O^t \hat K O =K_{2n},
$$
defined in formula (\ref{f.correc.Kn}). Thus $Sp^G(\BC)$ is
isomorphic to $O(\BC^{4n}, K_{2n})$. Similarly, $P^G$ is
isomorphic to $P_{2n}$ defined in formula (\ref{f.correc.PO}).
Thus, $Sp^G(\BC)/P^G\simeq {\Bbb O}^{2n}$ and $\BL^G\simeq {\Bbb
S}{\Bbb O}^{2n}$.$\Box$


\footnotesize
\begin{thebibliography}{99}
\bibitem{Alexander}
S. ALEXANDER,
{\it Some properties of the spectrum of the Sierpinski gasket in a
magnetic field,}
Phys. Rev., B29 (1984), 5504-5508.
\bibitem{BarlowP}
M.T. BARLOW, E. PERKINS,
{\it Brownian motion on the Sierpi\'nski gasket.}
Probab. Theory Related Fields 79 (1988), no. 4, 543--623.
\bibitem{BarlowK}
M. T. BARLOW and J. KIGAMI,
{\it Localized eigenfunctions of the Laplacian
on p.c.f. self-similar sets},
J. Lond. Math. Soc., 56 (2)(1997), n 2, 320-332.
\bibitem{Bellissard}
J. BELLISSARD, {\it Renormalization group analysis and quasicrystals.}
Ideas and methods in quantum and statistical physics (Oslo, 1988), 118--148,
Cambridge Univ. Press, Cambridge, 1992.
\bibitem{Berezin}
F. A. BEREZIN,
The method of second quantization. Translated from the Russian by Nobumichi Mugibayashi and Alan Jeffrey. Pure and
Applied Physics, Vol. 24 Academic Press, New York-London 1966 xii+228 pp.
\bibitem{LascouxB}
Marcel  BERGER, Alain LASCOUX,
Variétés Kähleriennes compactes. (French)
Lecture Notes in Mathematics, Vol. 154
Springer-Verlag, Berlin-New York 1970 vii+83 pp
\bibitem{Carlson}
David CARLSON, {\it What are Schur complements, anyway?} Linear
Algebra Appl. 74 (1986), 257--275.
\bibitem{CarmonaL}
R. CARMONA and J. LACROIX, Spectral Theory of Random Schr\"odinger
Operators, Probabilities and applications, Birkha\"user, Boston,
1990.
\bibitem{Collin1}
Y. COLIN DE VERDI\`ERE, {\it R\'eseaux \'electriques planaires I},
Commentarii Math. Helv., 69 (1994), 351-374.
\bibitem{Collin2}
Y. COLIN DE VERDI\`ERE, {\it D\'eterminants et int\'egrales de
Fresnel}, Ann. Inst. Fourier, 49, 3 (1999), 861-881.
\bibitem{Demailly1} Jean-Pierre DEMAILLY,
{\it Monge-Amp\`ere operators, Lelong numbers and intersection
theory}, complex analysis and geometry, Univ. Ser. Math., Plenum
Press, p.115-193, 1993.
\bibitem{DFavre}
Jeffrey DILLER, Charles FAVRE,
{\it Dynamics of bimeromorphic maps of surfaces},
Preprint
\bibitem{Favre1}
Charles FAVRE, {\it Dynamique des applications rationelles.}
PhD thesis, Universit\'e Paris-Sud-Orsay.
\bibitem{FavreG}
Charles FAVRE, Vincent GUEDJ, {\it Dynamique des  applications rationelles
des espaces multi-projectifs,}
To appear in Indiana Math. J.
\bibitem{Forman1}
R. FORMAN,
{\it Functional determinants and geometry},
Invent. Math., 88, 447-493 (1987).
\bibitem{FSibony2}
John Erik FORNAESS, Nessim SIBONY,
{\it Complex dynamics in higher dimension II.}
Modern methods in complex analysis (Princeton, NJ,
1992), 135--182, Ann. of Math. Stud., 137, Princeton Univ. Press, Princeton, NJ, 1995
\bibitem{Fukushima1}
M. FUKUSHIMA, Y. OSHIMA and M. TAKEDA,
Dirichlet forms and symmetric Markov processes,
de Gruyter Stud. Math. 19, Walter de Gruyter, Berlin, New-york, 1994.
\bibitem{Fukushima2}
M. FUKUSHIMA, {\it Dirichlet forms, diffusion processes and spectral
 dimensions
for nested fractals,} in : Ideas and Methods in Mathematical analysis,
Stochastics and Applications, Proc. Conf. in Memory of Hoegh-Krohn,
vol. 1 (S. Albevario et al., eds.),
Cambridge Univ. Press, Cambridge, 1993, pp 151-161.
\bibitem{FukuShima1}
M. FUKUSHIMA and T. SHIMA,
{\it On the spectral analysis for the Sierpinski gasket,}
Potential Analysis 1 (1992), 1-35.
\bibitem{FukuShima2}
M. FUKUSHIMA and T. SHIMA,
{\it
On the discontinuity and tail behaviours of the
integrated density of states for nested pre-fractals.},
Comm. Math. Phys., 163, 461-471 (1994).
\bibitem{Guedj1}
Vincent GUEDJ,
{\it Representation theorems for positive closed $(1,1)$-currents on
flag manifolds of $GL_m(\BC)$,} Preprint.
\bibitem{GriffithsH}
Phillip GRIFFITS, Joseph HARRIS,
Principles of algebraic geometry.
Wiley Classics Library. John Wiley \&
Sons, Inc., New York, 1994. xiv+813 pp
\bibitem{Hambly}
B. HAMBLY, {\it Brownian motion on a homogeneous random fractal.}
Probab. Theory Related Fields 94 (1992), no. 1, 1-38.
\bibitem{Hormander}
Lars H\"ORMANDER, Notions of convexity.
Progress in Mathematics, 127. Birkhäuser Boston, Inc., Boston, MA, 1994. viii+414 pp
\bibitem{Kigami1}
J. KIGAMI,  {\it Harmonic calculus on p.c.f. self-similar sets,}
Trans. Am. Math. Soc., 335:721-755, 1993.
\bibitem{Kigami3}
J. KIGAMI,
{\it Distribution of localized eigenvalues of Laplacians on post-critically
finite self-similar sets,}
J. Funct. Anal. 159 (1998), n 1, 170-198.
\bibitem{KigamiL}
J. KIGAMI and
M. L. LAPIDUS,
{\it Weyl's problem for the spectral distribution of Laplacians on p.c.f.
self-similar fractals,}
Commun. Math. Phys. 158, No 1,
(1993), 93-125.
\bibitem{Klein}
Abel KLEIN, {\it
Extended states in the Anderson model on the Bethe lattice.}
 Adv. Math. 133 (1998), no. 1, 163--184.
\bibitem{Koranyi}
A. KOR\'ANYI, {\it A Schwartz lemma for bounded symmetric
domains}, Proc. Am. Math. Soc. 17 (1966), 210-213.
\bibitem{Kusuoka} S.
KUSUOKA, {\it Dirichlet forms on fractals and products of random
matrices.}
 Publ. Res. Inst. Math. Sci. 25 (1989), no. 4, 659--680.
\bibitem{Lindstrom}
T. LINDSTR\O M.
{\it Brownian motion on nested fractals.}
Mem. Amer. Math. Soc., 420, 1990.
\bibitem{Metz2}
Volker METZ,
{\it Shorted operators: an application in potential theory.}
Linear Algebra Appl. 264 (1997), 439--455.
\bibitem{PasturF}
L. PASTUR, A. FIGOTIN,
Spectra of Random and Almost-Periodic Operators,
Grundlehren der mathematischen Wissenschaften,
297, Springer-Verlag, Berlin Heidelberg 1992.
\bibitem{Rammal}
R. RAMMAL,
{\it
Spectrum of harmonic excitations on fractals},
J. de Physique 45, 191-206 (1984).
\bibitem{RammalT}
R. RAMMAL and G. TOULOUSE, J. Phys. Lett., 44 (1983), L-13
\bibitem{Sabot1}
C. SABOT,
{\it
Existence and uniqueness of diffusions on
finitely ramified self-similar fractals,}
in Ann. Scient. Ec. Norm. Sup.,
4\`eme s\'erie, t. 30, 1997, p. 605 \`a 673.
\bibitem{Sabot2}
C. SABOT,
{\it Espaces de Dirichlet reli\'es par des points
et application aux diffusions sur les fractals finiment ramifi\'es.}
Potential Analysis, 11 (1999), n 2, 183-212.
\bibitem{Sabot3}
C. SABOT,
{\it
Integrated density of states of self-similar Sturm-Liouville
operators
and holomorphic dynamics in higher dimension},
Ann. Inst. H. Poincaré Probab.
Statist 37 (2001), no. 3, 275-311.
\bibitem{Sabot4}
C. SABOT,
{\it Pure point spectrum for the Laplacian on unbounded nested fractals.}
 J. Funct. Anal. 173 (2000), no. 2, 497--524.
\bibitem{Sabot6}
C. SABOT,
{\it
Schr\"odinger operators on fractal lattices
with random blow-up},
Prerint.
\bibitem{Sabot7}
C. SABOT, {\it Spectral Analysis of a self-similar Sturm-Liouville
operator}, preprint.
\bibitem{SankaranV}
P. SANKARAN, P.  VANCHINATHAN,
{\it Small resolutions of Schubert varieties in symplectic and orthogonal Grassmannians.}
 Publ.
Res. Inst. Math. Sci. 30 (1994), no. 3, 443--458
\bibitem{Seneta}
E. SENETA,
Nonnegative matrices and Markov chains. Second edition.
Springer Series in Statistics. Springer-Verlag, New York, 1981.
xiii+279 pp.
\bibitem{Serre}
J.P. SERRE,
Linear Representations of Finite Groups,
Graduated Texts in Mathematics,
Springer-Verlag.
\bibitem{Sibony1}
Nessim SIBONY, {\it Dynamique des applications rationnelles de
$\bold P\sp k$.} (French) Dynamique et géométrie complexes (Lyon,
1997), ix--x, xi--xii, 97--185, Panor. Synthèses, 8, Soc. Math.
France, Paris, 1999.
\bibitem{Siegel}
Carl Ludwig SIEGEL, {\it Symplectic geometry}. Amer. J. Math. 65,
(1943), 1--86.
\bibitem{Wang} J.
SJ\"OSTRAND, W.M. WANG, {\it Exponential decay of averaged Green
functions for random Schrödinger operators. A direct approach.}
Ann. Sci. École Norm. Sup. (4) 32 (1999), no. 3.
\bibitem{Strichartz1}
R. S. STRICHARTZ, {\it Fractals in the large}, Canad. Math. J., 50 (
1998), n 3,
638-657.
\bibitem{Teplyaev}
A. TEPLYAEV, {\it Spectral Analysis on infinite Sierpinski Gasket},
J. Funct. Anal., 159 (1998), n 2, 537-567.
\bibitem{Terras}
A. TERRAS, {Harmonic analysis on symmetric spaces and
applications, I, II,} Spinger-Verlag New-York Inc.
\bibitem{Wang2}
W. M. WANG,
{\it
Localization and universality of Poisson
statistics for the multidimensional Anderson model at
weak disorder,}
Invent. Math. 146,
365-398 (2001).
\end{thebibliography}
\end{document}